\documentclass[12pt,preprint]{aastex}

\usepackage{amsmath,amssymb}

\begin{document}
\title{Attenuation caused by a Distant Isothermal Turbulent Screen}

\author{J\"org Fischera \& Michael Dopita}

\affil{Research School of Astronomy \& Astrophysics,
Institute of Advanced Studies, The Australian National University,
Cotter Road, Weston Creek, ACT 2611 Australia}
\email{fischera,mad@mso.anu.edu.au}

\begin{abstract}
We analyse in detail the attenuation caused by an isothermal turbulent distant foreground dust screen. The attenuation curve is well determined by two parameters, the absolute-to-relative attenuation ratio $\rm R^A_V=A_V/E(B-V)$ and the absolute attenuation $\rm A_V$. We show quantitatively how these two observable quantities depend on the statistical properties of the local density and the mean attenuation $\left<\rm A_V\right>$ and how they vary with the thickness of the screen measured in units of the largest turbulent scale. The attenuation through a turbulent medium is characterised by higher transparency and a flatter attenuation curve in comparison with a homogeneous dust screen. In general, the effect of the turbulent medium on the attenuation increases with slice thickness. In the limit of thick slices, typically larger than the maximum turbulent scale, $\rm R^A_V$ asymptotically approaches a maximum value. 
\end{abstract}

keywords{ISM:dust---dust:extinction}


\section{\label{intro}Introduction}

Dust grains in the interstellar medium attenuate the star light, in particular at UV wavelengths, and make it difficult to directly measure intrinsic parameters such as the star-formation rate of galaxies or to study highly obscured astrophysical objects like compact HII-regions. To quantify the star formation rate as a function of redshift $z$ through UV-observations, a correction for attenuation is essential but is still very uncertain. One of the reasons is that the dust obscuration at high red-shift is not well understood, partly because of the lack of a satisfying physical model of the dust attenuation.

Due to its turbulent motion, most likely dominated by MHD-turbulence, the diffuse interstellar medium shows a very inhomogeneous fractal structure which has important consequences for the dust attenuation of star light from galaxies. This attenuation is qualitatively different
to the extinction measured for single stars where the attenuation is simply proportional to the
column density and is caused by both absorption and scattering by dust grains.

By contrast the physics of the attenuation for the collected star light from galaxies is much more complex. It depends in general on the distribution of dust and gas, the variation of dust properties
and the orientation of the galaxies with respect to the observer. A first approach, shown by \citet{Calzetti2001} to be a very successful one, provides a model where all stars are seen though a foreground dust screen which can be assumed to be distant from stars. This is the model discussed in this paper. 

In case of an inhomogeneous medium the extinction varies over the observed field. 
Regions of high column densities may be optically thick in the UV but optical thin in the IR while
regions of low column densities may be optical thin over the whole wavelength range. 
Therefore, the effective extinction curve or attenuation curve for all stars seen through a 
turbulent dust screen is generally flatter than the extinction curve of single stars. 
Furthermore, since the light can escape efficiently through regions of low optical depth the attenuation of a turbulent screen is generally smaller than one would expect in case of an homogeneous ISM.

The flattening of the attenuation curve can be described by the absolute-to-relative attenuation ratio $\rm R^A_{V} = A_V/E(B-V)=\tau_V/(\tau_B-\tau_V)$ which
is determined by the effective optical depths $\tau_{\rm V}$ and $\tau_{\rm B}$ in $\rm V$ ($0.548~\mu{\rm m}$) and $\rm B$ ($0.44~\mu{\rm m}$). A flatter attenuation curve is characterised by a higher $\rm R_V^A$.
This value has to be distinguished from the $\rm R_V$-value of pure extinction curves
which is related to the extinction coefficients $k_\lambda$ by ${\rm R_V}=k_{\rm V}/(k_{\rm B}-k_{\rm V})$ and therefore can be used to infer some constraints on the dust properties.
A larger $\rm R_V$-value for example points to a larger grain population. In the limit
of a homogenous distant screen both values are the same. In a turbulent medium
the $\rm R_V^A$ will in general be larger than $\rm R_V$ of the underlying extinction curve.

Assuming that the dust properties have no spatial variation in the ISM, then the attenuation curve of the screen is determined by the probability distribution function (PDF) of the column density 
$N$ alone. As found by \citet{Ostriker2001} using MHD-simulations the PDF of the column density
of an isothermal medium should be approximately log-normal, the functional form also found
for the PDF of the local density \citep{Vazquez1994, Padoan1997, Passot1998}.

Based on this hypothesis we have shown \citep{Fischera2003} (Paper I) that the overall 
curvature of the attenuation curve, empirically derived for star-burst galaxies and known as the
`Calzetti-extinction-curve', can be naturally explained in terms of a turbulent dust screen. Below $2200$~\AA{}
the empirical curve could be reproduced and the $\rm R_V^A$-value
was consistent with the value of the `Calzetti-curve'. Furthermore, the lower values
of the Mach-numbers derived from the fit was found to be in agreement with the Mach-numbers 
of our galaxies.

To obtain a deeper understanding of the attenuation caused by a turbulent foreground screen in Paper~II we investigated how the PDF of the column density relates to the density distribution and the correlation function (the one point and the two point statistics) of the local density. 
We described how the standard deviation of the column densities varies with slice thickness and showed
that for moderate log-normal density distributions of the local density $\rho$, where its standard deviation 
$\sigma_{\rho/\left<\rho\right>}$ is not much larger than $\sim 2.5$, the PDF can well be approximated 
by a log-normal density distribution. For $\sigma_{\rho/\left<\rho\right>}>1$ the approximation 
becomes less accurate, tending to systematically underestimate the probability of encountering high column densities and therefore tending to underestimate the standard deviation.

In this paper we will analyse in detail the attenuation curve caused by an isothermal 
turbulent dust screen using the analytical formula we derived in Paper~II. We will describe
how the attenuation curve depends on the statistical properties of the
local density and how the attenuation curve varies with slice thickness.

\section{Model of the turbulent dust screen}

\label{model}
To analyse the attenuation caused by a distant turbulent screen we use an idealised
model based on the one point and two point statistic as outlined in \cite{Fischera2004}. 
The turbulent medium is taken to be isotropic and we assume that 
the power spectrum $P(\mbox{\boldmath$k$})$=$|\rho(\mbox{\boldmath$k$})|^2$ 
of the local density $\rho(\mbox{\boldmath$r$})$
is a simple power law $P(k)\propto k^{n}$ with power $n$ where the scales extend from a minimum scale $L_{\rm min}$ to a maximum scale $L_{\rm max}$. As in our previous paper we will consider a power spectrum with $n=-10/3$
and $n=-3$ which corresponds to Kolmogorov turbulence and to the isotropic MHD-model
discussed by \citet{Iroshnikov1964} and \citet{Kraichnan1965}, respectively. The turbulent medium is additionally described by the standard deviation $\sigma_{\rho/\left<\rho\right>}$ of the local density, $\rho$, where $\left<\rho\right>$ is its mean value.

In general the width of the density distribution increases as 
consequence of higher compression resulting from higher Mach numbers $M$.
In particular for non-magnetised forced turbulence it has been found  
by using 3D-simulation \citep{Padoan1997, Nordlund1999} 
that the standard deviation of the density is almost linearly correlated with Mach number:
\begin{equation}
   \label{densmachcorr}
   \sigma_{\rho}\approx \beta M\left<\rho\right>.
\end{equation}
where $\beta\approx 0.5$. The more general case appears to be more complicated as
for magnetised turbulence no simple correlation between density contrast and
Mach number has been found. However, as might be expected from the 
additional pressure support to the plasma provided by magnetic fields, it seems
that the density contrast becomes weaker when magnetised
turbulence is considered \citep{Nordlund1999, Ostriker2001}.

The actual standard deviation may change throughout the whole interstellar medium as the
physical conditions vary.
As shown in Paper~I, the dispersion velocity of the cold neutral medium (CNM) and the warm neutral medium (WNM) of our galaxy suggests a Mach number of approximately 12 and 1.8, respectively. If the relation given in equation~\ref{densmachcorr} is true the standard deviation should vary in the range $\sim 0.9$ to $\sim 6$.
More violent systems like star burst galaxies are likely to show higher Mach numbers. Their
turbulent medium may therefore be characterised by a wider distribution of the local density.

The standard deviation $\sigma_{N/\left<N\right>}$ of the normalised column density $N$ is determined by 
the thickness $\Delta/{L_{\rm max}}$ of the turbulent screen and the power spectrum of the local density. For simplicity we assume that the PDF of the column density $N$ through a turbulent 
isothermal medium is described by a simple log-normal density distribution and therefore
by the same functional form as the PDF of the local density. The PDF
of the normalised column density $\xi=N/\left<N\right>$ is then described by
\begin{equation}
  \label{eqlognormal}
  p(\ln\xi) = \frac{1}{\sqrt{2\pi}\sigma_{\ln N}}e^{-x^2/2\sigma_{\ln N}^2}
\end{equation}
with $x = \ln\xi-\ln\xi_0$ where $\ln\xi_0=-\onehalf \sigma^2_{\ln N}$.
The standard deviation $\sigma_{\ln N}$ of the log-normal density distribution is connected with the standard deviation 
$\sigma_{N/\left<N\right>}$ by:
\begin{equation}
  \label{eqstandard deviation}
  \sigma_{N/\left<N\right>}^2=e^{\sigma_{\ln N}^2}-1.
\end{equation}

For small fluctuations of the local density up to $\sigma_{\rho/\left<\rho\right>}=1$ the assumption
of a log-normal PDF seems to be very accurate. If the medium is more turbulent the PDF of the column density may show
some deviations from the log-normal distribution even though the overall shape
seems to be well preserved. We will discuss the consequences of this  in Sect.~\ref{sectdiscussaccuracy1}.

As in Paper~I we assume that the dust properties are not varying throughout the
whole turbulent density structure so that the optical depth is simply given by $\tau_\lambda=N k_\lambda$ where $k_\lambda$ is the wavelength dependent extinction coefficient. The underlying
extinction curve is assumed to be the mean extinction curve of our galaxy where we 
adopted the curve obtained by \citet{Weingartner2001}. As this curve has been derived
by fitting a dust model to the mean extinction curve it also provides further information
about the scattering properties. As this paper describes the attenuation
of a distant dust screen, dust scattered light does not contribute to the measured light.

The main parameter determining the attenuation of a distant turbulent dust screen is the PDF of
the optical depths. Due to the log-normal density distribution the light of an extended source or a distribution of stars behind the screen can suffer a large point-to-point variation of extinction, $e^{-\tau}$. The amplitude of this variation will depend on the strength of the fluctuation of the column densities. 
On average the turbulent screen leads to an effective extinction $\tau_{\rm eff}$ given by:
\begin{eqnarray}
  \label{eqtaueff}
   \tau_{\rm eff}&=& -\ln{\left(\int{\rm 
d}\ln(\xi)\,p(\ln(\xi))\,e^{-\xi\left<\tau\right>}\right)}\\
&=& -\ln\left(\int{\rm d}y\,p(y)\,e^{-e^y\left<\tau\right>}\right),
\end{eqnarray}
 where $e^y = \xi=\tau/\left<\tau\right>$.
 
\begin{figure*}
 \begin{center}
  \includegraphics[width=0.9\hsize]{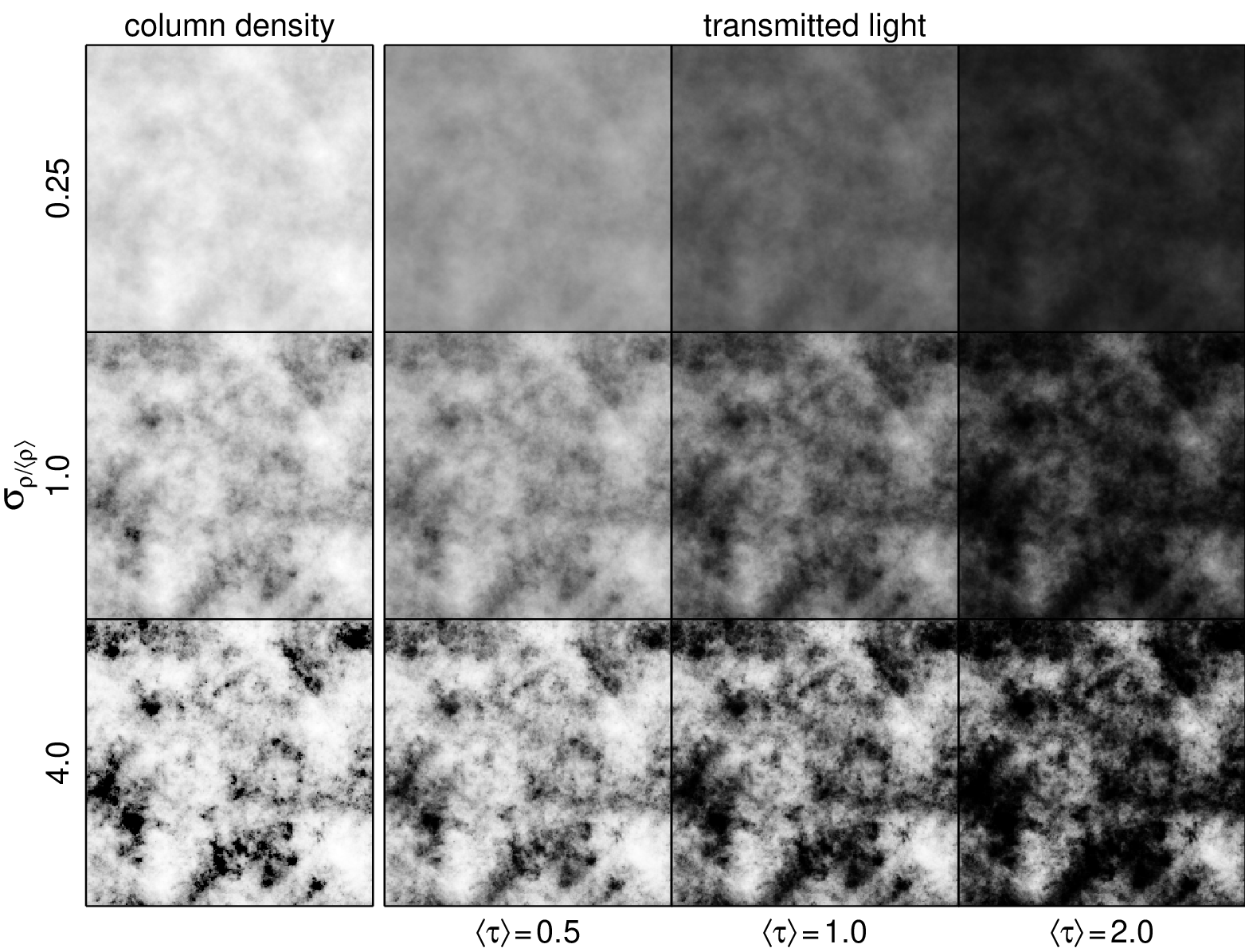}
  \caption{\label{slabattenvisu}
  Visualisation of the attenuation caused by a turbulent distant screen. The turbulent
  density structure has been derived using a volume of $216^2\times108$ pixels. The
  thickness of the screen is taken to be one maximum scale $L_{\rm max}$
  and the width of the screen area is chosen to be two maximum scales. The power
  spectrum of the normal distributed density values is taken to be Kolmogorov.
  The first column shows the
  column density per pixel for three different values of the standard deviation 
  $\sigma_{\rho/\left<\rho\right>}$ of the local density.
  Bright regions are locations of low column densities, dark regions of high column densities. The values are
  linearily scaled from the minimum to the maximum value of the column densities obtained for a density structure with $\sigma_{\rho/\left<\rho\right>}=1$. The transmitted light of the screens is shown
  in column 2 to 4 for three different values of the mean optical depth $\left<\tau\right>$. The grey tone
  represents the amount of transmitted light. Bright regions correspond to small extinction, dark regions to
  high extinction.}
  \end{center}
\end{figure*}

In Fig.~\ref{slabattenvisu} the attenuation caused by a turbulent dust screen is visualised as
function of mean optical depth $\left<\tau\right>$ and standard deviation $\sigma_{\ln N}$ of the log-normal
density distribution. 
By increasing $\sigma_{\rho/\left<\rho\right>}$ the column density in certain areas is enhanced while the regions in between becomes almost free of material. The effect of the column density fluctuation on extinction increases both with $\sigma_{\rho/\left<\rho\right>}$ and with mean optical depth of the screen.
While the screen with $\sigma_{\rho/\left<\rho\right>}=0.25$ seems to be rather uniform, for higher values of
$\sigma_{\rho/\left<\rho\right>}$ extinction at certain areas becomes rather high. These regions appear as dark
clouds. As the regions of low optical depth transmit light very efficiently, the transmission of 
a turbulent screen, or general of all clumpy media, is higher than in the case of a homogeneous medium. Therefore, for the same
mean optical depth the effective extinction decreases with $\sigma_{\rho/\left<\rho\right>}$. 

If we compare
the transmission for different mean optical depths but the same fluctuations of column densities
then the structure appears more clearly towards higher values of $\left<\tau\right>$. This reflects the fact that the difference of the extinction of regions with different optical depth increases. This can be easily seen if two regions with optical depths $\tau_1$ and $\tau_2$ with $\tau_2>\tau_1$ are considered where the ratio in extinction
is $e^{\tau_2-\tau_1}$. If we scale the mean optical depth by a factor $f$ then the ratio becomes
$e^{f(\tau_2-\tau_1)}$.
As consequence the contrast between high transmission regions and low transmission regions in a turbulent screen becomes larger for higher values of the mean optical depth.
If we consider the effective extinction as function of wavelength then a flattening of the 
curve is expected either when the mean optical depth or when the fluctuations of the column density increases.

In this paper we will quantitatively analyse these effects of the turbulent dust screen on the attenuation curve. In Sect.~\ref{attenuation} we will discuss attenuation curves as function of the standard deviation $\sigma_{\ln \tau}$ and the mean attenuation 
${\rm \left<A_V\right>}=2.5\log_{10}e^1\left<\tau_{\rm V}\right>$ of the turbulent dust screen. We will
show how these physical quantities relate to the observed quantities, the absolute-to-relative attenuation ratio ${\rm R^A_V=A_V/E(B-V)}$ and the attenuation ${\rm A_V}$.
In Sect.~\ref{sectAV} and Sect.~\ref{sectRV} we describe how ${\rm A_V}$ and ${\rm R_V^A}$ vary with slice thickness $\Delta/{L_{\rm max}}$ and mean attenuation $\left<{\rm A_V}\right>$ and how the attenuation curves are effected by the statistical properties (the standard deviation $\sigma_{\rho/\left<\rho\right>}$ and the power spectrum $P(k)$) of the local density. 
Finally, we discuss which conclusions can be made about the turbulent density structure by measuring the attenuation curve.

\section{Attenuation of a distant turbulent screen}

\label{attenuation}
\begin{figure*}
	\includegraphics[width=0.49\hsize]{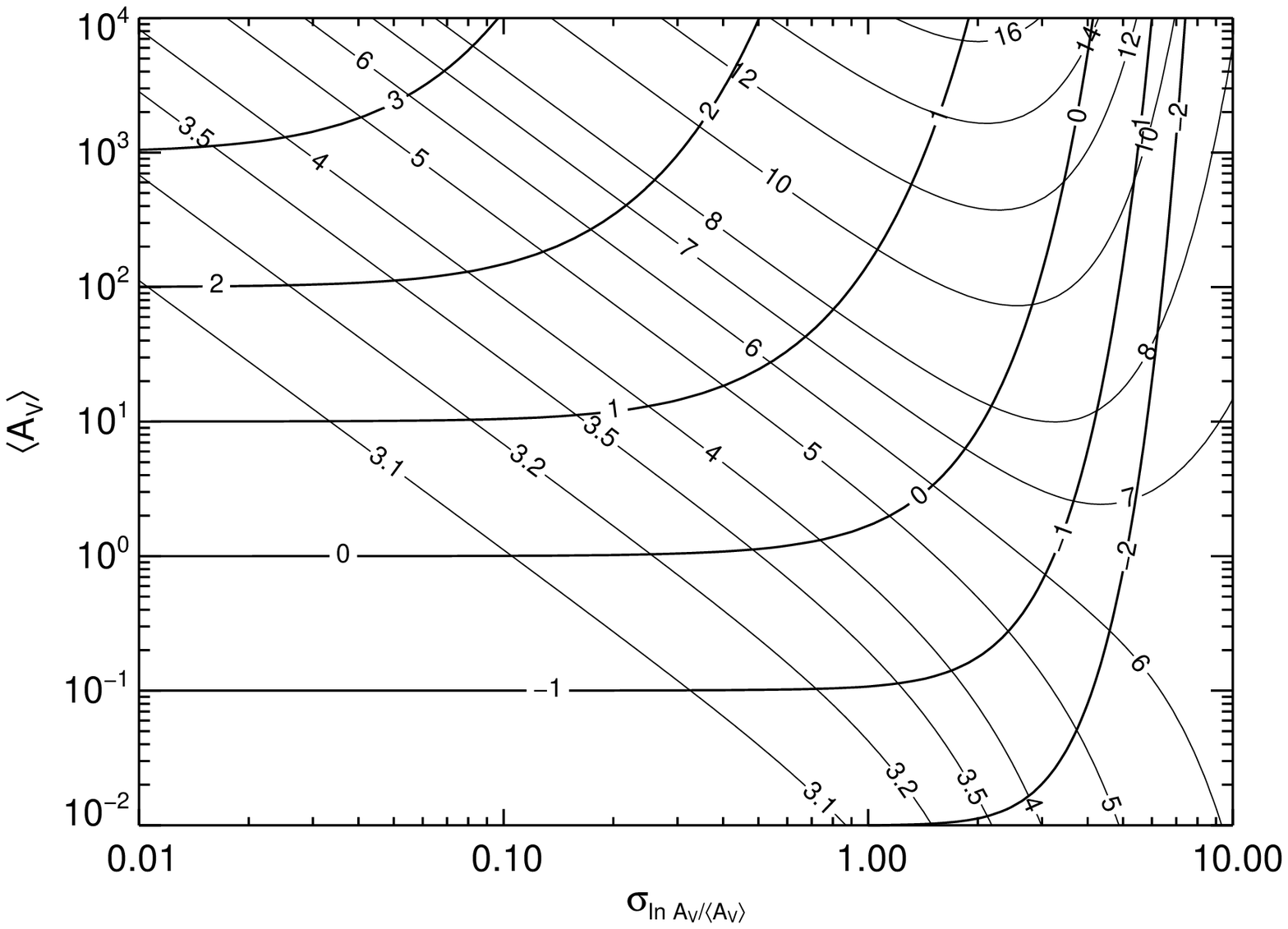}
	\hfill
	\includegraphics[width=0.49\hsize]{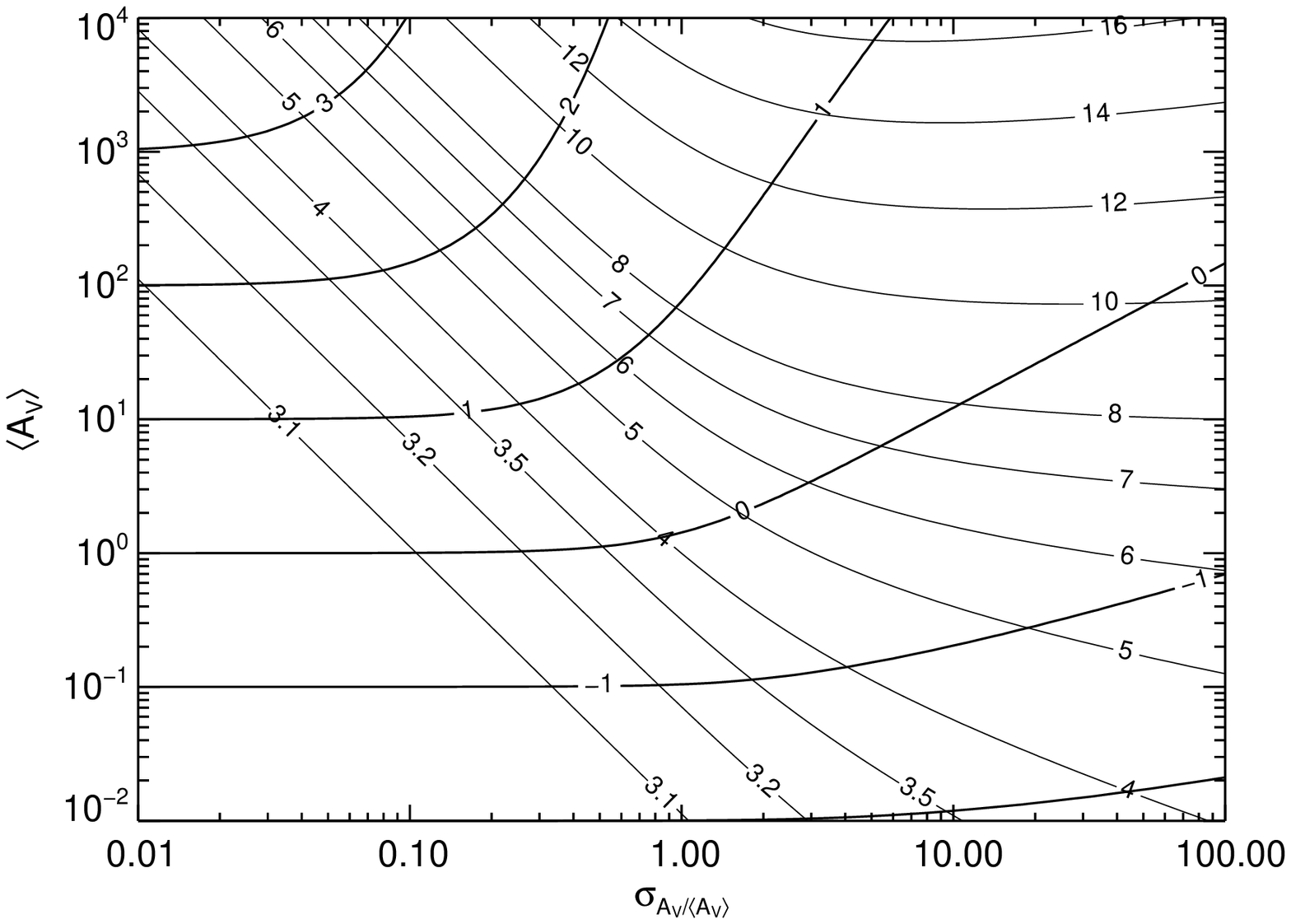}

	\caption{\label{reffcontourplus} 
	Attenuation ${\rm A_{V}}$ (thick solid lines) as a function of the averaged attenuation 
	$\left<{\rm A_V}\right>$ and the standard deviation 
	$\sigma_{\ln{\rm A_V}/\left<{\rm A_V}\right>}=
		\sigma_{\ln \tau_{\rm V}/\left<\tau_{\rm V}\right>}$ (left hand figure) and as a function
	of the averaged attenuation 
	$\left<{\rm A_V}\right>$ and the standard deviation 
	$\sigma_{{\rm A_V}/\left<{\rm A_V}\right>}=
		\sigma_{\tau_{\rm V}/\left<\tau_{\rm V}\right>}$ (right hand figure).
	The label values correspond to 
	$\log{\rm A_{V}}$. 
	Also shown are lines of constant values of the absolute-to-relative attenuation ratio
	${\rm R_{V}^A=A_{V}/E(B-V)=\tau_V/(\tau_B-\tau_V)}$.
	} 
\end{figure*}

In Fig.~\ref{reffcontourplus} the observable quantities ${\rm A_V}$ and ${\rm R_V^A}$
are shown in the physical plane defined by the standard deviation $\sigma_{\rm A_V/\left<A_V\right>}=\sigma_{\tau/\left<\tau\right>}$ of the fluctuation of optical depths and the mean attenuation $\left<{\rm A_V}\right>$. In the limit of both small attenuation $\left<{\rm A_V}\right>$ and small fluctuations the attenuation approaches the limit of the homogeneous dust screen where $\rm R_V^A=R_V=3.08$.

As expected the attenuation curve becomes flatter by increasing either the optical depth or the fluctuations of the optical depth. This results in higher values of ${\rm R_V^A}$. In the limit of small fluctuations, as will be verified in Sect.~\ref{sect_approx}, the contours of constant ${\rm R_V^A}$-values are described by $\sigma^2_{\ln \tau}\left<\tau\right>={\rm constant}$. The same lines also have ${\rm A_V/\left<A_V\right>}={\rm constant}$. 

The curves with ${\rm R_V^A}={\rm const.}$ describe essentially identical attenuation curves
${\rm A}_\lambda/{\rm A}_V$.  This is particularly true when the relation ${\rm A}_{\lambda}/\left<{\rm A}_\lambda\right>={\rm const.}$ is valid over the whole wavelength range. 

The maximum flattening that can be produced by the turbulent screen is determined
by the physical conditions.
If we consider the Mach numbers in the ISM of our galaxy the standard deviation of the column density
should be smaller than 10. In case of an attenuation $\rm A_V=1$ we would expect attenuation curves with $\rm R_V^A$-values smaller than 10. The flattening produced in more optical thick systems can be much more pronounced.

Naively one would expect that in the limit of small fluctuations of the optical depth 
the attenuation can be described by a homogeneous screen. But as the flattening is a function
of $\sigma_{\ln \tau}^2\left<\tau\right>$ a screen with very small fluctuations can effectively
produce the same attenuation curve as a screen with strong fluctuations. The mean 
optical depths however is very different. 

\begin{figure*}
	\includegraphics[width=0.49\hsize]{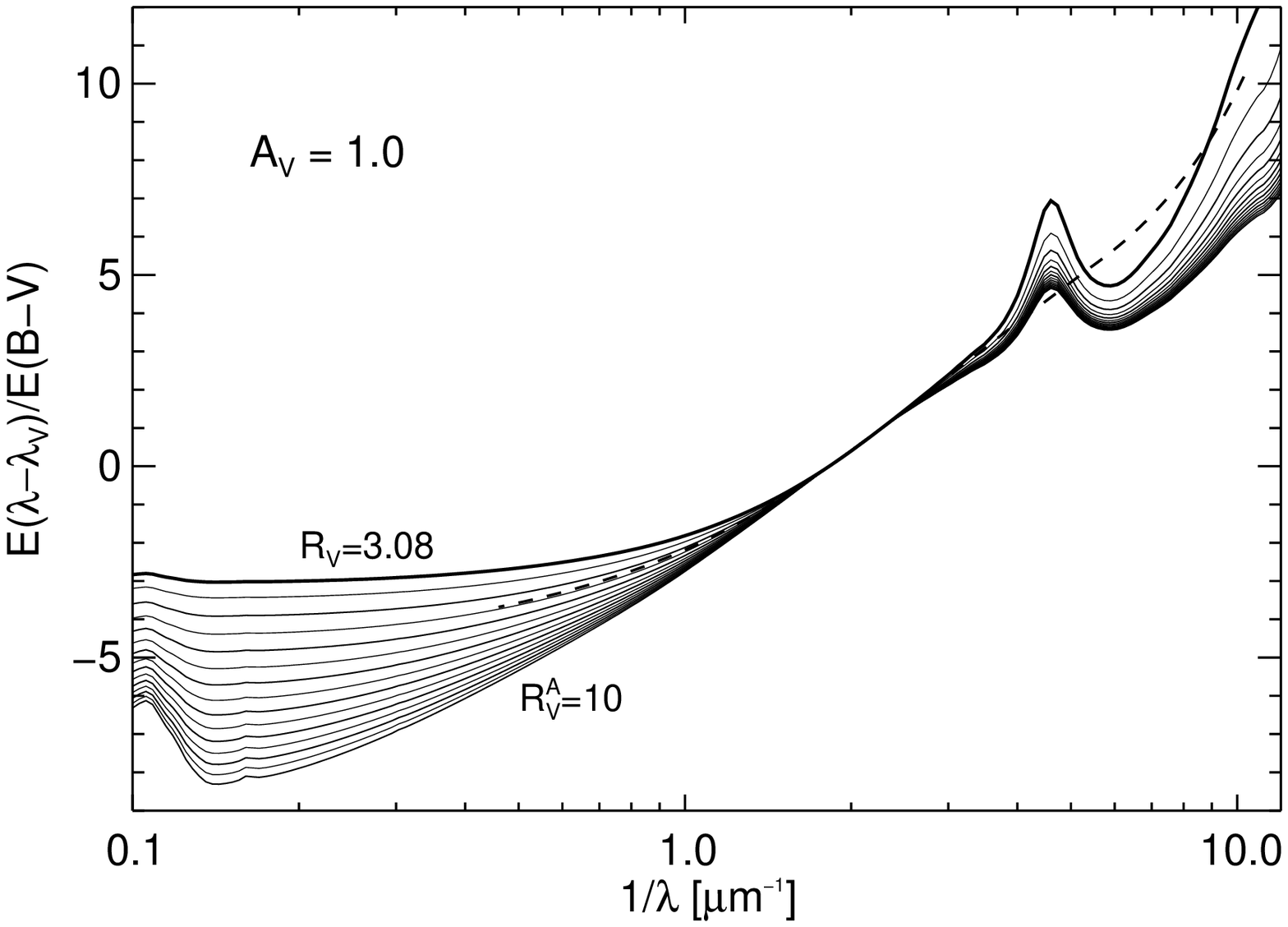}
	\hfill
	\includegraphics[width=0.49\hsize]{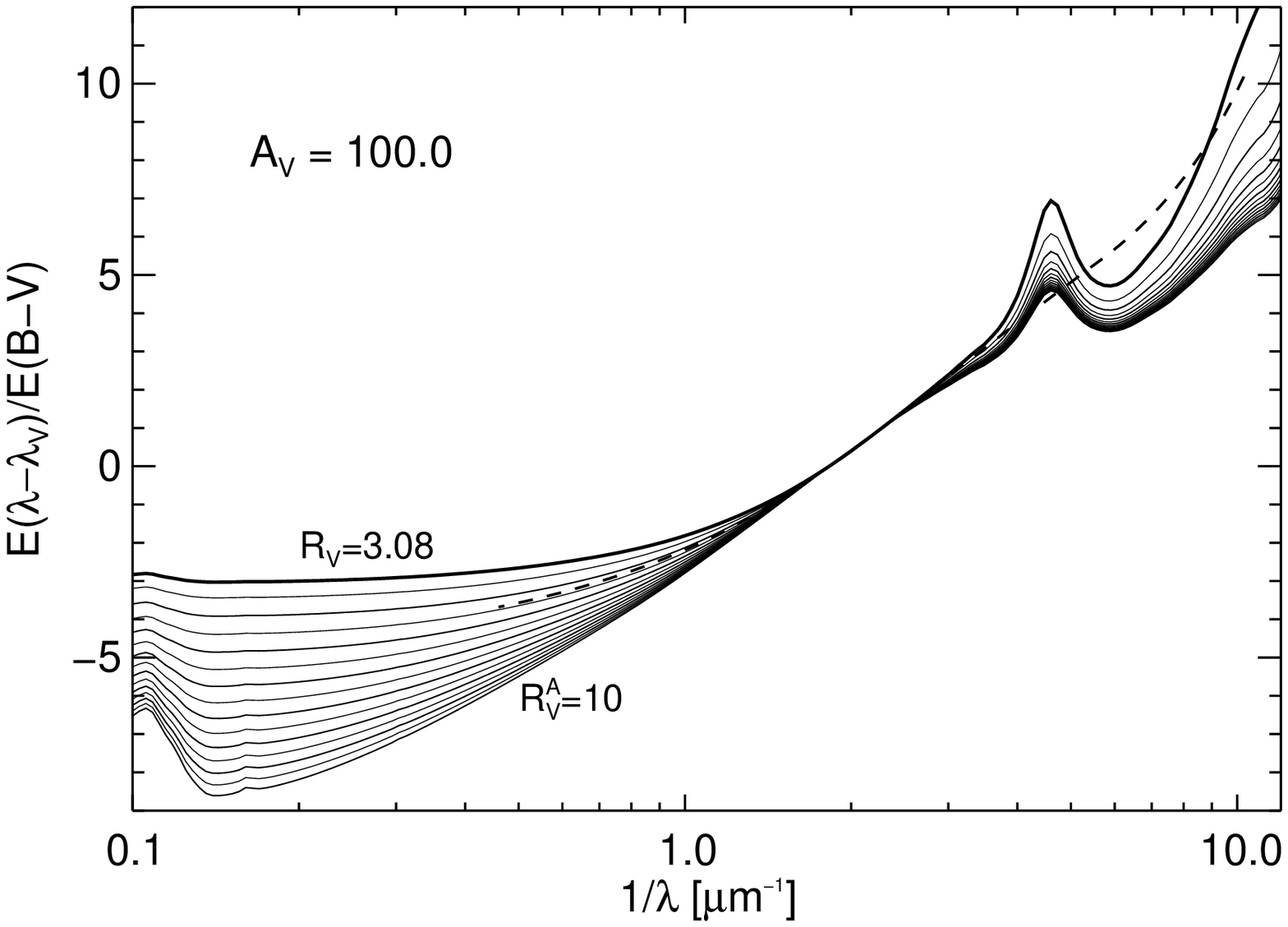}

	\caption{\label{figattenuation_rv1} 
	Relative attenuation curves ${\rm E(\lambda-\lambda_V)/E(B-V)}$ 
	for various values $\rm R_V^A$ varied from 3.5 up to 10 with $\Delta \rm R^A_V=0.5$.
	The attenuation ${\rm A_V}$ is constant chosen to be 1.0 and 100.0, respectively. 
	As can be seen in Fig.~\ref{reffcontourplus} the conditions of the foreground screen (the
	mean attenuation $\rm \left<A_V\right>$ and the contrast $\sigma_{\ln \tau}$ of the optical
	depths) which lead to the same $\rm R_V^A$ in the left and in the right hand figure
	are different.
	The thick solid line is the initial extinction curve provided by \citet{Weingartner2001} 		where $\rm R_V=3.08$. For comparison the attenuation curve derived for star burst 
	galaxies (broken line) is shown (\cite{Calzetti2001}).
	} 
\end{figure*}
\begin{figure*}
  \includegraphics[width=0.49\hsize]{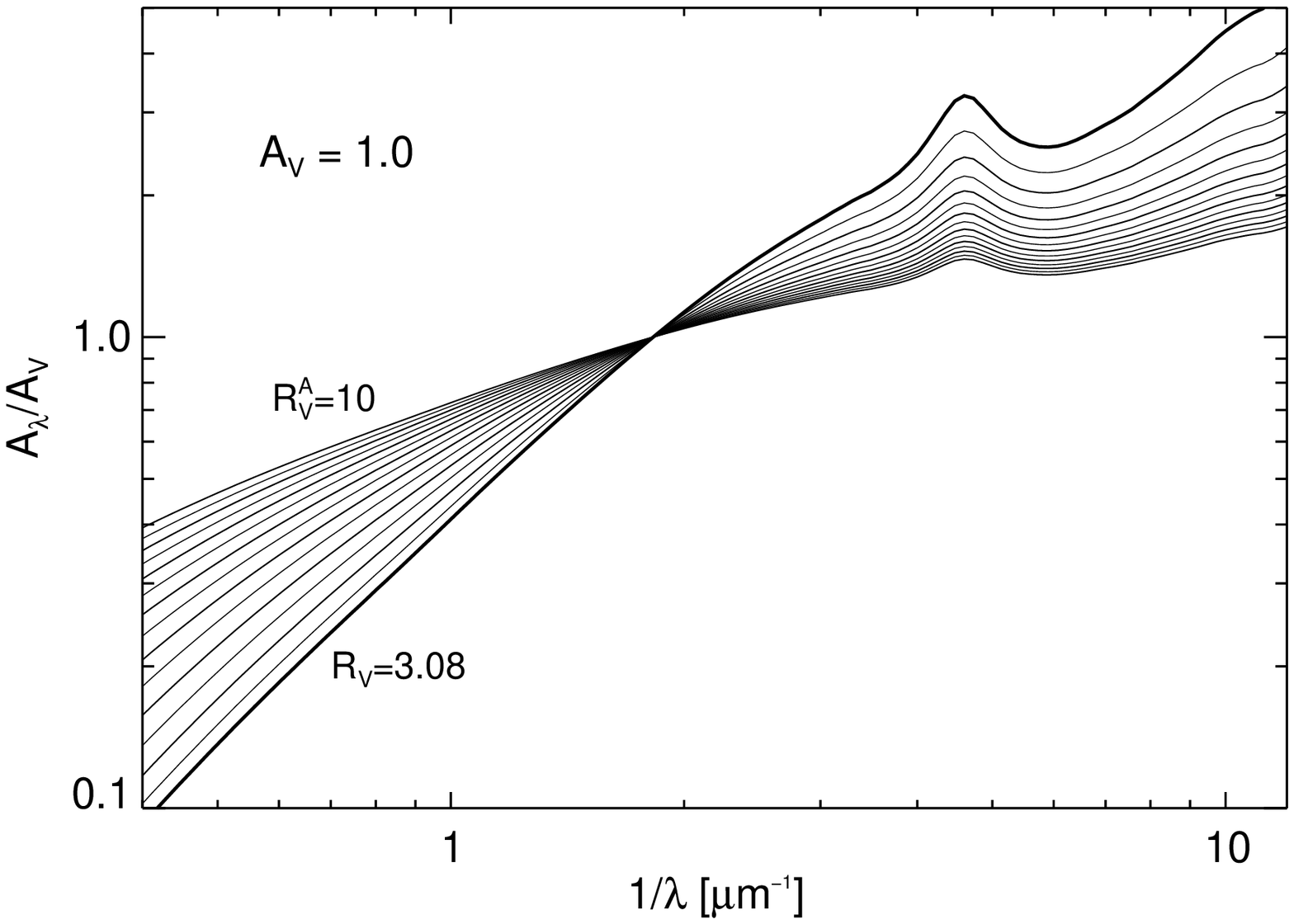}
  \hfill
  \includegraphics[width=0.49\hsize]{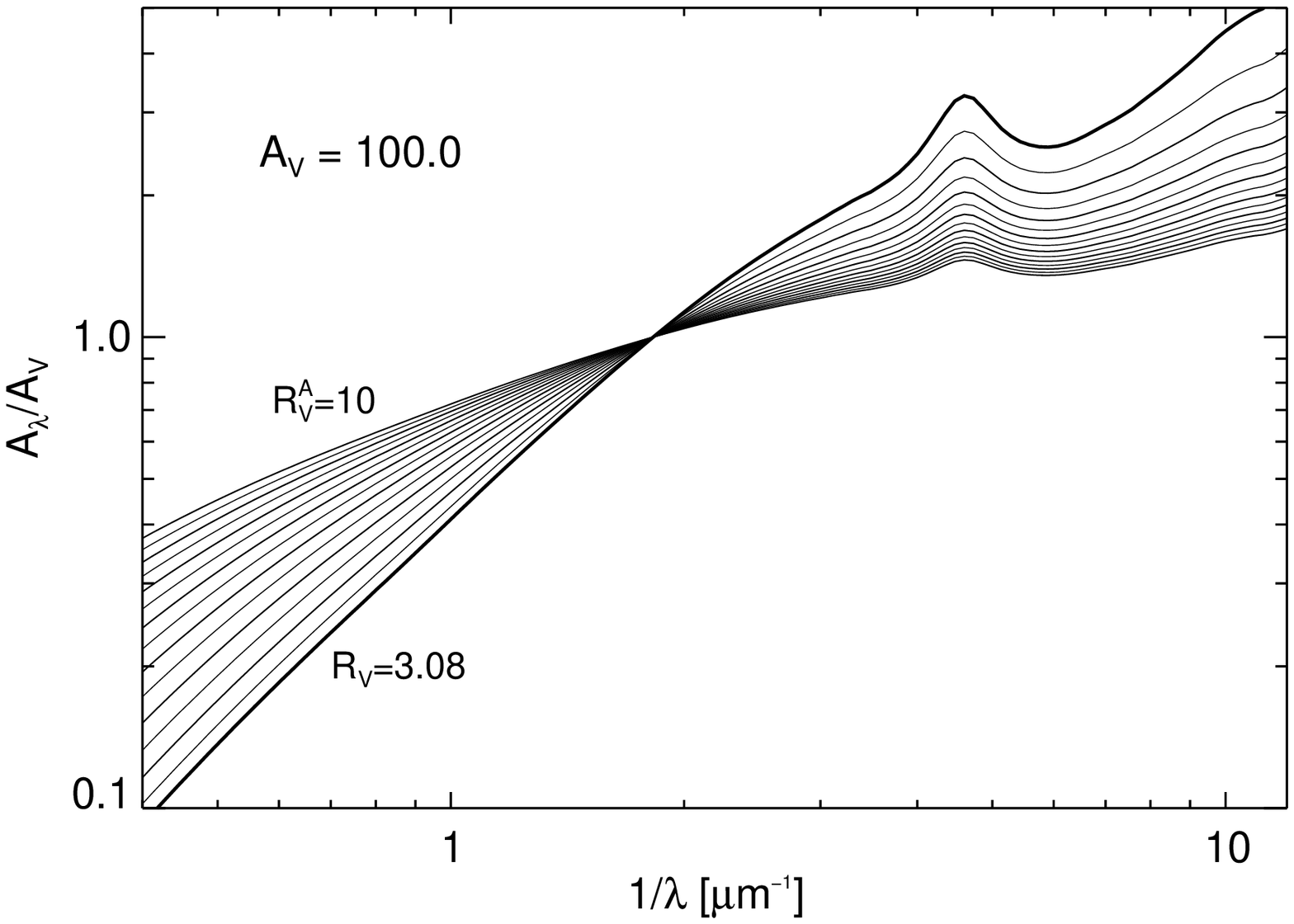}
  \caption{
  \label{figattenuation_rv2}
  Attenuation curves $\rm A_\lambda/A_V$ caused by a distant turbulent dust screen 
  varying $\rm R_V^A$ from 3.5 to 10. The difference in $\rm R^A_V$ between
  neighboured curves is taken to be 0.5. Also shown is the initial mean extinction curve
  with $\rm R_V=3.08$ (thick solid line).
  }  
\end{figure*}

In Fig.~\ref{figattenuation_rv1} and Fig.~\ref{figattenuation_rv2} we show the variation of the attenuation curves as a function of ${\rm R_V^A}$. In Fig.~\ref{figattenuation_rv1} they are given as a function of the relative attenuation
\begin{equation}
  \mathcal{E}_{\lambda} = {\rm \frac{A_{\lambda}-A_V}{A_B-A_V} = \frac{E(\lambda-V)}{E(B-V)}}.
\end{equation}
The attenuation ${\rm A_V}$ is held fixed with ${\rm A_V}=1$ and ${\rm A_V}=100$,
respectively. The curves obtained for the same $\rm R_V^A$-value for the two choices of the attenuation $\rm A_V$ are essentially the same even though
they imply very different physical conditions of the turbulent screen 
In the case of ${\rm A_V}=1$ an ${\rm R_V^A}$-value of 10 would correspond to a screen with
strong fluctuations of the optical depth with $\sigma_{\tau/\left<\tau\right>}\approx50$. For ${\rm A_V}=100$ the fluctuations would be very small with $\sigma_{\tau/\left<\tau/\right>}$ less than $\approx 0.4$.

As can be seen in the figure the curvature itself allows a determination of the $\rm R_V^A$-value.
The curves are compared with the curve empirically derived for star-burst galaxies, known as
the `Calzetti-extinction-law'. At wavelengths below the extinction bump at 2200~\AA{} the `Calzetti-curve' is obtained for ${\rm R_V^A}\approx4.3$. 
As is shown below, in reality variations of ${\rm R_V^A}$ are expected by comparing the attenuation
of different galaxies. In star-burst galaxies the bump at 2200~\AA{} seems to be absent. As we show elsewhere \citep{Dopita2005} this can be understood if the carriers of the 2200~\AA{} absorption feature are the PAH molecules, which are photo-dissociated in the strong EUV radiation field of these galaxies.

In Table~\ref{tableattenuation} the attenuation $\mathcal{E}_{\lambda}$ of an isothermal turbulent dust screen is given for several values of $\rm R_V^A$. The attenuation in $\rm V$ is taken to be ${\rm A_V}=1$ and ${\rm A_V}=10$, respectively. The attenuation $\rm A_\lambda$ can be obtained by using the following formula:
\begin{equation}
  {\rm A}_{\lambda} = \left(\mathcal{E}_{\lambda} / {\rm R_V^A}+1\right)\,{\rm A_V}.
\end{equation}

\subsection{General attenuation curve}

In the previous section we discussed the effect of the turbulent screen on the attenuation curve using the mean extinction curve of our galaxy. But in general the underlying extinction curve of a turbulent screen might not be the same. The extinction curve might show a different wavelength dependence and might be characterised by another $\rm R_V$-value. It is therefore appropriate to use a more general approach where these effects can be easily taken into account.

For this we considered the attenuation $\rm A_{\lambda}/A_V$ of the turbulent screen as function of $x=\left<{\rm A}_\lambda\right>/\left<{\rm A_V}\right>=k_\lambda/k_{\rm V}$ rather than as function of wavelength. To characterise the attenuation curves we used the absolute-to-relative attenuation ratio$\rm \alpha_V^A=A_V/E(\tilde \lambda-V)$ where the wavelength $\tilde \lambda$ is defined by the absolute-to-relative extinction $\alpha_{\rm V}=k_{\rm V}/(k_{\tilde \lambda}-k_{\rm V})=3.08$. The variation of $\alpha_{\rm V}^{\rm A}$ is therefore taken to be identical to the variation of $\rm R_V^A$ of the mean extinction curve of our galaxy.

\begin{figure}
  \includegraphics[width=\hsize]{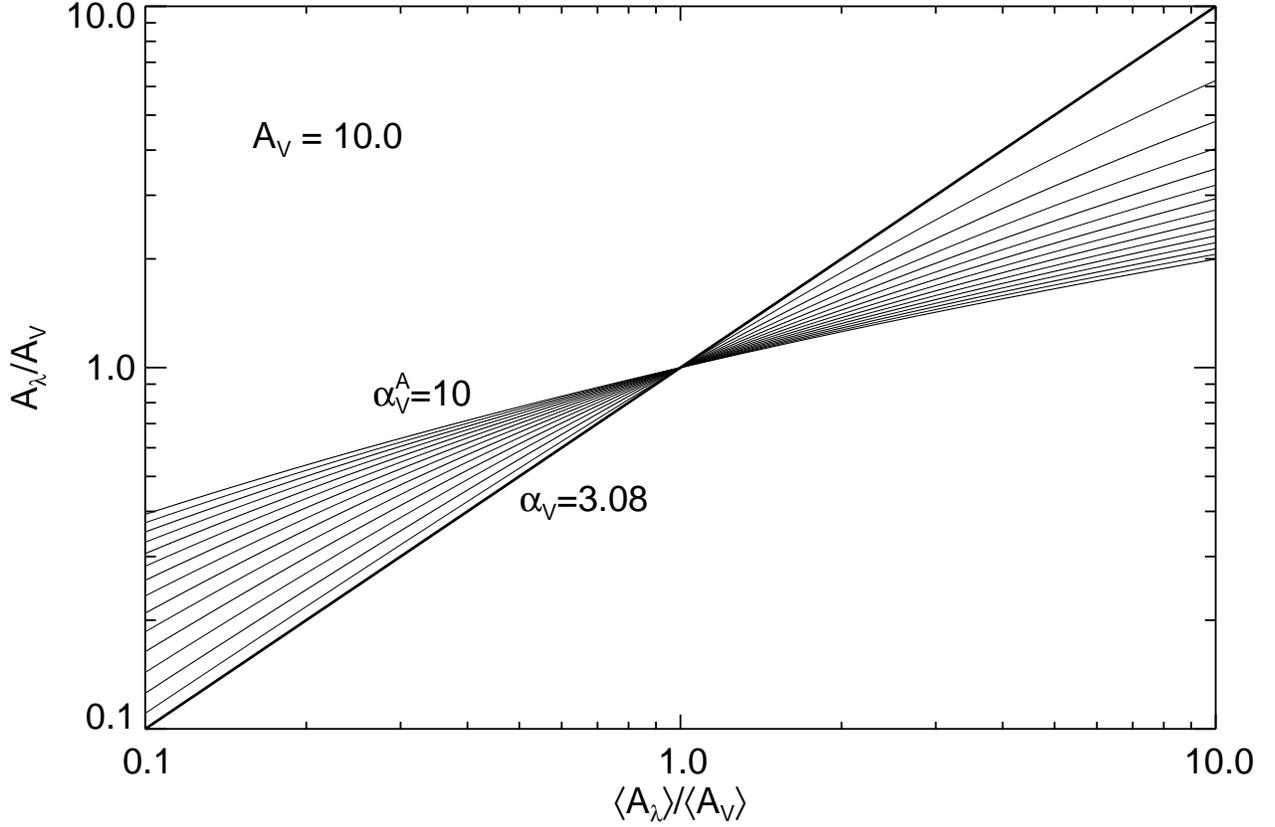}
  \caption{\label{figatttau}
  Attenuation $\rm A_\lambda/A_V$ of a turbulent screen as function of the extinction 
  ${\rm \left<A_\lambda\right>/\left<A_V\right>}=k_\lambda/k_{\rm V}$.
  The absolute attenuation has been chosen to be $\rm A_V=10$. The result, however,
  depends primarily on the ratio on the absolute
  to relative attenuation $\rm \alpha^A_V=A_V/E(\tilde \lambda-V)$ where the
  extinction coefficient at wavelength $\tilde \lambda$ has to fulfil 
  $\alpha_V=k_{\rm V}/(k_{\tilde \lambda}-k_{\rm V})=3.08$.
  $\rm \alpha_V^A$ is varied from 3.5 to 10 in steps of 0.5. 
}  
\end{figure}

\begin{figure}
  \includegraphics[width=\hsize]{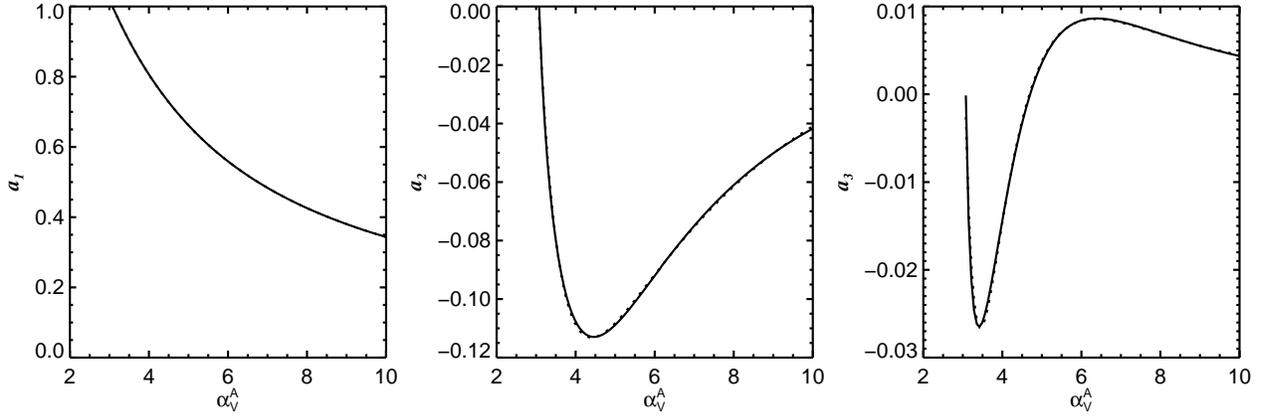}
  \caption{\label{figattfitpara}
  Coefficients $a_1$, $a_2$, and $a_3$ of the approximation of $\log \rm A_\lambda/A_V$ as function
  of the absolute-to-relative attenuation ratio$\alpha_{\rm V}^{\rm A}$ (solid lines) using a polynomial function of third order (Eq.~\ref{eqattapprox}). The dotted line, which can be barely distinguished from the solid line in these panels, is a fit using the polynomial function given in Eq.~\ref{eqfitcoeff}.
  The attenuation chosen for this figure is $\rm A_V=1$. However, the functional form obtained for $\rm A_V=10$ is very similar.
  }
\end{figure}

We considered the ratio of the extinction coefficients $k_\lambda/k_{\rm V}$ in the range $0.1$ to $10$.
As shown in the previous sections the attenuation is well determined by the $\alpha_{\rm V}^{\rm A}$-value, largely independent
of the actual conditions, the mean attenuation $\rm \left<A_V\right>$ and the contrast $\sigma_{\tau_{\rm V}/\left<\tau_{\rm V}\right>}$. The dependence on the absolute attenuation $\rm A_V$, on the other hand, is only weak. To obtain a general expression for the attenuation curve we hold the absolute attenuation $\rm A_V$ constant and considered all attenuation curves up to $\alpha_{\rm V}^{\rm A}=10$. The absolute attenuation
$\rm A_V$ has been chosen to be 1 and 10. The derived attenuation curves, shown in Fig.~\ref{figatttau} for $\rm A_V=10$,
can be fitted by a simple polynomial function
\begin{equation}
 \label{eqattapprox}
 \log {\rm A_{\lambda}/A_{V}} = \sum\limits_{i=1}^3 a_i\, (\log x)^i
\end{equation}
where the constants $a_i$ depend on the absolute-to-relative attenuation ratio$\alpha_{\rm V}^{\rm A}$. 
The accuracy of all fitted curves obtained for $\rm A_V=1$ and $\rm A_V=10$ is better than $0.55\%$ and $0.65\%$, respectively. The variation of the parameter
$a_i$ as function of $\alpha_{\rm V}^{\rm A}$ are shown in Fig.~\ref{figattfitpara}. 

We performed an individual polynomial fit to the coefficients with
\begin{equation}
  \label{eqfitcoeff}
  a_{i}(z) = \sum\limits_{j=0}^{N_i} b_j\,z^j
\end{equation}
where $z=1/({\alpha_{\rm V}^{\rm A}}-1)$ and $N_1=3$ and $N_2=N_3=4$. The result is also shown 
in Fig.~\ref{figattfitpara}. The fitted coefficients $b_j$ are given in Table~\ref{tableattfitcoeff}.

If applied to situations where the absolute attenuation $\rm A_V$ is either $1$ or
$10$, the approximations are better than $1.3\%$. In general the accuracy is somewhat lower. If the approximation obtained for $\rm A_V=1$
is used for a turbulent medium where $\rm A_V=10$ the accuracy is better than $\sim 5\%$. The
largest errors occurs at low optical depths. For $x \ge 1$ the approximation is better than $\sim 2\%$.

\subsubsection{Attenuation curves for different $\rm R_V$}

The approximations above can be directly applied for extinction curves with
an absolute-to-relative extinction $\rm R_V\approx \alpha_{\rm V}=3.08$. For extinction curves
with a different $\rm R_V$ one can use a simple linear transformation.
The correct attenuation curve for a given $\rm R_V^A$ is obtained by applying the relation
\begin{equation}
  \alpha_{\rm V}^{\rm A} = \frac{1}{m}\left({\rm R_V^A}-{\rm R_V}\right)+\alpha_{\rm V}
\end{equation}
where $m$ is a linear function of $\rm R_V$ with
\begin{equation}
  \label{eqfitm}
  m = 1+0.275\,\left(\rm R_V-\alpha_{\rm V}\right).
\end{equation}

The simple correction has been obtained by varying $\rm R_V$ from
2 to 6 and by using the approximation for $\rm A_V=1$ and $\rm A_V=10$. For the considered
parameter range of $\rm R_V$ and $\rm \alpha_V^A$ the accuracy of the approximation is better than $\sim 0.6\%$.

\subsection{Analytical approximation of $\tau_{\rm eff}$}

\label{sect_approx}
In the limit of an optical thick slice with $\left<\tau\right>\gg1$ and in the limit of small fluctuations
of the optical depth with $\sigma_{\tau/\left<\tau\right>}\ll1$ an analytical expression for the effective
optical depth can be given. This approximation is particularly useful in deriving the attenuation
of thick slices of the turbulent medium.

The function $e^{-e^y\left<\tau\right>}$ stays almost constantly $1$ for $y<-\ln\left<\tau\right>$ 
and drops sharply at larger values. In the limit $\ln\left<\tau\right>\ll y_0$ the function
$p(y)e^{-e^y\left<\tau\right>}$ becomes a narrow distribution around a maximum at $\tilde y$.
By using the approximation
\begin{equation}
   e^y\approx e^{\tilde y}\left(1+(y-\tilde y)+\onehalf(y-\tilde y)^2\right)
\end{equation}
it is straight forward to show that the effective extinction is given by
\begin{equation}
	\label{eqtaueffapprox}
	\tau_{\rm eff} = \onehalf \ln \gamma +\onehalf \left<\tau\right>e^{\tilde y}\left(1+\gamma\right)
\end{equation}
where $\gamma=1+\left<\tau\right>\sigma_{\ln \tau}^2\,e^{\tilde y}$ and
where $\tilde{y}$ is the solution of the equation:
\begin{equation}
   \label{eqymax}
  \tilde{y}-y_0 = -\sigma^2_{\ln \tau}\left<\tau\right>e^{\tilde{y}}
\end{equation}

In the limit of small fluctuations of the optical thickness and $\left<\tau\right>\ll 1$
the maximum is at $\tilde y\approx y_0=-\onehalf \sigma_{\ln \tau}^2\ll 1$ so that
$\gamma\approx 1$. Therefore, the approximation is still valid and gives the correct 
result $\tau_{\rm eff}\approx \left<\tau\right>$. 

\begin{figure}
\includegraphics[width=\hsize]{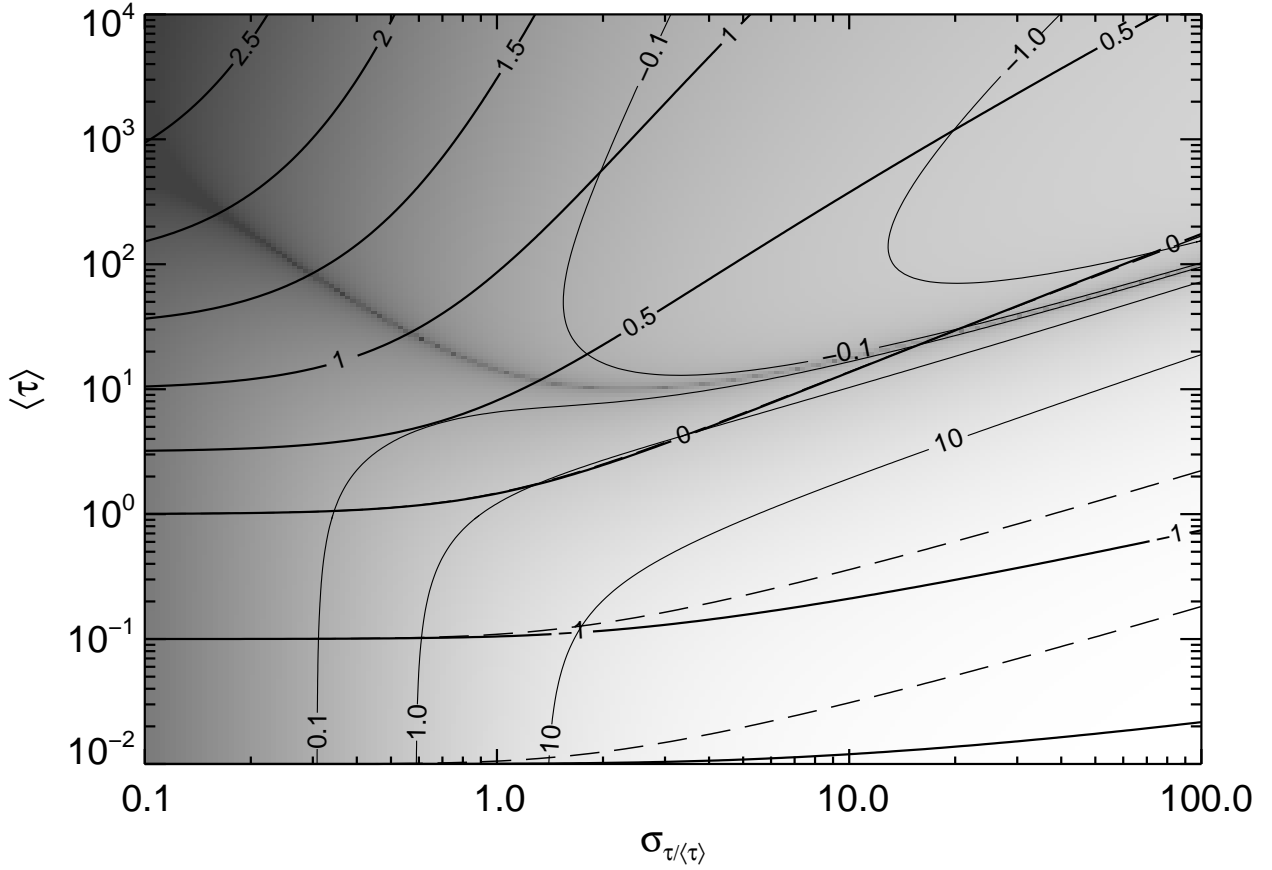}
\caption{\label{figaccuracy_a} Accuracy of the approximation Eq.~\ref{eqtaueffapprox}  of the effective optical depth $\tau_{\rm eff}$ shown in grey scale where the accuracy increases
towards darker colour. The accuracy of 0.1, 1.0, and $10.0\%$ is shown as thin contour lines.
Along the dark stripe the approximation gives the correct value. Above this curve
the correct values are generally smaller while they are larger otherwise.
The thick and dashed lines refer to constant effective optical depths derived by calculating
the integral~\ref{eqtaueff} and by using the approximation~\ref{eqtaueffapprox}. 
} 
\end{figure}

In Fig.~\ref{figaccuracy_a} we compare the effective optical depth obtained by solving
the integral~\ref{eqtaueff} and by using the approximation~\ref{eqtaueffapprox}. As
shown the analytical solution is accurate for small fluctuations for all optical depths and
can be used at higher optical depths also for larger fluctuations.
In case of an effective optical depth $\tau_{\rm eff}=1$ the uncertainty by using
the approximation is at most $1\%$.

If only small fluctuations of the optical depths are considered the first part of the
approximation~\ref{eqtaueffapprox} can be neglected. Furthermore $y_0=\onehalf\sigma_{\ln \tau}^2\approx 0$ in Eq.~\ref{eqymax} so that $\tau_{\rm eff}$ becomes a function of 
the product $\sigma_{\ln \tau}^2\left<\tau\right>$ only.

In the limit $\sigma_{\ln \tau}\ll 1$ the attenuation is given by
\begin{equation}
  \label{eqapprox_av}
  {\rm A_\lambda} = \onehalf\left<{\rm A_\lambda}\right>\,e^{\tilde y_\lambda}\left(1-\onehalf\tilde y_\lambda\right)
\end{equation}
so that $\rm A_\lambda/\left<A_\lambda\right>=constant$ for $\sigma_{\ln \tau}^2\left<\tau\right>=\rm constant$.

In addition the absolute-to-relative attenuation ratio$\rm R_V^A$ in the limit of small 
fluctuations of the optical thickness becomes only a function of $\sigma_{\ln \tau}^2\left<\tau\right>$
and is given by the expression
\begin{equation}
  \label{eqapprox_rv}
  {\rm R_V^A} = \frac{{\rm A_V}}{{\rm A_B-A_V}}=\frac{\tilde y_{\rm V}(2+\tilde y_{\rm V})}{\tilde y_{\rm B}(2+\tilde y_{\rm B})-\tilde y_{\rm V}(2+\tilde y_{\rm V})}.
\end{equation}

As the width of the PDF of the column density decreases with thickness and the fluctuations
will therefore always be small at a certain point both approximations are in particular helpful to describe the attenuation caused by dust screens which
are larger than the maximum scale $L_{\rm max}$. We will discuss this in Sect.~\ref{sectapprox_rvthick}.


\section{Attenuation with slice thickness}

\label{attenuation2}

In the following we analyse in some detail how the attenuation curve depends
on the statistical properties of the local density of the turbulent density structure
and how the attenuation curve is varying with the thickness of the distant screen where
we consider the ratio $\Delta/L_{\rm max}$ of its thickness $\Delta$ 
and the maximum turbulent scale $L_{\rm max}$. The effects on the attenuation curves are discussed by considering ${\rm R_V^A}$ and $\left<{\rm A_V}\right>$ as these parameters determine the attenuation sufficiently well. The attenuation through the turbulent medium is furthermore characterised
by the mean optical depth $\left<\tau\right>_{L_{\rm max}}$ or the corresponding attenuation 
$\rm \left<A_V\right>_{L_{\rm max}}$ at one maximum turbulent scale $L_{\rm max}$.

\subsection{Variation of $\rm A_\lambda$ with slice thickness}

\label{sectAV}

\begin{figure*}
  \includegraphics[width=0.5\hsize]{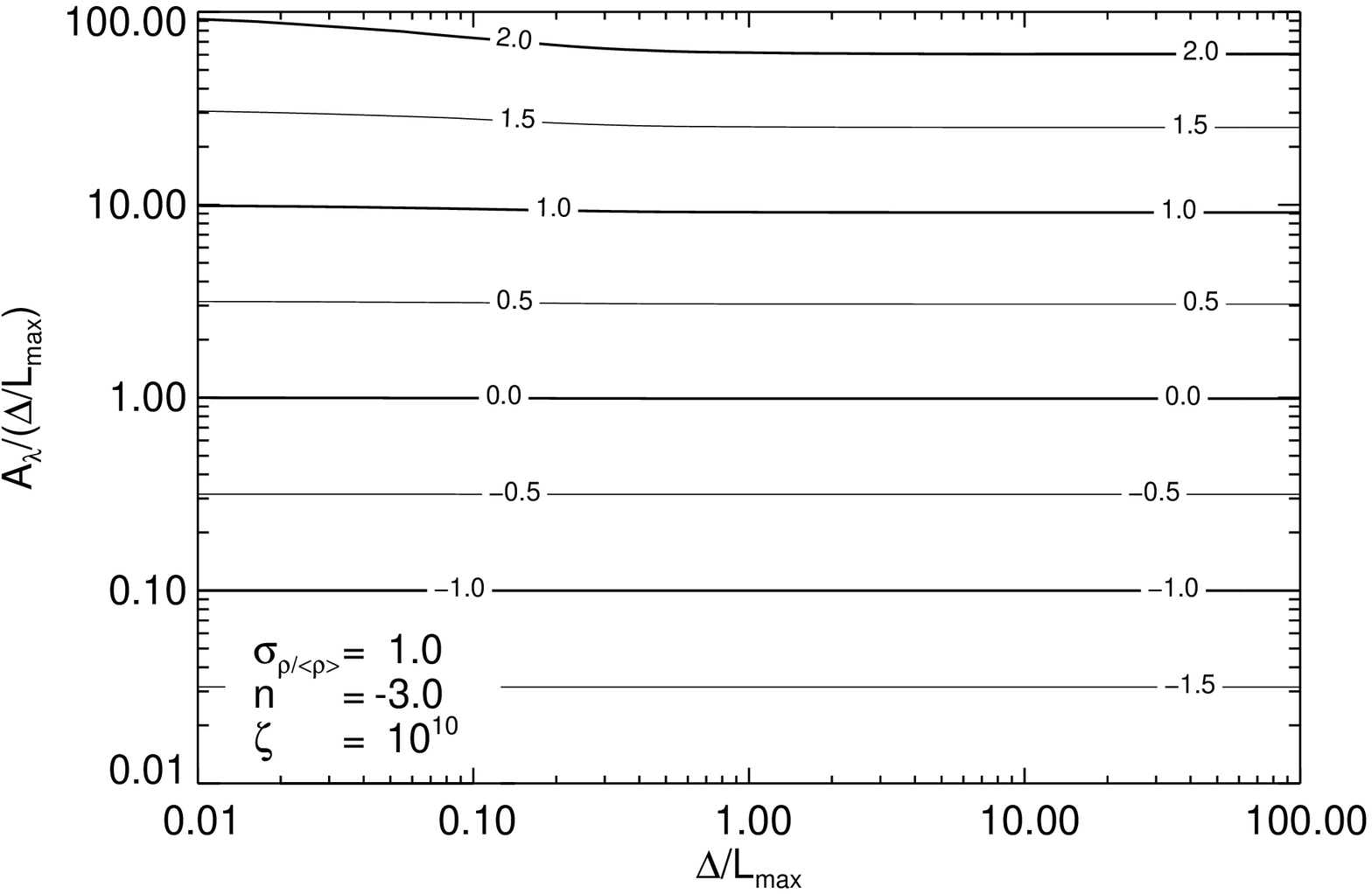}
  \includegraphics[width=0.5\hsize]{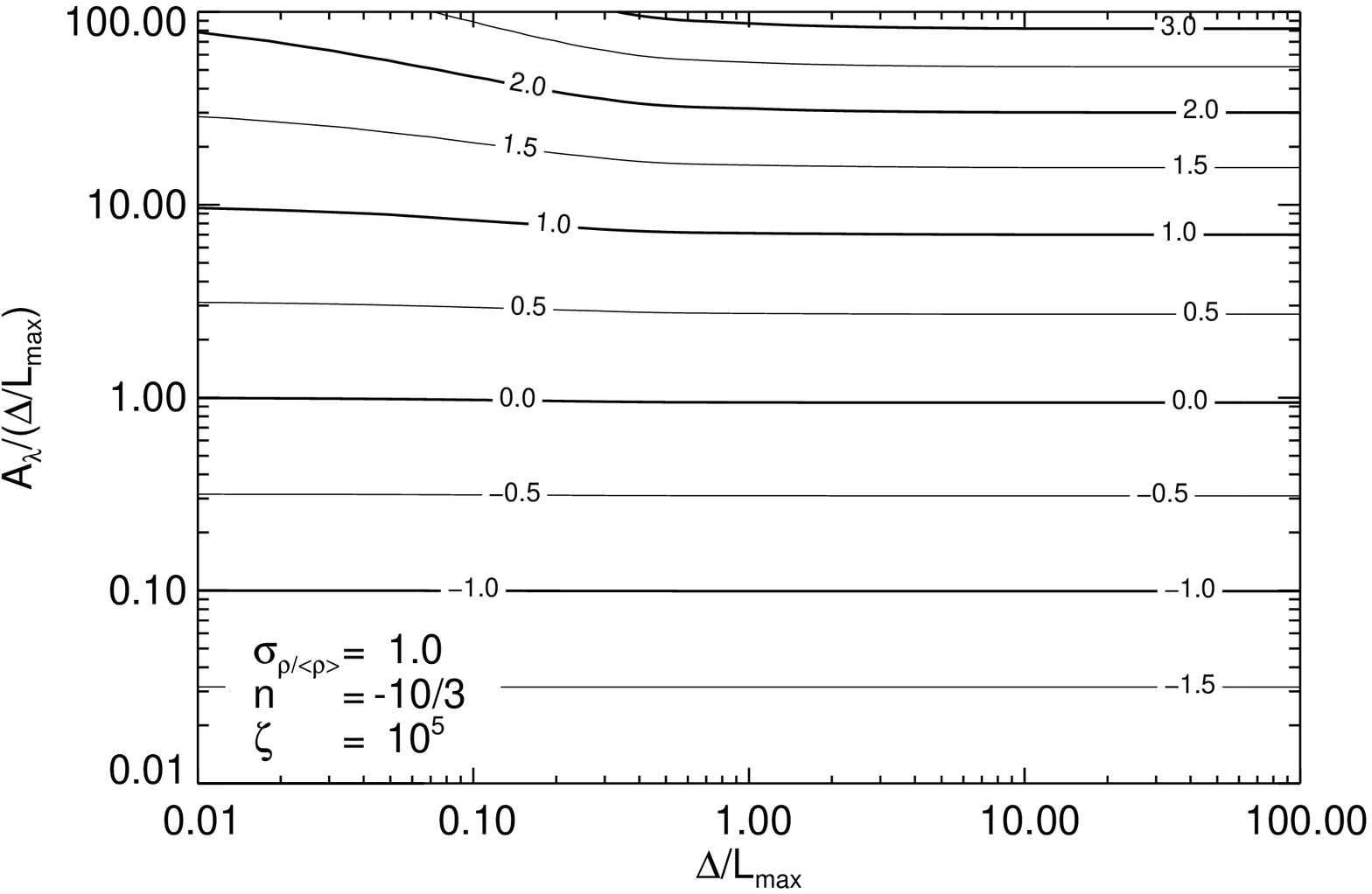}

  \includegraphics[width=0.5\hsize]{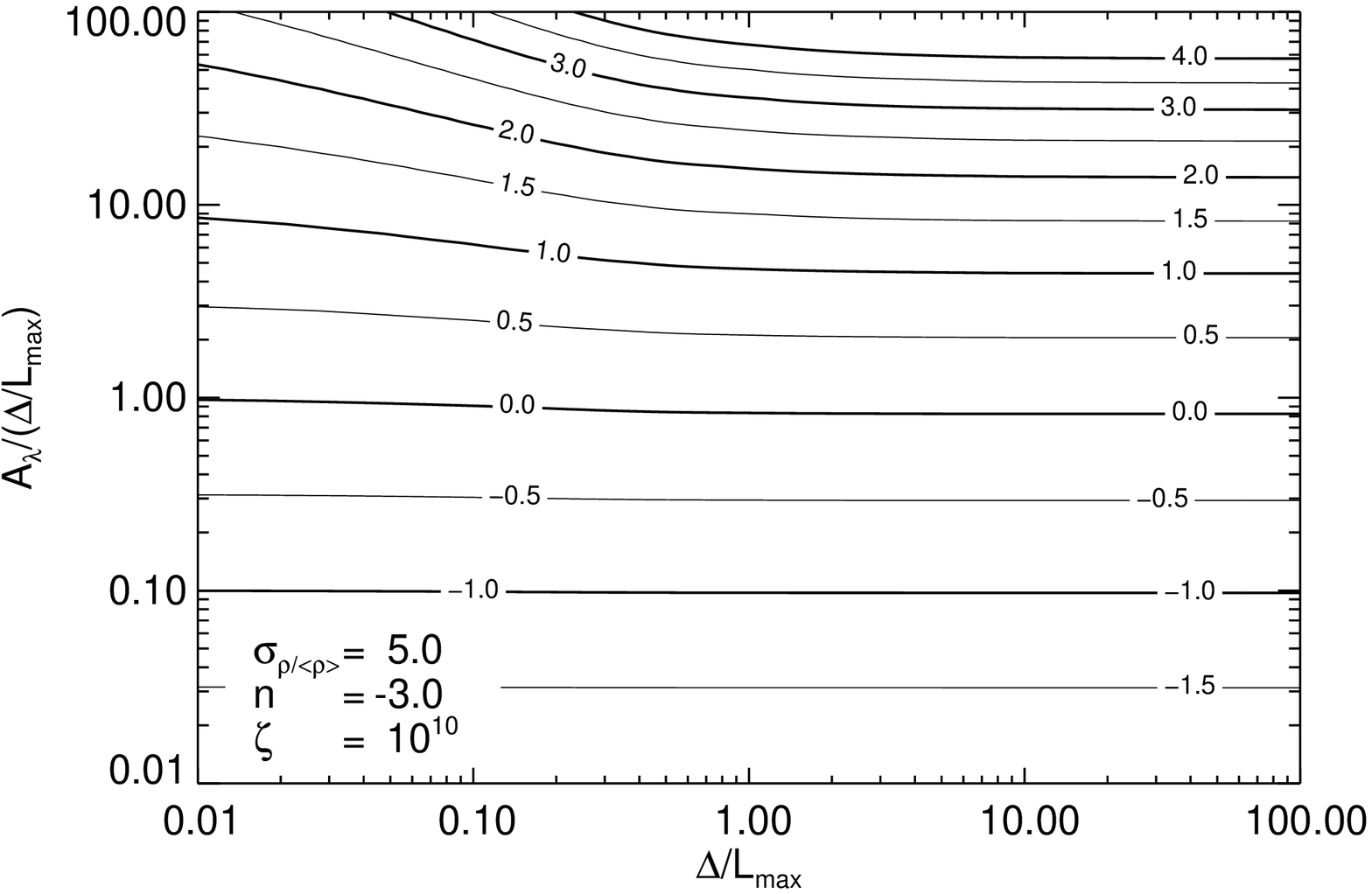}
  \includegraphics[width=0.5\hsize]{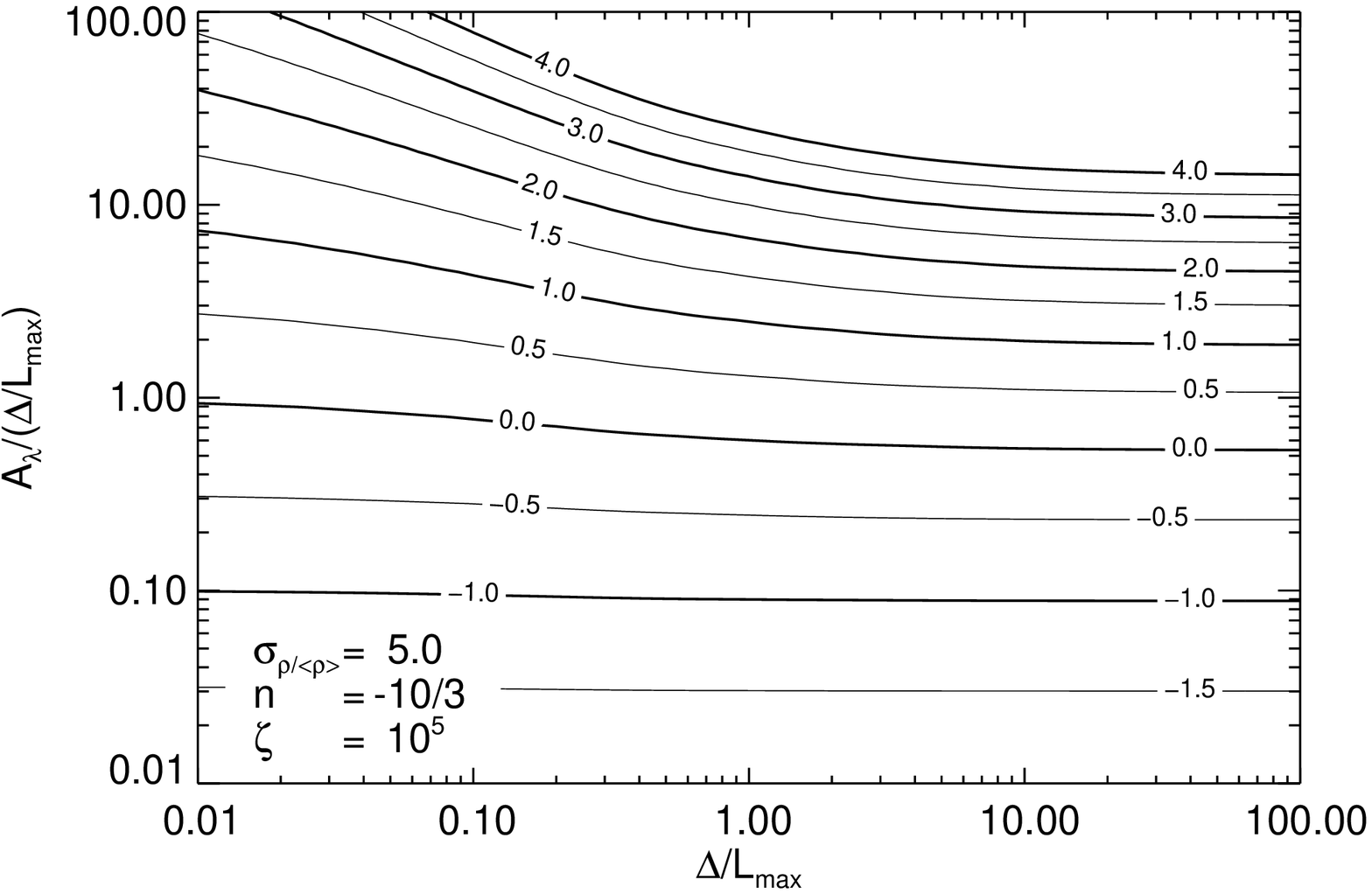}

  \includegraphics[width=0.5\hsize]{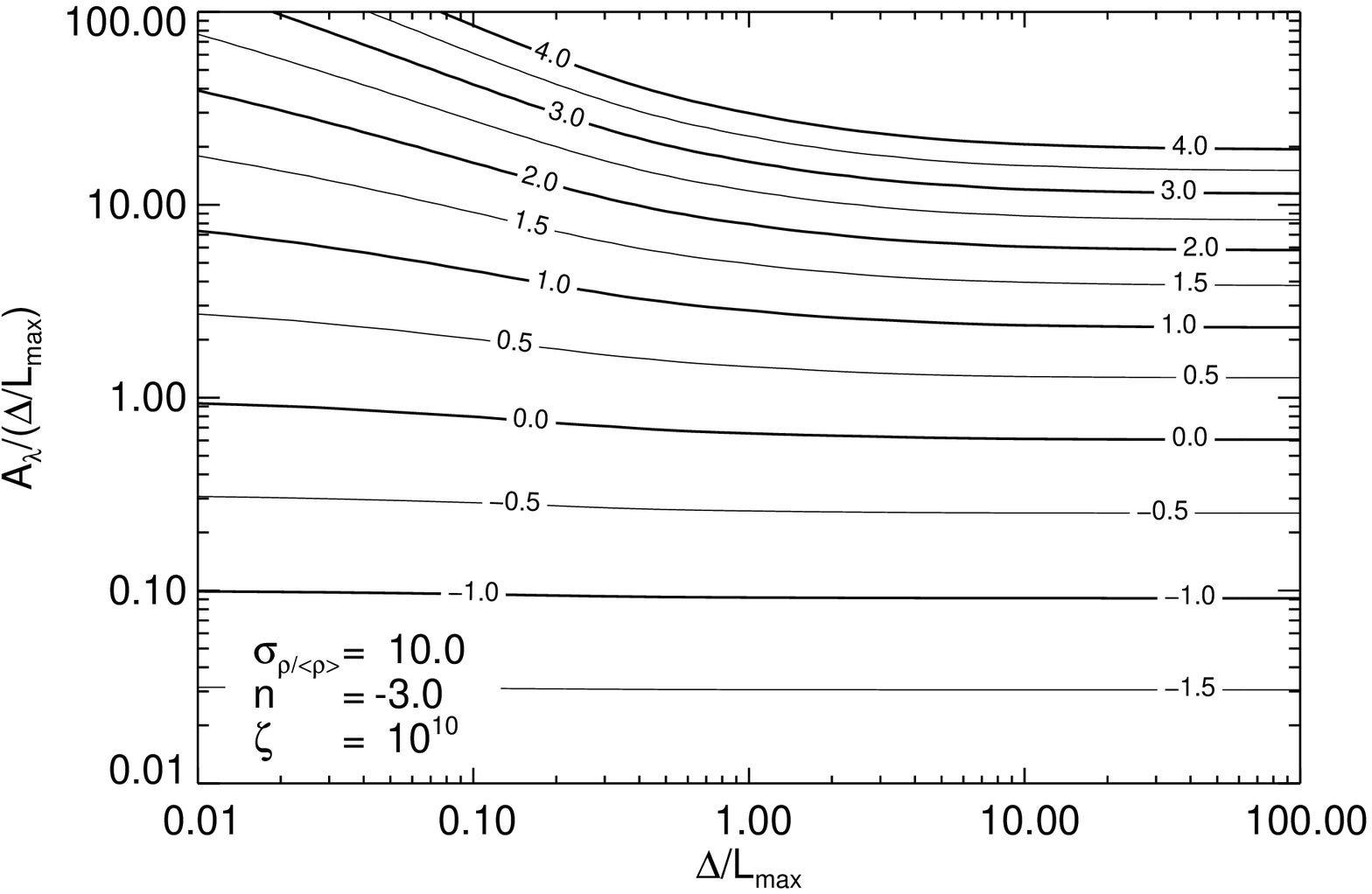}
  \includegraphics[width=0.5\hsize]{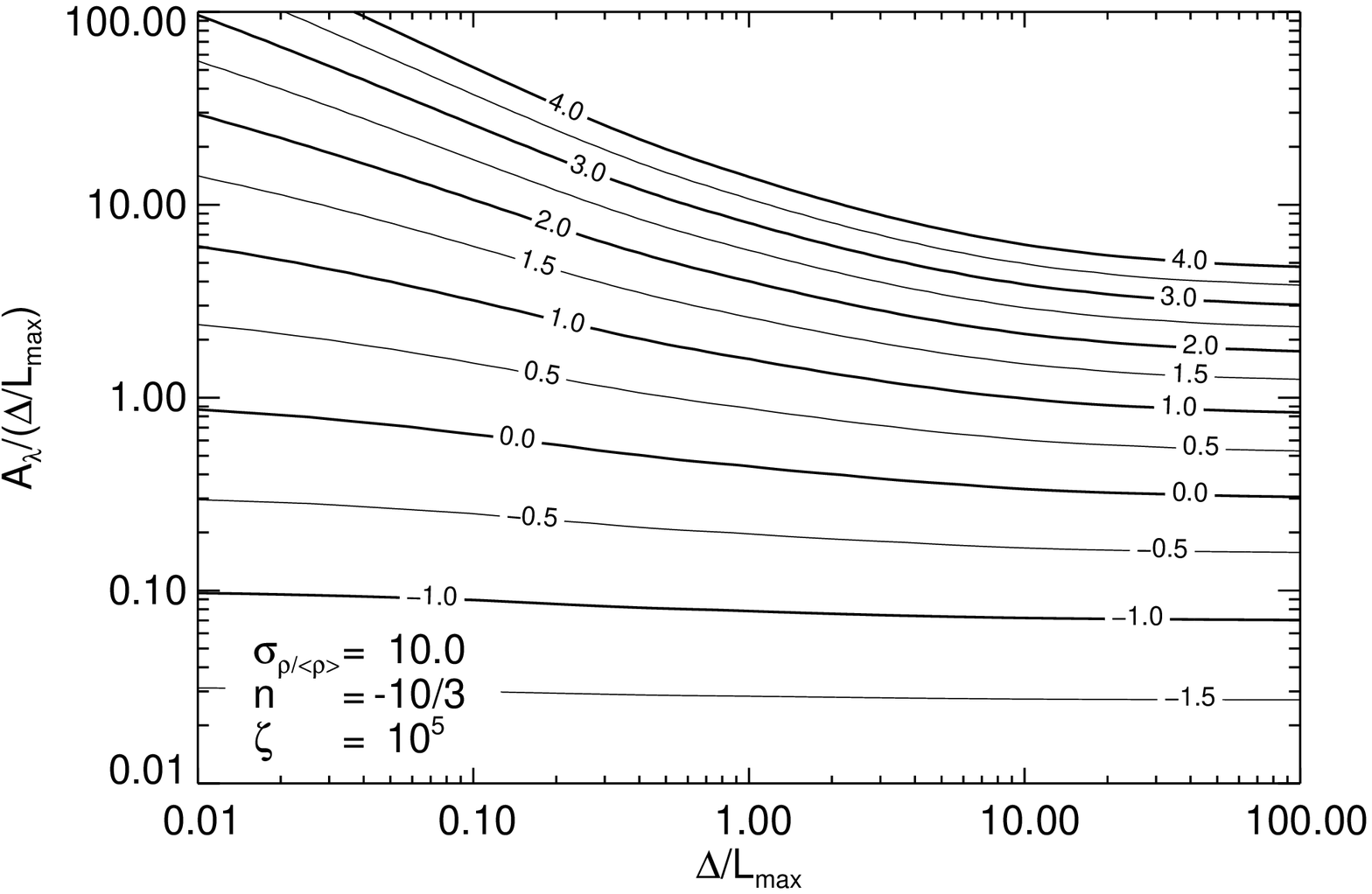}
  
  \caption{\label{figAlambda}
	Attenuation ${\rm A}_{\lambda}/(\Delta/{L_{\rm max}})$ as function of slice thickness
	for a foreground turbulent screen with $\sigma_{\rho/\left<\rho\right>}=1.0$, $5.0$,
	and $10.0$. 
	The different curves correspond to different cases of the 
	mean attenuation $\left<{\rm A}_\lambda\right>_{L_{\rm max}}$
	varied from $10^{-1.5}$ to $10^4$.
	The power index of the power spectrum is taken to
	be $n=-3$ and $n=-10/3$ with $\zeta=10^{10}$ and $\zeta=10^5$, respectively. 
  }
\end{figure*}

A very thin slice through the turbulent density structure is obviously also optically
thin. In this case $\left<{\rm A}_\lambda\right>={\rm A}_{\lambda}$. By increasing the thickness, some parts start to become optically thick and
the effective extinction or the attenuation becomes smaller than the value
of an homogeneous screen. To visualise the effect we considered the corresponding
attenuation at a thickness of one maximum scale ${L_{\rm max}}$. The variation of ${\rm A_{\lambda}}/(\Delta/L_{\rm max})$ with slice thickness for different assumptions of the mean attenuation
$\left<\rm A_{\lambda}\right>_{L_{\rm max}}$ is shown
in Fig.~\ref{figAlambda}. The standard deviation $\sigma_{\rho/\left<\rho\right>}$ of the local density 
is considered to be $1.0$, $5.0$, and $10.0$.
For Kolmogorov turbulence we have chosen $\zeta={L_{\rm max}/L_{\min}}=10^5$ high enough
to have an effect on the results. For a turbulent medium with $n=-3$ we haven chosen
a scaling relation ranging over 10 magnitudes.

As expected the effect of the turbulent screen increases with $\sigma_{\rho/\left<\rho\right>}$ and
$\left<{\rm A}_\lambda\right>_{L_{\rm max}}$. In case of slices thinner than the maximum scale the attenuation ${\rm A_\lambda}/(\Delta/{L_{\rm max}})$ drops relatively strongly with slice thickness. In the limit of slices typically larger than the maximum scale the attenuation approaches asymptotically a lower limit. As
will be shown, this limit is, for a given power spectrum, determined by the single expression
$\sigma_{\rho/\left<\rho\right>}^2\left<\tau\right>_{L_{\rm max}}$. The screen will be significantly different from the homogeneous screen if this expression is larger than $\sim 1$ independent of the actual fluctuations of the column density.

The effect of a turbulent medium on the attenuation is weaker in case of a power spectrum
with $n=-3$. This is expected as there is more power in small density fluctuations. Due
to averaging effects the standard deviation $\sigma_{N/\left<N\right>}$ of such a medium is smaller than the corresponding value of a Kolmogorov turbulent medium (see Paper II).

\subsection{Variation of $\rm R_V^A$ with slice thickness}

\label{sectRV}

To study the effect of the slice thickness of the screen on ${\rm R_V^A}$ we 
considered the same turbulent density structures as in Sect. \ref{sectAV}.
The variation of ${\rm R_V^A}$ with slice thickness is shown in Fig.~\ref{figrv1}.

%

The dependence of ${\rm R_V^A}$ with slice thickness reflects the 
behaviour of ${\rm A_V}$.
In general the attenuation curves become flatter with increasing
thickness of the dust screen.
The $\rm R_V^A$-value increases strongly with slice thickness
if the slice is thin in comparison with the largest scale ${L_{\rm max}}$
but approaches a maximum value for thick slices once the thickness
is larger than $L_{\rm max}$. The flattening is stronger in more turbulent
media characterised by a larger standard deviation $\sigma_{\rho/\left<\rho\right>}$
and in more optical thick media.
As expected from the results found for $\rm A_V$ the flattening is much weaker 
in the case of $n=-3$.

\subsection{$\rm R_V^A$ as function of $\rm A_V$}

In Fig.~\ref{figrv2} we show how the relation of the two observable quantities $\rm R_V^A$ and 
$\rm A_V$ is affected by different conditions of the turbulent medium.
As can be seen in the figure, for given attenuation $\rm A_V$ the attenuation curve flattens for higher values of the mean attenuation $\left<{\rm A_V}\right>_{L_{\rm max}}$. 
The flattening as function of $\rm A_V$ for given $\left<{\rm A_V}\right>_{L_{\rm max}}$ is strongest in a region where $\rm A_V<1$.

One has to be aware that the $\rm R_V^A$-value for a certain attenuation $\rm A_V$ depends on the slice thickness. As an example we can consider a turbulent medium with standard deviation $\sigma_{\rho/\left<\rho\right>}=1$ and $n=-10/3$. If $\rm A_V=1$ then $\rm R_V^A<3.3$ unless the thickness is smaller than $\Delta/L_{\rm max}=1$.

The relation shown in Fig.~\ref{figrv2} has consequences for the interpretation
of the change by dust obscuration of the intrinsic spectral energy distribution from galaxies. 
A stellar spectrum well described over a certain wavelength range by a simple power
law $I_\lambda\propto \lambda^\beta$ \citep{Calzetti94} with power $\beta$ will have a different
functional form if seen through a dust screen. 
If the spectrum is observed at two different wavelengths $\lambda_1$ and $\lambda_2$ the 
inferred power $\tilde \beta$ would be different from $\beta$ by
\begin{eqnarray}
  \Delta\beta &= &\beta-\tilde \beta=\frac{\tau_{\lambda_1}-\tau_{\lambda_2}}{\ln\lambda_1-\ln\lambda_2}\nonumber\\ &=&
  	\frac{1}{\rm \tilde R_{\lambda_2}^{\rm A}}\frac{\tau_{\lambda_2}}{\ln\lambda_1-\ln\lambda_2}.
\end{eqnarray}
Here, we introduced the absolute-to-relative attenuation ratio
$\rm \tilde R_{\lambda_2}^{\rm A}=\tau_{\lambda_2}/(\tau_{\lambda_1}-\tau_{\lambda_2})$. $\tau_{\lambda_1}$ and $\tau_{\lambda_2}$ are the effective optical depths at wavelength $\lambda_1$ and $\lambda_2$.

As can be seen in Fig.~\ref{figrv2} the variation $\Delta\beta$ depends linearly on the optical depth only in the limit of a thick slice with $\Delta/L_{\rm max}\gg 1$ or in case of a
homogeneous screen where $\rm \tilde R_{\lambda_1}^{\rm A}=\tilde R_{\lambda_1}$ and 
is determined by the extinction coefficients $k_{\lambda_1}$ and $k_{\lambda_2}$. 
In case of a turbulent screen the attenuation curve is more flat and therefore $\Delta\beta$ smaller in comparison with a homogeneous screen with the same attenuation $\rm A_V$.

\subsection{Attenuation for certain thickness}

In Fig.~\ref{2dslice} we show both the attenuation $\rm A_V$ and the absolute-to-relative attenuation ratio
${\rm R_V^A}$ in the parameter plane
defined by $\sigma_{\rho/\left<\rho\right>}$ and $\left<{\rm A_V}\right>$ for
three slices through the turbulent density structure with varied thickness
$\Delta/{L_{\rm max}}$ chosen to be 0.1, 1.0, and 10.0. The plot is similar
to the one shown in Fig.~\ref{reffcontourplus} where the attenuation is shown as
function of $\sigma_{\tau/\left<\tau\right>}$ and $\left<{\rm A_V}\right>$.


The lines of constant ${\rm R_V^A}$ are defined by $\sigma_{\rho/\left<\rho\right>}^2\left<\tau\right>={\rm constant}$ in the region where they become parallel. In this region constant $\rm R_V^A$ also defines ${\rm A_V/\left<A_V\right>}={\rm constant}$.
As is shown in Sect.~\ref{sectapprox_rvthick} this characteristic behaviour appears in the limit
of thick slices with $\Delta/{L_{\rm max}}\gg 1 $ and where $\sigma_{\ln \tau}\ll 1$.

\subsection{Approximation for thick slices}

\label{sectapprox_rvthick}
In the limit of thick slices with $\Delta/{L_{\rm max}}\gg 1$ the effect of a turbulent
screen on the attenuation curve can be described analytically for the case where
the fluctuations of the optical depths become sufficiently small.
As discussed in Paper~II the standard deviation of the column density of a thick slice 
through the turbulent medium with standard deviation $\sigma_{\rho/\left<\rho\right>}$ 
and a power spectrum with $n \ne -3$ and $n \ne -2$ is given by 
 \begin{equation}
  \sigma_{N/\left<N\right>}^2=\onehalf\sigma_{\rho/\left<\rho\right>}^2
  	\frac{(n+3)(1-\zeta^{n+2})}{(n+2)(1-\zeta^{n+3})}\left(\frac{\Delta}{L_{\rm max}}\right)^{-1}
\end{equation}
and for $n=-3$ by
\begin{equation}
  \sigma_{N/\left<N\right>}^2 = \sigma_{\rho/\left<\rho\right>}^2\frac{1-1/\zeta}{2\ln \zeta}
  \left(\frac{\Delta}{L_{\rm max}}\right)^{-1}.
\end{equation}
If $\sigma_{\ln N}\ll 1$ then Eq.~\ref{eqstandard deviation} can be replaced by
$\sigma_{\ln N}\approx \sigma_{N/\left<N\right>}$. Therefore, the product $\sigma_{\ln N}^2\left<\tau\right>$
becomes a constant. For a power spectrum with $n<-3$ and $\zeta\gg 1$ this expression is given by:
\begin{equation}
    \sigma_{\ln N}^2\left<\tau\right> = \frac{1}{2}\frac{n+3}{n+2}\,\sigma^2_{\rho/\left<\rho\right>}\left<\tau\right>_{L_{\rm max}}
\end{equation}
and for $n=-3$ and $\zeta\gg 1$ by
\begin{equation}
  \sigma_{\ln N}^2\left<\tau\right> = \frac{1}{2}\frac{1}{\ln \zeta} \,\sigma_{\rho/\left<\rho\right>}^2\left<\tau\right>_{L_{\rm max}}
\end{equation} 
where $\left<\tau\right>_{L_{\rm max}}$ is the mean optical depth for a slice with thickness $\Delta={L_{\rm max}}$. 
In the limit of a thick slice the attenuation $\rm A_\lambda/\left<A_\lambda\right>$ and the absolute-to-relative attenuation ratio$\rm R_V^A$ become simply a function of the product 
$\sigma_{\rho/\left<\rho\right>}^2\left<\tau_\lambda\right>_{L_{\rm max}}$ and can be derived by
using the approximations of equations \ref{eqapprox_av} and \ref{eqapprox_rv}.

The dependence of $\rm A_\lambda/\left<A_\lambda\right>$ and
${\rm R_V^A}$ on $\sigma_{\rho/\left<\rho\right>}^2\left<\tau_\lambda\right>_{L_{\rm max}}$ 
in the limit of thick slices is shown in Fig.~\ref{av&rv_limit}.
The power $n$ is
chosen to be $-11/3$, $-10/3$, and $-3$. For all cases a scaling relation extending over 10 magnitudes is assumed. 

The effect of the turbulent dust screen increases towards steeper power laws.
For $n=-10/3$ or $n=-11/3$ 
the attenuation starts to deviate significantly from the homogeneous screen for $\sigma_{\rho/\left<\rho\right>}^2\left<\tau\right>_{L_{\rm max}}\approx 1$. In case of flatter power laws the transition from a homogeneous to a clumpy screen is shifted towards higher values.

For a known power spectrum the combined information of 
${\rm R_V^A}$ and ${\rm A_V}$ allows a determination
of $\left<{\rm A_V}\right>$. If these curves are applied in cases of turbulent screens which are not necessarily thick the derived value has to be considered as \emph{lower} limit.

In Fig.~\ref{rvavlimitcontour} the attenuation caused by a thick turbulent screen is shown in
the plane defined by $\sigma_{\rho/\left<\rho\right>}$ and $\left<{\rm A_V}\right>/(\Delta/{L_{\rm max}})$.
As seen all contour lines defining different attenuation curves (or $\rm R_V^A$-values) are now parallel.
The curves showing different
values of the attenuation ${\rm A_V}/(\Delta/{L_{\rm max}})$ can be obtained by shifting a single curve
along those parallel lines.


%

\section{Discussion}

\subsection{Accuracy of the model}

\subsubsection{Error caused by the representation of the PDF by a log-normal density function}

\label{sectdiscussaccuracy1}

The largest uncertainty in the calculations as presented here is caused by the simplification of the PDF of the column densities by a log-normal density distribution. As we have shown in Paper~II, deviations exist when turbulent density structures with $\sigma_{\rho/\left<\rho\right>}>1$ are considered and these deviations become more prominent for a wider PDF of the local density. 
Therefore, the correlation between the standard deviation $\sigma_{N/\left<N\right>}$ and the standard deviation $\sigma_{\ln N}$ of the log-normal density distribution as given in Eq.~\ref{eqstandard deviation} is not strictly valid in case of broad distributions of the local density.
By using this correlation, we might overestimate the probabilities
at low column densities but underestimate those at high column densities. 

As the attenuation is mainly determined by the probabilities of low optical depths a better result would be obtained by using a log-normal density distribution with smaller standard deviation. 
This is particularly true in the case of turbulent media with high mean optical depth.

As stated in Paper~II that the difference between the actual PDF and the log-normal density distribution becomes smaller for thick slices and should disappear altogether in the limit of slices with a thickness much larger than the maximum scale ${L_{\rm max}}$. In this case the PDF becomes a narrow Gaussian centred around unity. Therefore, the results presented for very thick slices will be accurate. The same applies for very thin slices where the PDF of the column density is very similar to the PDF of local density. 

The largest difference will appear for slices slightly larger than one
maximum scale and for wide density distributions of the local density.
As the PDF overestimates the probabilities at low column densities the effect caused by
the turbulent screen would be smaller than presented here. The actual transparency would be
lower and the ${\rm R_V^A}$-value smaller.

To assign an uncertainty to the data presented here we simply assume that the actual form is still represented by a log-normal density distribution but with a smaller standard deviation 
$\sigma_{N/\left<N\right>}=\sigma_{A_{\rm V}/\left<A_{\rm V}\right>}$.
The derived uncertainties of the quantities ${\rm A_V}$ and ${\rm R_V^A}$ is shown
in Fig.~\ref{figaccuracy_b}. As can be seen the effect due to uncertainties in the
standard deviation depends on the mean attenuation $\left<{\rm A_V}\right>$ and the standard deviation of the column density. The error of the attenuation increases
generally with optical thickness and can be quite substantial in case of turbulent media with very high mean optical depth. The accuracy of ${\rm R_V^A}$ is somewhat smaller. 

In  Paper~II we analysed the uncertainty by approximating the PDF of the column
density by a simple log-normal density distribution. The results can not be simply generalised
as they may depend on the dynamic range of the turbulent density structure which in these calculation was only $\zeta=5.4$. However, for a standard deviation of $\sigma_{\rho/\left<\rho\right>}=2.5$ the maximal error of the standard deviation of the column density was found to be slightly larger than 12\% which appears at a slice thickness of $\Delta/{L_{\rm max}}\approx 2$. Assuming a linear increase of the error then the maximal error for $\sigma_{\rho/\left<\rho\right>}=5$ is roughly 25\% and the maximal error for $\sigma_{\rho/\left<\rho\right>}=10$ already $\sim 50\%$.

But even for these extreme turbulent media with broad density distributions of the local
density the $\rm R_V^A$-values derived here will have less than $\sim 10\%$ error
if the screen has a mean attenuation $\left<{\rm A_V}\right>=1$. For a screen with ten times higher mean attenuation the accuracy is still better than $\sim 15\%$.

\subsubsection{Dependence on $\zeta$}

In general the attenuation with slice thickness depends not only on the number of turbulent scale length but also on the dynamic range $\zeta$ of the turbulent density structure. For comparison
we calculated the attenuation $\rm A_V$ and $\rm R_V^A$ as function of slice thickness assuming a relatively low value $\zeta=10$. 
The standard deviation of the local density is chosen to be $\sigma_{\rho/\left<\rho\right>}=5$.
The results are shown in Fig.~\ref{figAlambda_2} and Fig.~\ref{figrv3}. 
If the turbulence extends over a smaller dynamic range, then the averaging effect along the line of sight is smaller, and the distribution of the column densities is therefore wider. As consequence the screen will be more transparent and the flattening of the attenuation curve greater, or its ${\rm R_V^A}$-value larger.

As seen in the figure the effect for Kolmogorov turbulence is relatively small but for $n=-3$ the differences are quite substantial. In fact the results obtained for both different assumptions of the scaling relation are now quite similar as the standard deviation of the column density is now almost identical, as can be seen in Fig.~3 in Paper~II.

Simulations are necessarily quite limited and are not able to reproduce a scaling relation over several magnitudes as measured for the ISM. As mentioned already in Paper~II the accuracy of the attenuation derived by using simulations
and the conclusion about the statistical properties of the medium will depend on the actual power law of the turbulent density structure. For $n<-3$ the error might be quite small as in case of Kolmogorov
turbulence but for $n\geq -3$ this effect will obviously be important.

\subsection{Application of the model}

Despite the simplicity of the model presented here, which does not take into account
scattered light, it may well prove useful in helping to understand the
attenuation of the light from highly obscured objects such as young compact HII-regions or even of entire galaxies. 
As the model is based on the physical properties of the turbulent
density structure it may lead to realistic corrections which will be helpful
to determine intrinsic important parameters as such as emission line ratios 
or star-formation rates.
As discussed in Paper~I the model already has been proven to be a natural explanation
of the attenuation curve (the `Calzetti-extinction-law') derived empirically for star burst galaxies.

The model connects the observable quantities ${\rm A_V}$ and $\rm R_V^A$ with the
physical parameters $\sigma_{\rho/\left<\rho\right>}$, $\tau_{L_{\rm max}}$, and $\Delta/{L_{\rm max}}$ and provides therefore a method to provide insight into the turbulent
density structure through measurement of the attenuation curve alone. A major complication
in the interpretation is certainly the fact that the turbulent medium is characterised by
three parameters (for known power spectrum) while the attenuation curve only provides
two. 

A conservative lower limit of the standard deviation of the local density can be obtained by considering
a turbulent screen with a thickness much larger than the maximum scale ${L_{\rm max}}$.
A better estimate of the turbulence can be achieved by analysing the variation of the
attenuation curve with slice thickness for different assumptions of $\sigma_{\rho/\left<\rho\right>}$
and $\left<{\rm A_V}\right>_{L_{\rm max}}$ as shown in Fig.~\ref{figrv1}.

It is possible that the turbulence varies from galaxy to galaxy. In
a generalised model it prove useful to describe the turbulence by the mean value of the standard deviation $\sigma_{\rho/\left<\rho\right>}$ and the optical thickness
$\left<\tau\right>_{L_{\rm max}}$ at one maximum scale. In this case the attenuation would
be only a function of the slice thickness $\Delta/{L_{\rm max}}$.
A small attenuation is connected with a thin dust layer, a large attenuation with a thick
dust layer. As shown in the paper the curvature will change with slice thickness through the
turbulent medium. Therefore, the attenuation curve of a galaxy showing a relatively small attenuation should be characterised by a smaller ${\rm R_V^A}$-value.
Equally the attenuation curves of galaxies with high attenuation should be relatively flat with high $\rm R_V^A$.

If we consider spiral galaxies the apparent thickness of the dust screen will vary with viewing angle. Therefore, the attenuation curve of a galaxy seen edge on should be flatter than the attenuation curve of a galaxy seen face on. The effect should be stronger for an ISM which is both more dense and more turbulent.
It seems to be reasonable to assume that the largest turbulent scale $L_{\rm max}$ is connected
with the scale height of the galaxy so that the thickness of the dust screen of a galaxy seen face on should be around one maximum scale. In this case the thickness should vary approximately as $\Delta\approx L_{\rm max}/\cos i$ where $i$ is the viewing angle of the galaxy.

In general the situation for star-bursts and normal galaxies might be different as the star light in star-bursts
is stronger related to HII-regions. In case of normal galaxies it might be important to consider a more realistic
distribution of stars and dust. In spiral galaxies the old stars do, in general, have a greater scale height while young stars are more concentrated towards the galactic disc. In case of a young stellar population a foreground screen model might still be a reasonable approximation. For an old stellar population on the other hand it depends on the actual distribution of the dust and it is possible that a model where the stars are homogeneously mixed inside the turbulent medium is more appropriate. An additional complication might be the attenuation of 
the star light from the bulge. The importance of all these effects will dependent on the contribution of the different stellar populations to the intrinsic star light of the galaxy.

\subsubsection{The 'Calzetti-extinction-law'}

Let us consider the empirical curve derived for star-burst galaxies. We
assume a turbulent density structure with $n=-10/3$ and a scaling relation extending over several
magnitudes.
As shown in Sect.~\ref{attenuation} the relative attenuation $\mathcal{E}_\lambda$ suggests
$\rm R_V\sim 4.3$. We ignore the discrepancy at short wavelengths as this
is possibly due to the destruction of PAH-molecules which are believed to be the carriers of the 2200~\AA{} bump in star bursts. We already discussed this issue in Paper~I and in more detail in \citet{Dopita2005}.

The observations of star-bursts show that the effective optical depth varies quite strongly from
galaxy to galaxy. Following \citet{Calzetti2001} the attenuation ${\rm A_V}$ lies in the range from 
0.20 to 2.23 with a mean of 0.63\footnote{We have taken the attenuation ${\rm A_B}$ (Table 4, \citet{Calzetti2001})
and used an ${\rm R_V^A}$-value of 4.05 as stated in the text to obtain ${\rm A_V}$.}.

Using the approximation of a thick slice
and assuming that the dust screen has a thickness 
of at least one maximum scale ${L_{\rm max}}$ then the lower limit of the standard deviation $\sigma_{\rho/\left<\rho\right>}$ is in the range $2$ to $6$. If the turbulence of these galaxies is similar the standard deviation of the local density should be at least $\sim6$.

\begin{figure*}
  \includegraphics[width=0.49\hsize]{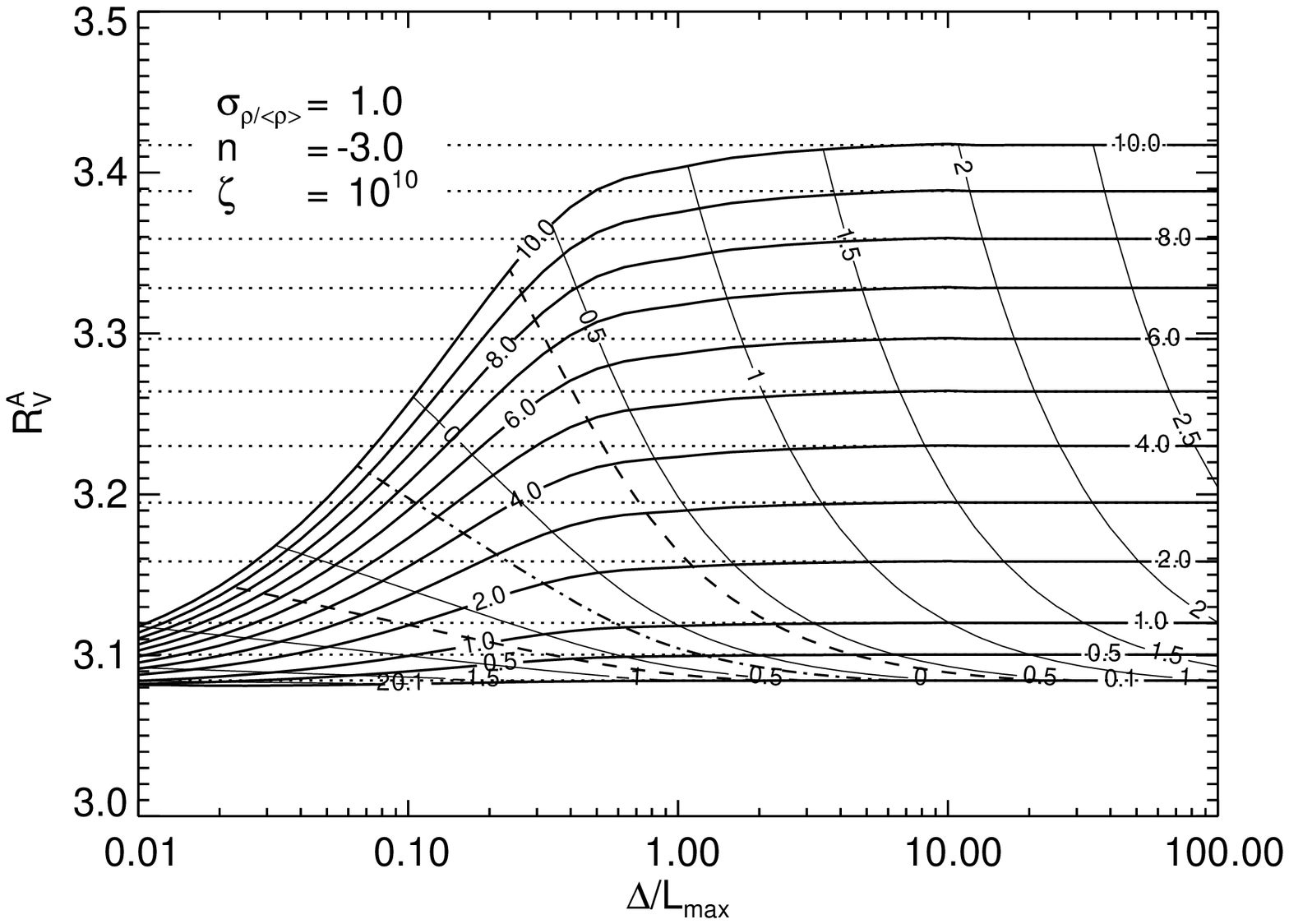}
  \hfill
  \includegraphics[width=0.49\hsize]{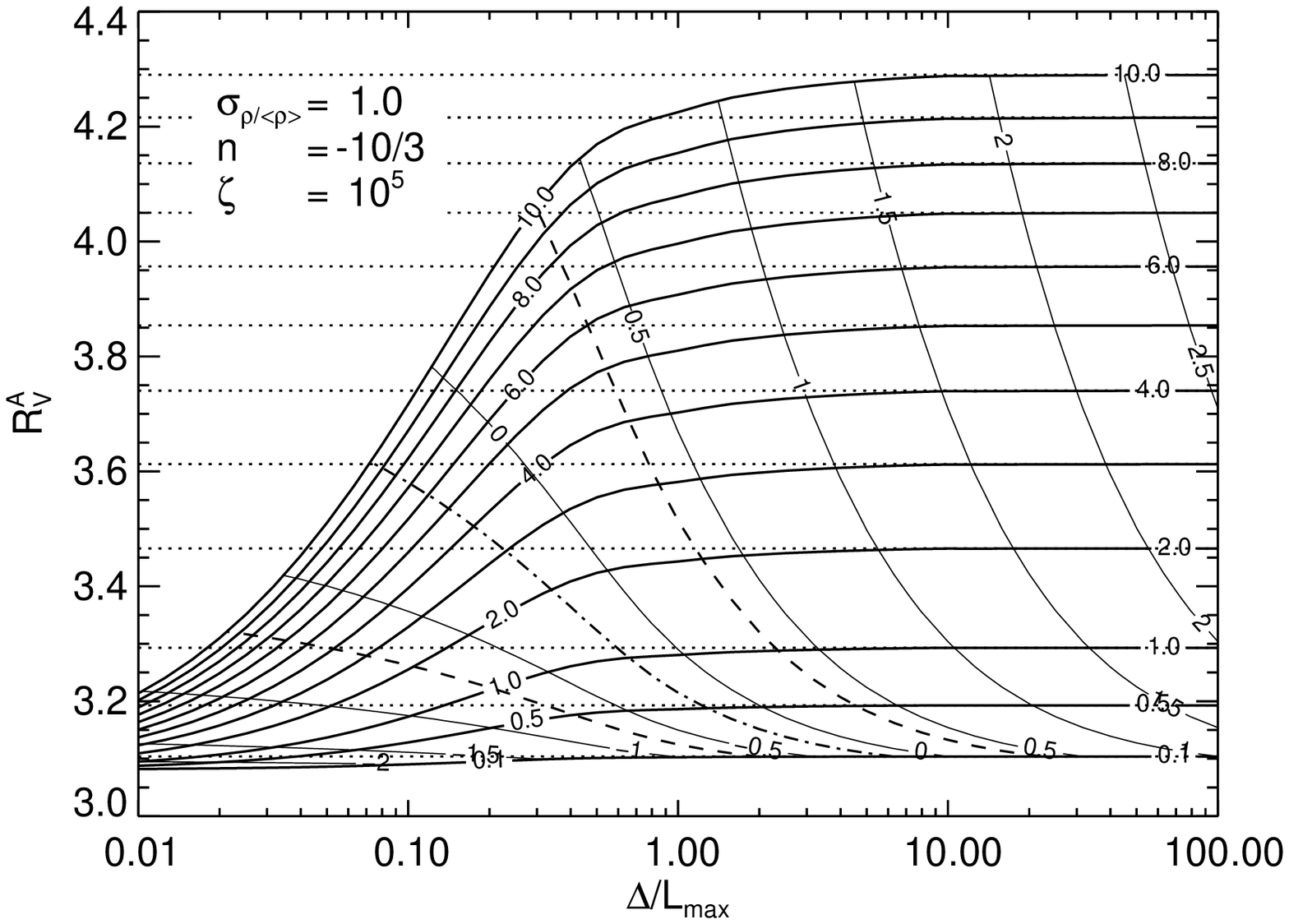}
  \vspace{0.1cm}

  \includegraphics[width=0.49\hsize]{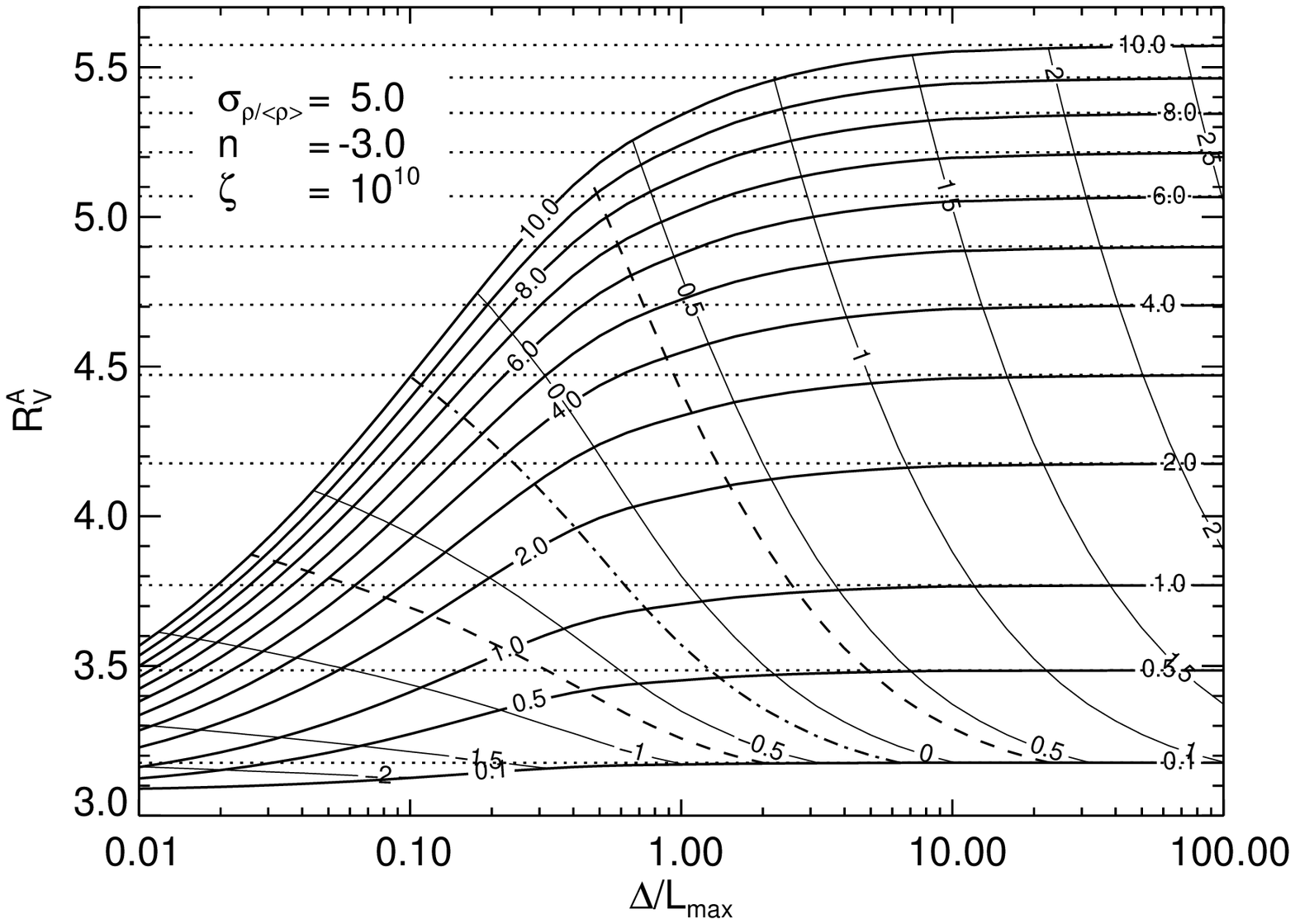}
  \hfill
  \includegraphics[width=0.49\hsize]{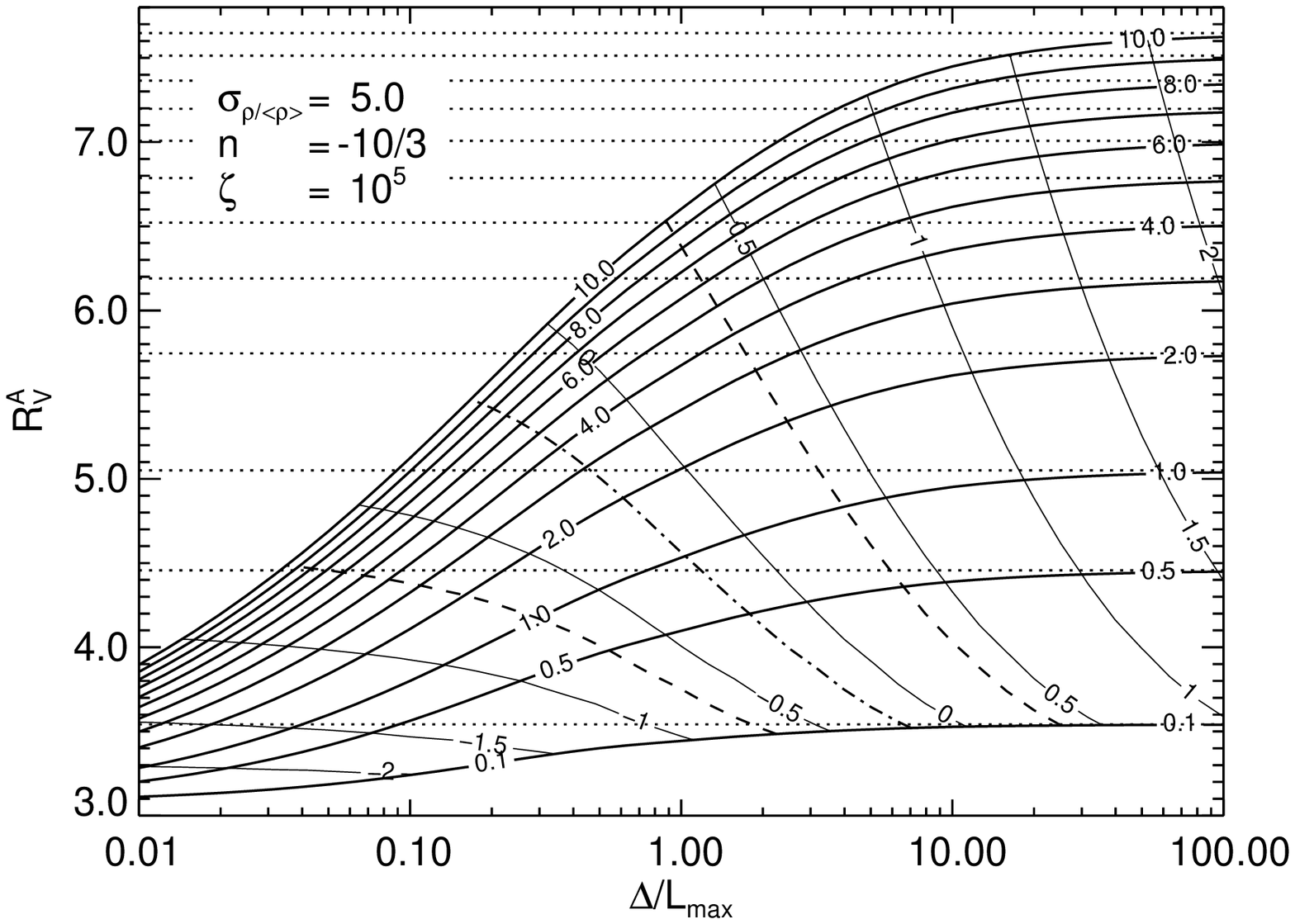}
  \vspace{0.1cm}

  \includegraphics[width=0.49\hsize]{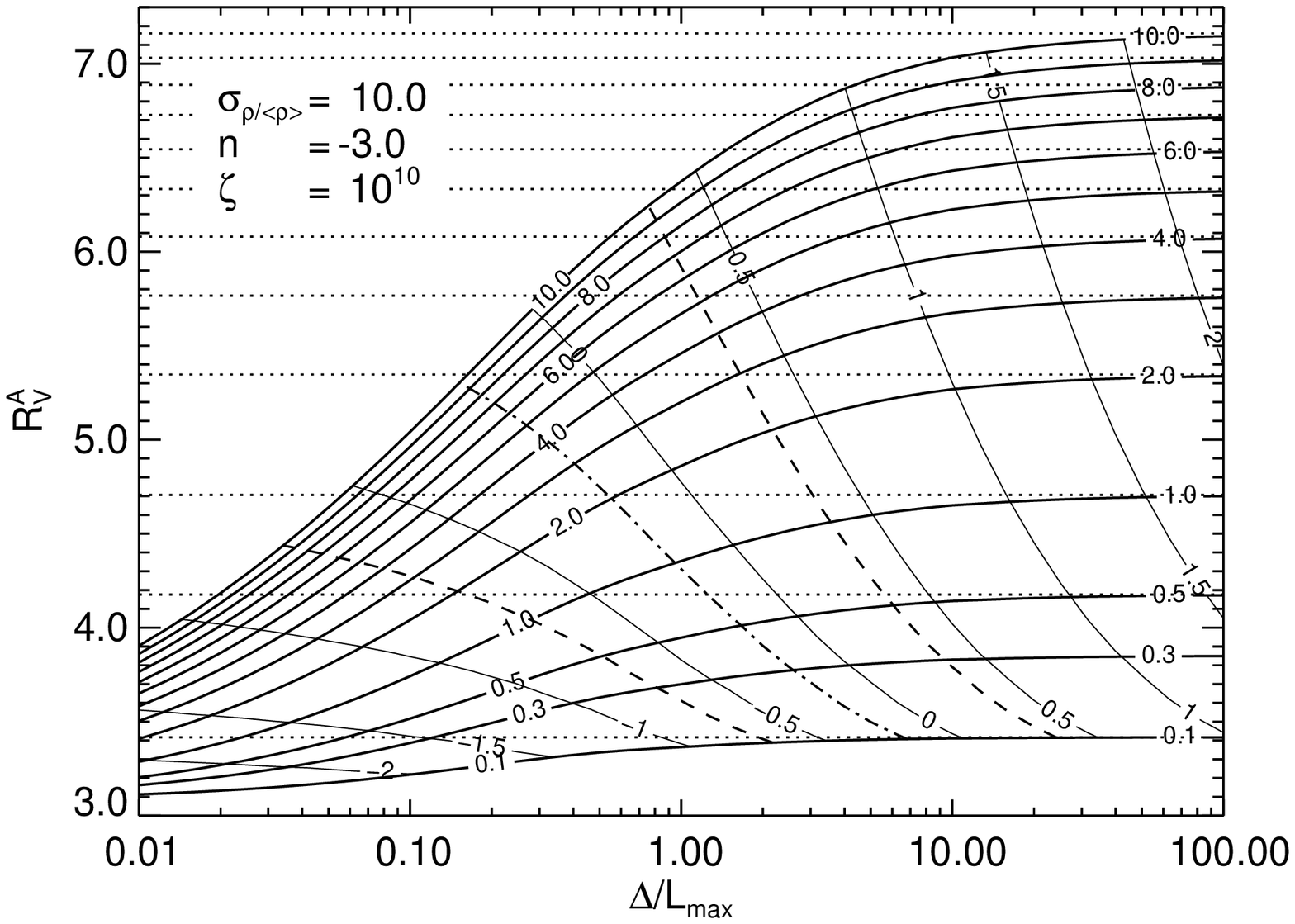}
  \hfill
  \includegraphics[width=0.49\hsize]{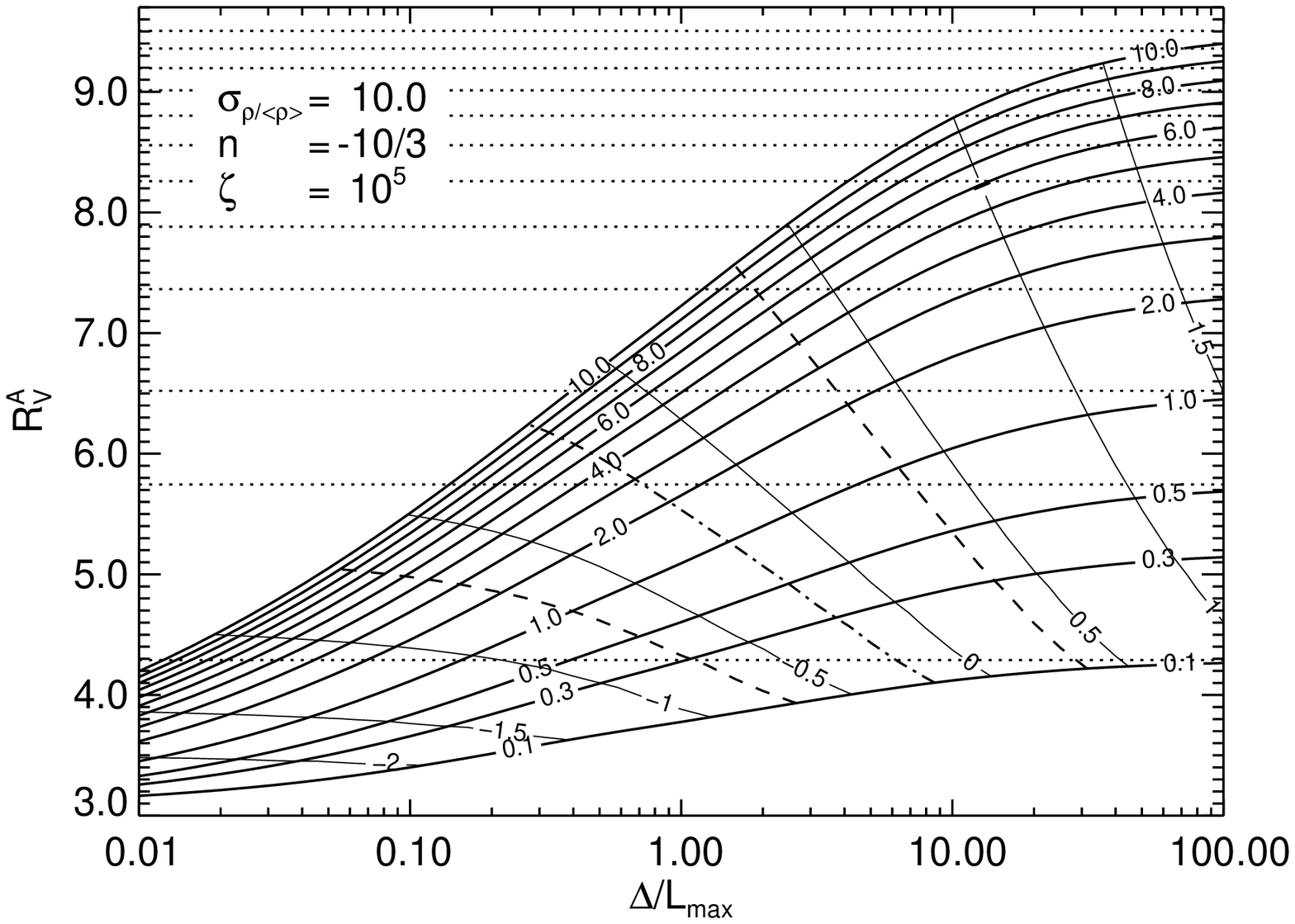}  

  \caption{\label{figrv1}
  Variation of the absolute-to-relative attenuation ratioÊ$\rm R_V^A$ with slice thickness 
  $\Delta/{L_{\rm max}}$ (thick solid lines). The curves are labelled with the
  assumed attenuation $\left<{\rm A_V}\right>_{L_{\rm max}}$ at one maximum scale
  $L_{\rm max}$ ranging from 0.1 up to 10.0. 
  The attenuation ${\rm A_V}$ is shown as thin solid lines. Those
  curves are labelled with $\log {\rm A_V}$. 
  Also shown as dashed, dashed dotted, and dashed curves are a minimum, mean 
  and maximum value of $\rm A_V$ as derived for star-burst galaxies \citet{Calzetti2001}.
  The dotted lines visualise the
  $\rm R_V^A$-value in the limit of thick slices using equation~\ref{eqapprox_rv} .
  The power spectrum is again assumed to have $n=-3$ and
  $n=-10/3$ with $\zeta=10^{10}$ and $\zeta=10^5$, respectively.
  The standard deviation of the local density is varied from 1.0, 5.0 to 10.0.
  }
\end{figure*}

\begin{figure*}
  \includegraphics[width=0.5\hsize]{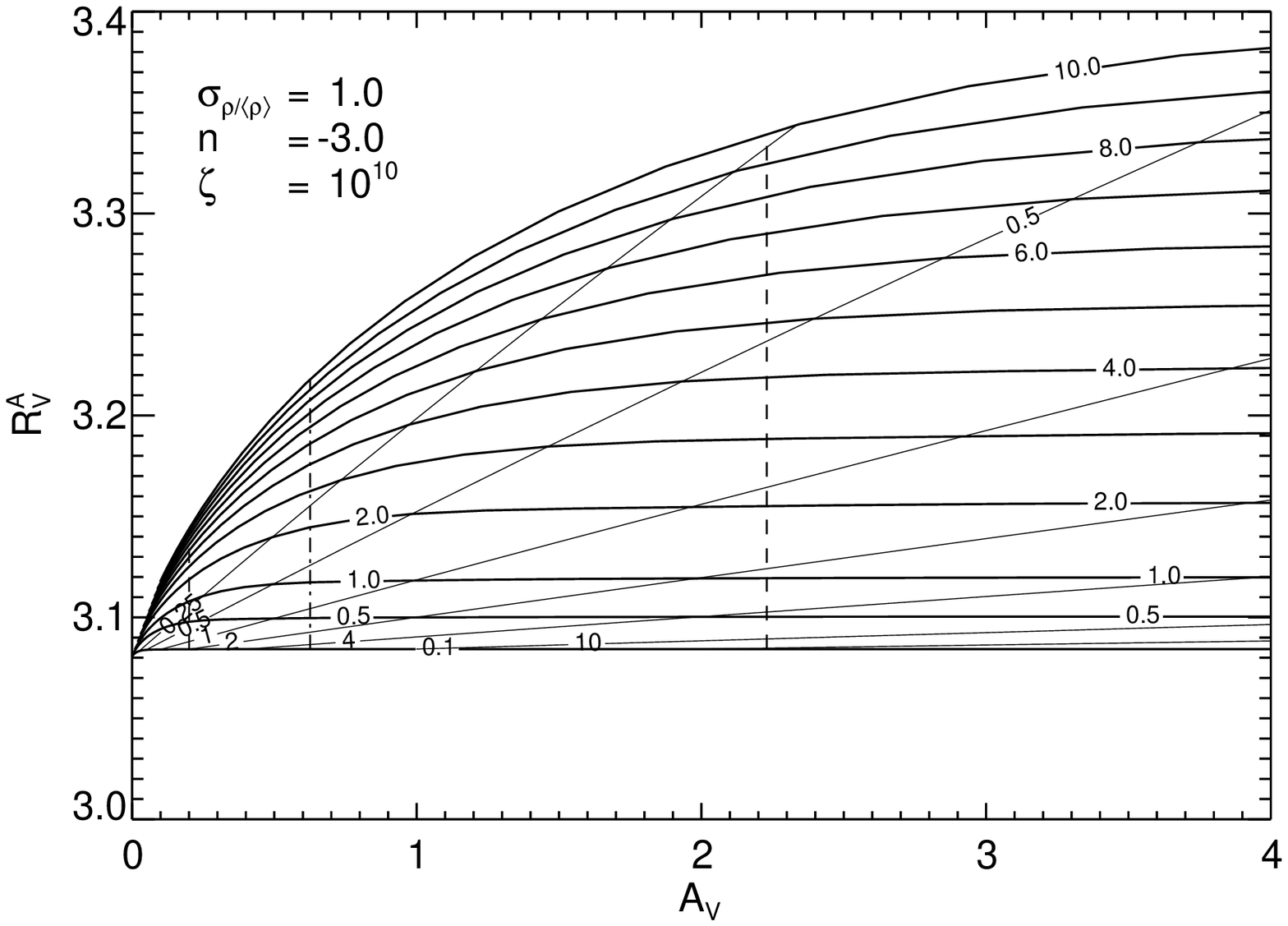}
  \includegraphics[width=0.5\hsize]{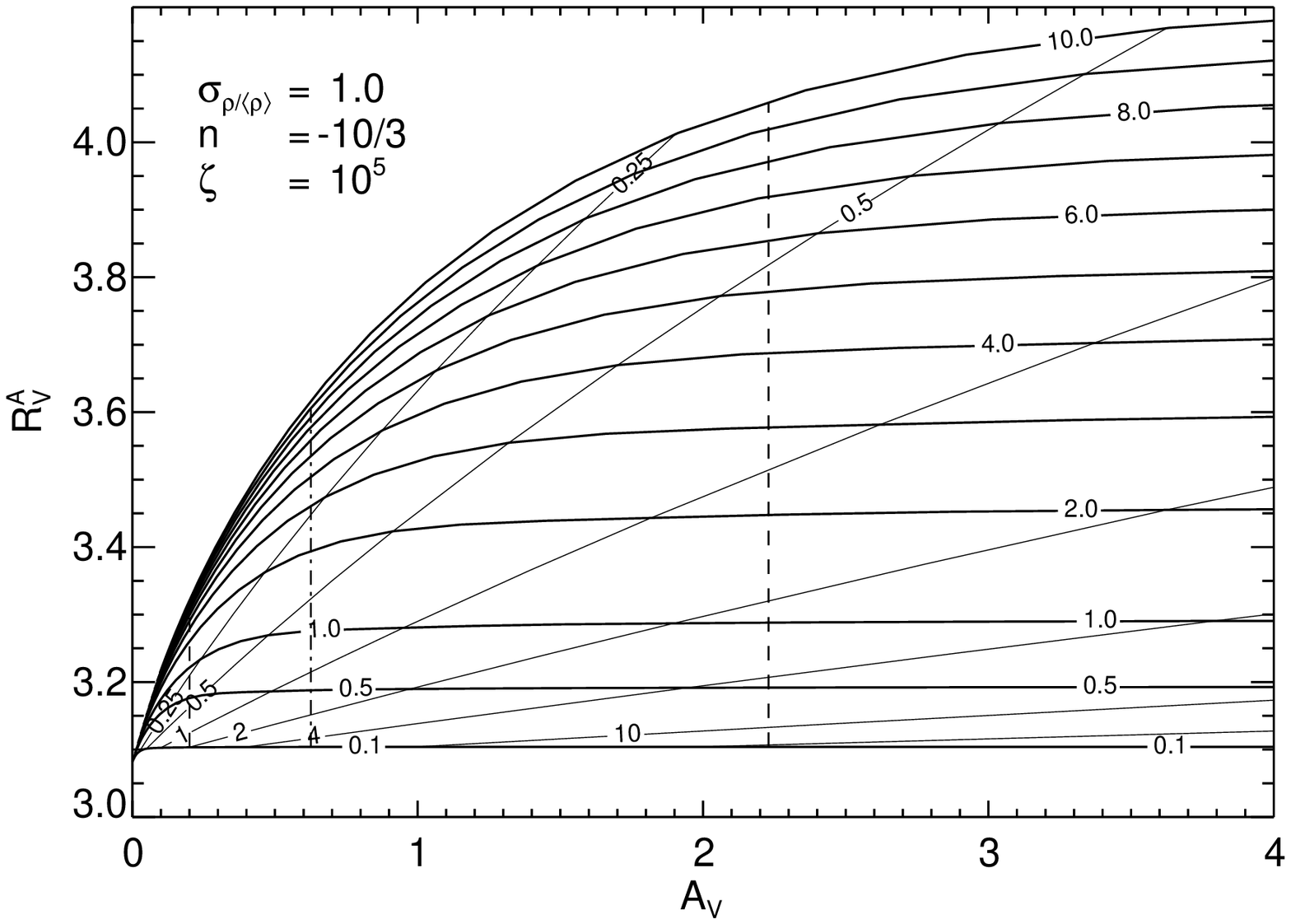}
  \vspace{0.1cm}

  \includegraphics[width=0.5\hsize]{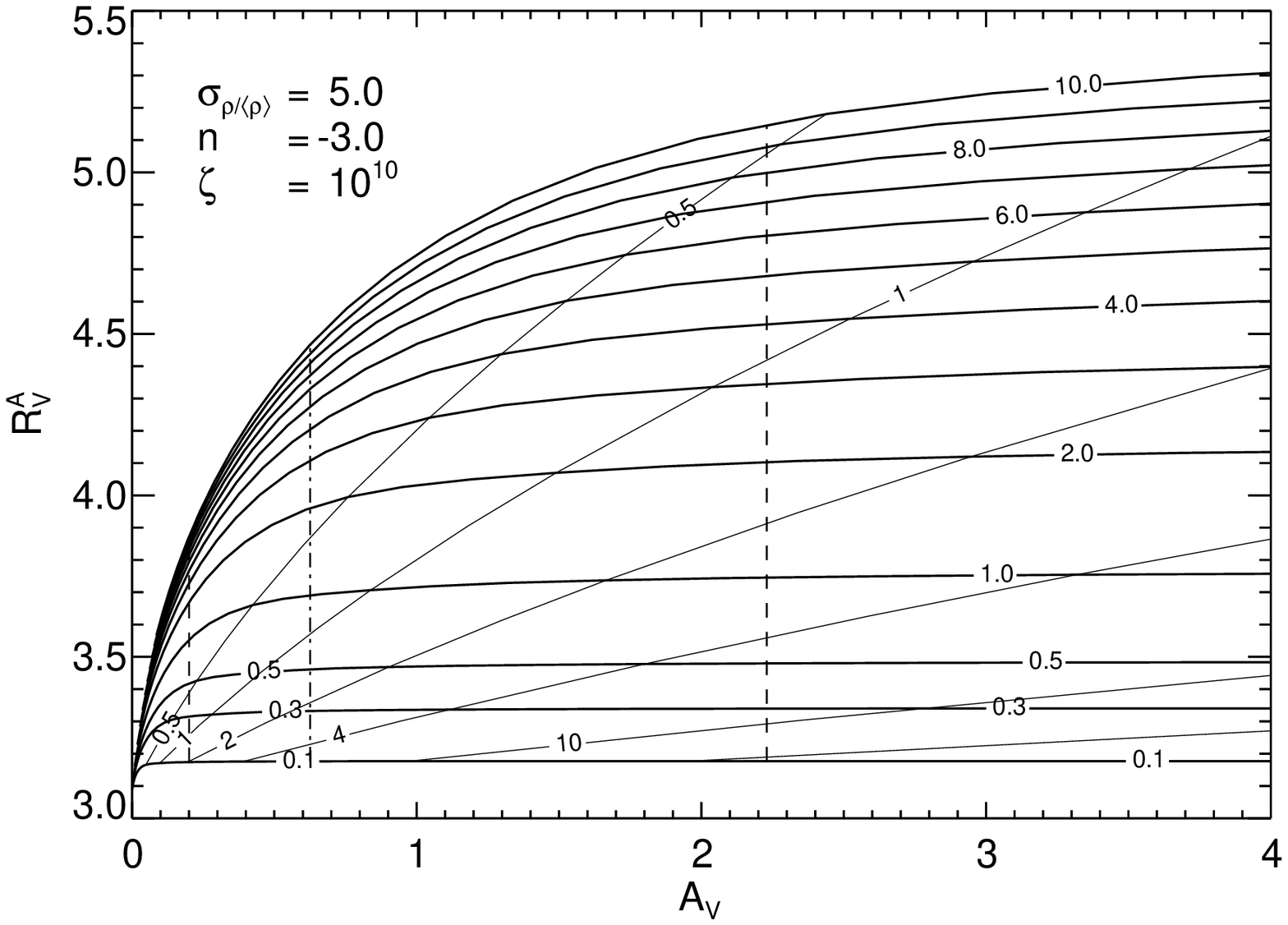}
  \includegraphics[width=0.5\hsize]{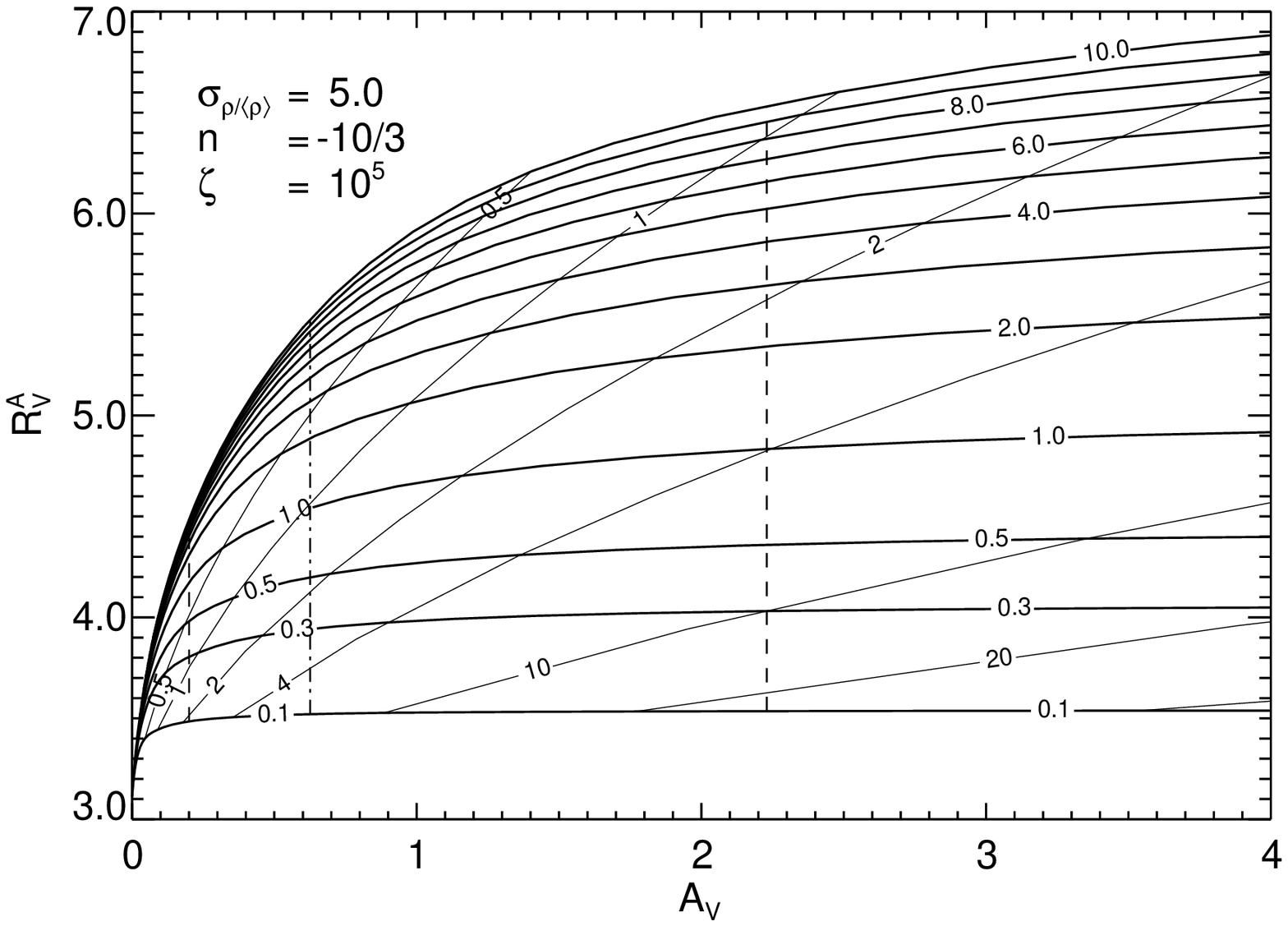}
  \vspace{0.1cm}

  \includegraphics[width=0.5\hsize]{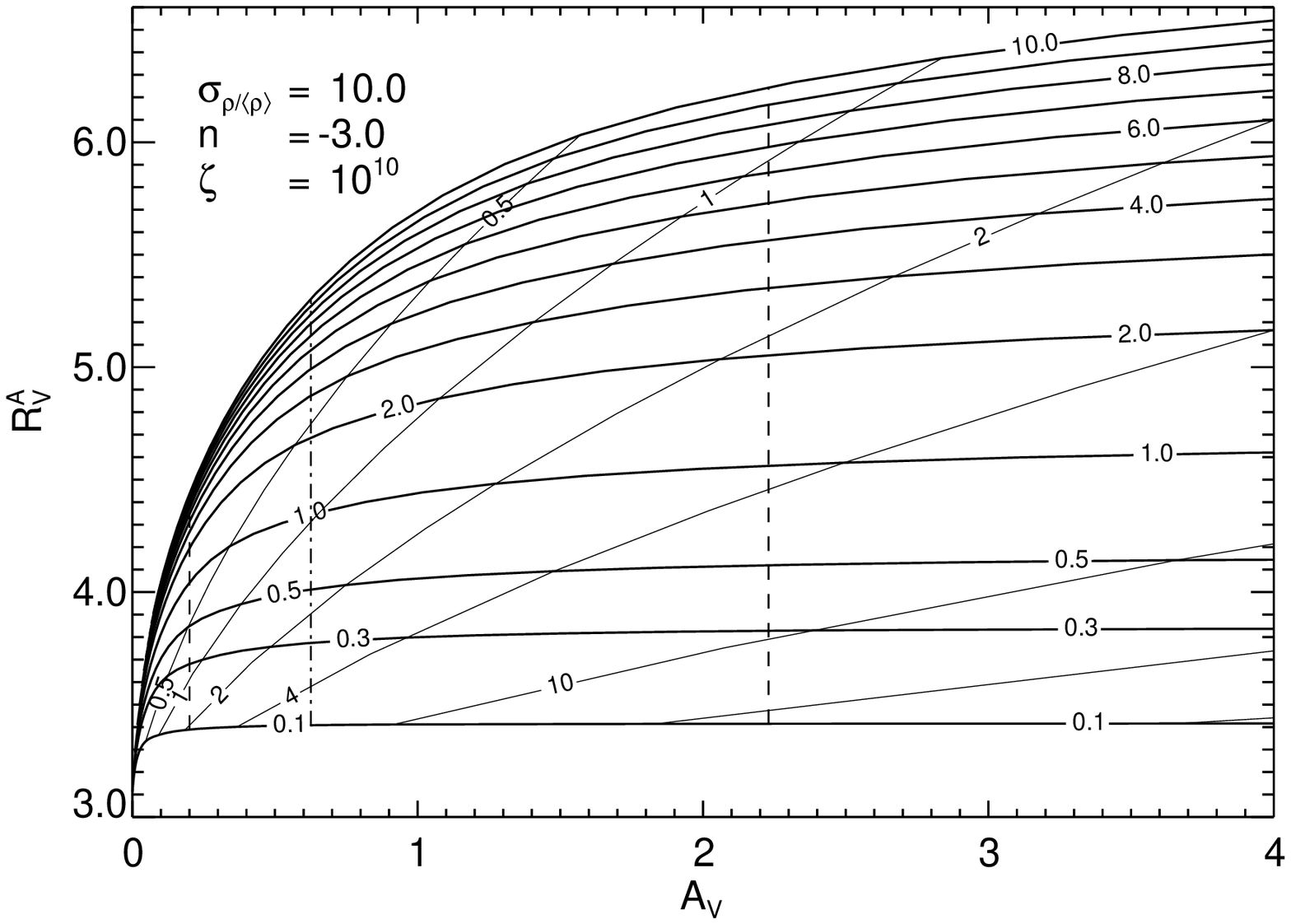}
  \includegraphics[width=0.5\hsize]{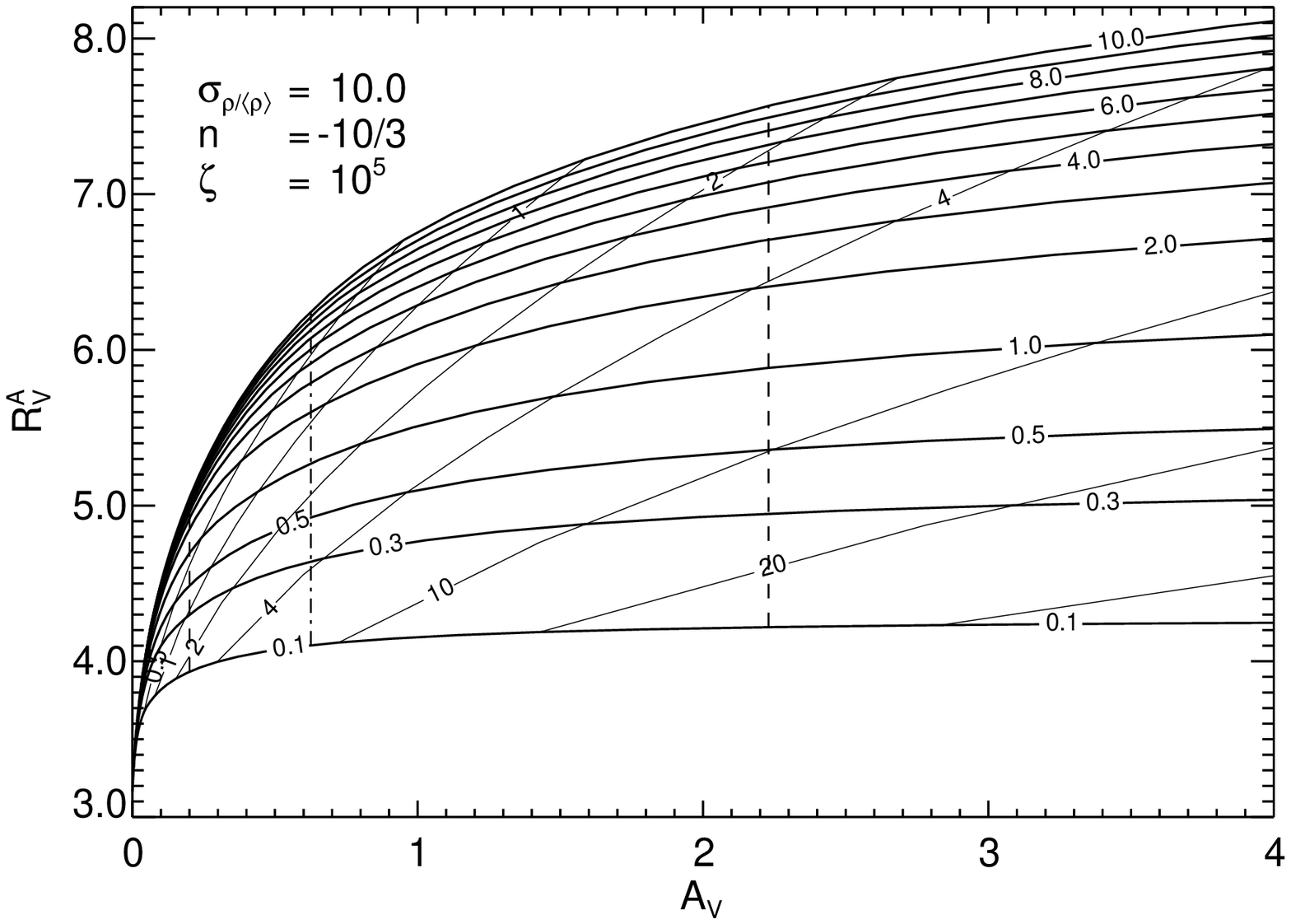}  

  \caption{\label{figrv2}
	Absolute-to-relative attenuation ratio $\rm R^A_V$ as function of the measurable
	attenuation 
	$\rm A_V$ for different assumptions of the distant turbulent screen. The curves shown 
	as thick lines correspond to certain values of the mean attenuation 
	$\left<\rm A_V\right>_{L_{\rm max}}$ at one maximum turbulent scale varied from 
	0.1 to 10. The thickness $\Delta/L_{\rm max}$ of the screen is shown as thin lines. 
	The power is again assumed to be $-3$ or $-10/3$ with a scaling relation ranging over
	$10$ and $5$ magnitudes. The standard deviation of the local density is again chosen to be 1, 5, 
	and 10. For comparison also the range of the attenuation measured for star burst galaxies 
	is shown, given as broken lines. The dashed-dotted line is its mean value.  }
\end{figure*}

\begin{figure*}
  \includegraphics[width=0.49\hsize]{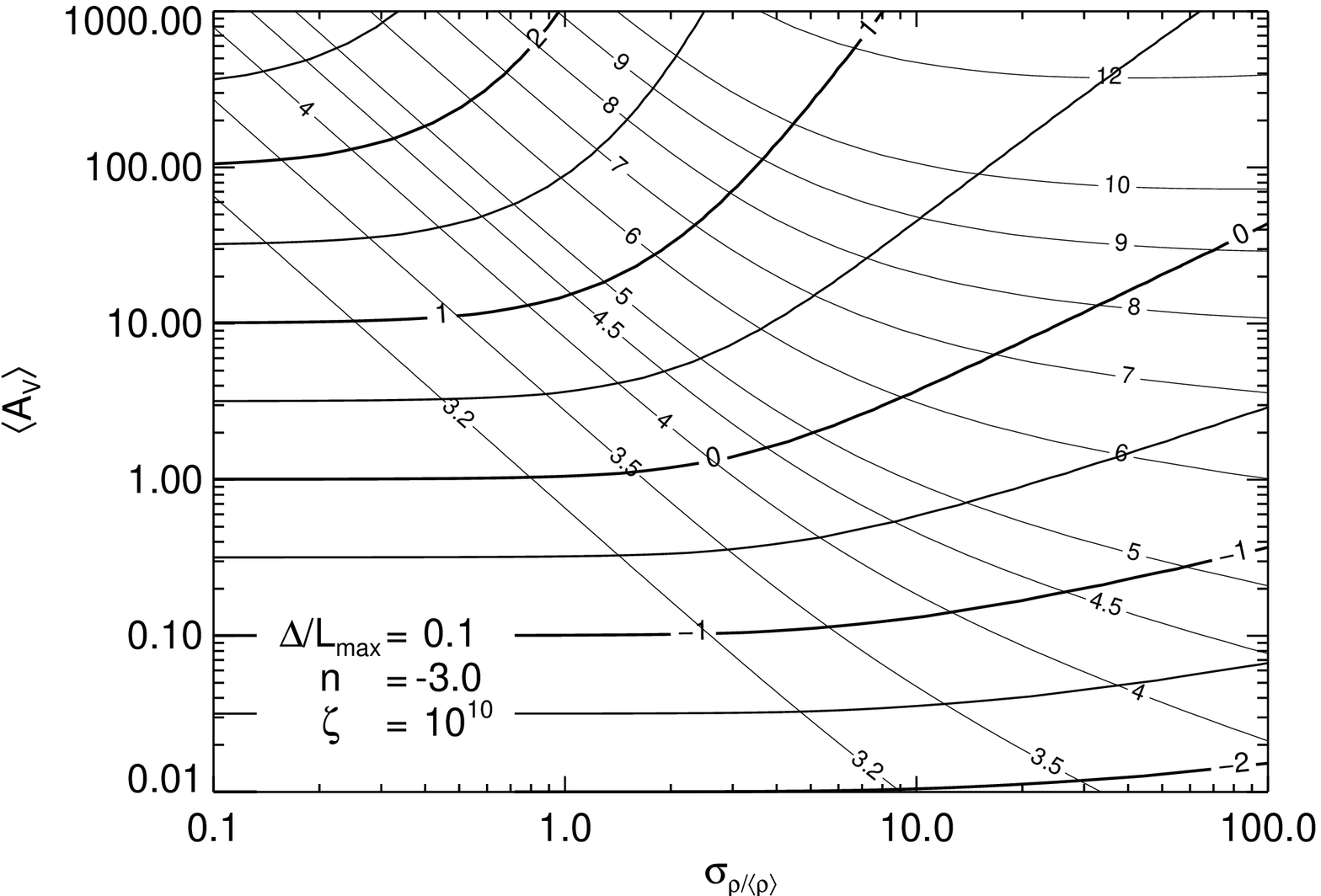}
  \hfill
  \includegraphics[width=0.49\hsize]{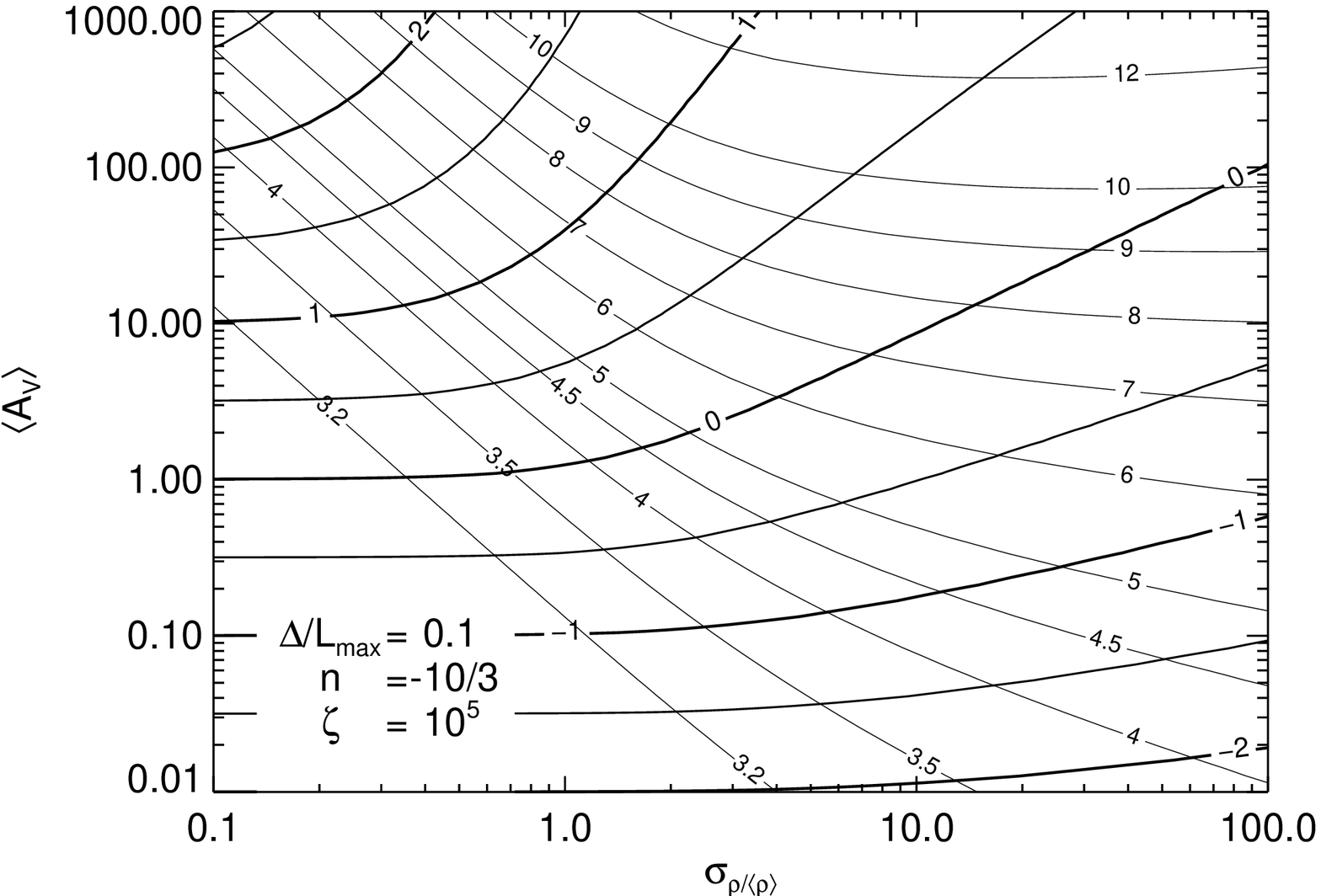}
  \vspace{0.1cm}

  \includegraphics[width=0.49\hsize]{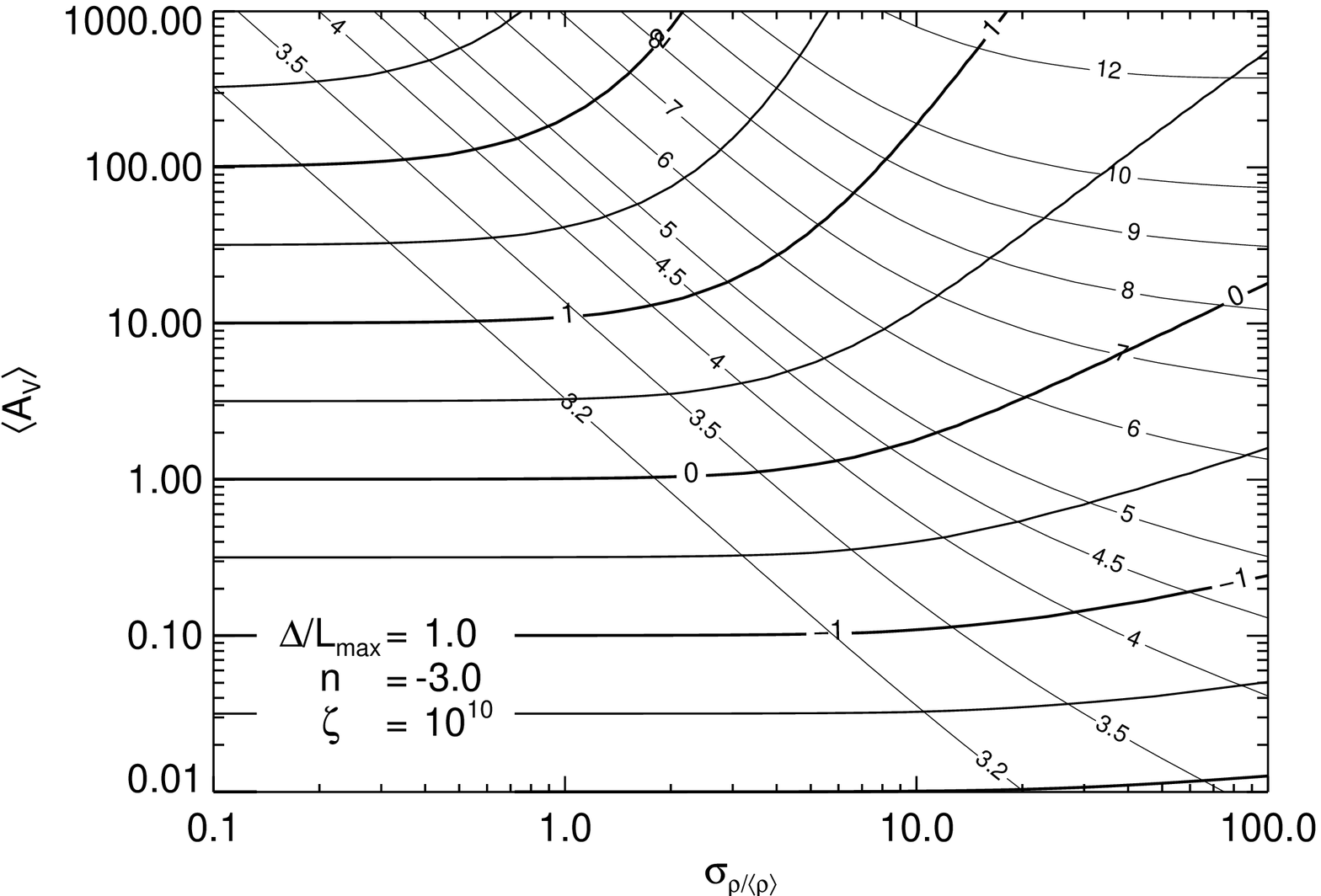}
  \hfill
  \includegraphics[width=0.49\hsize]{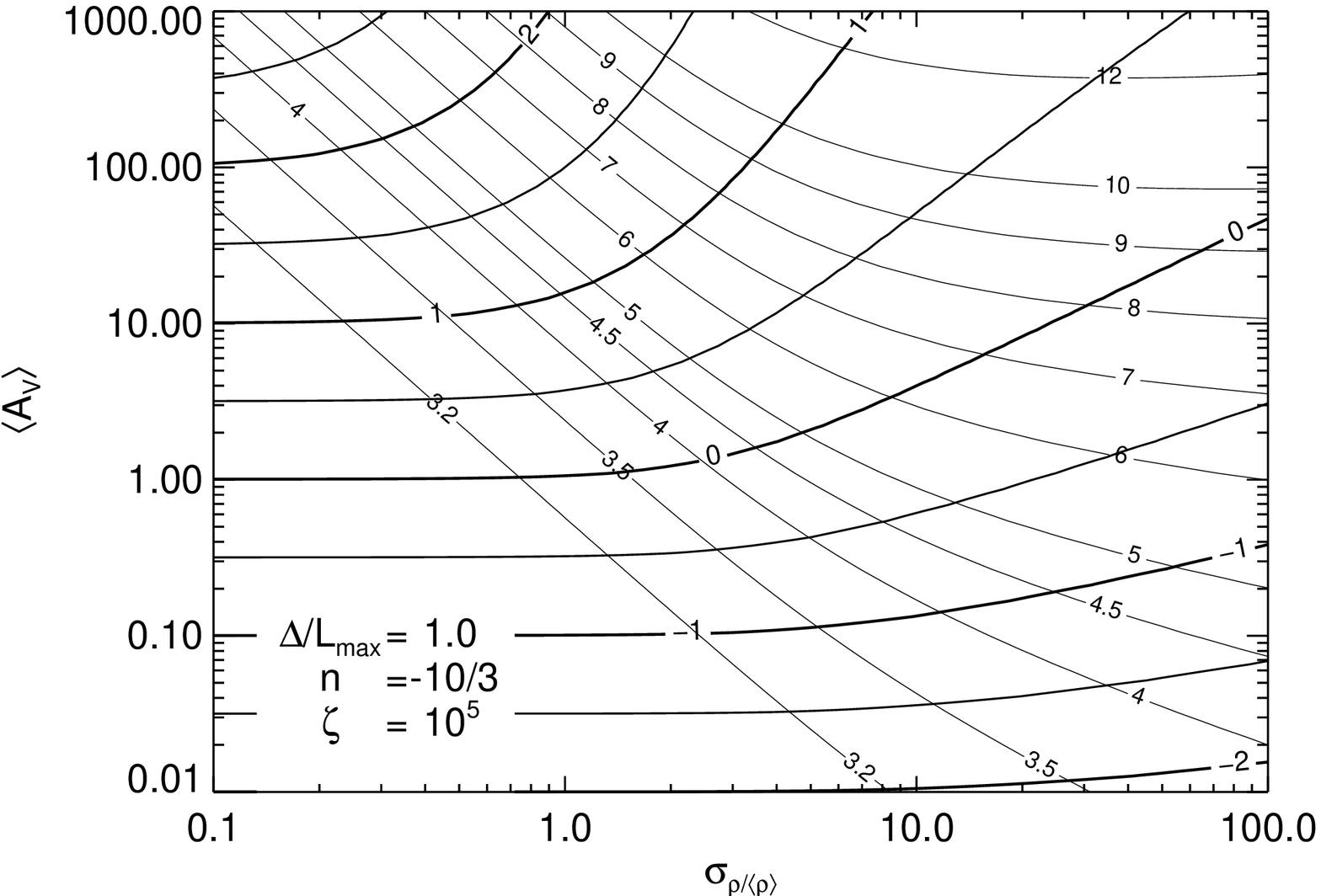}
  \vspace{0.1cm}

  \includegraphics[width=0.49\hsize]{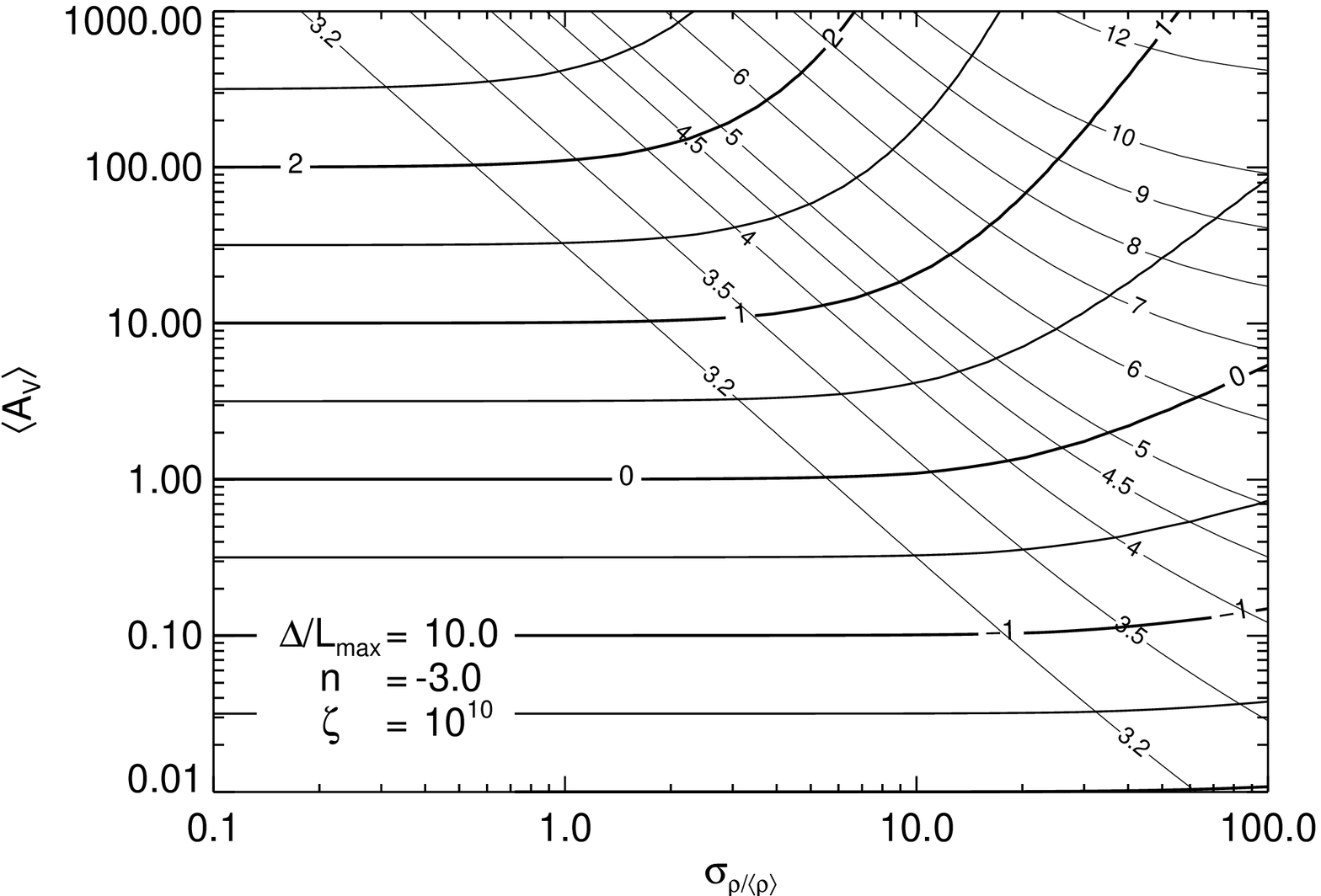}
  \hfill
  \includegraphics[width=0.49\hsize]{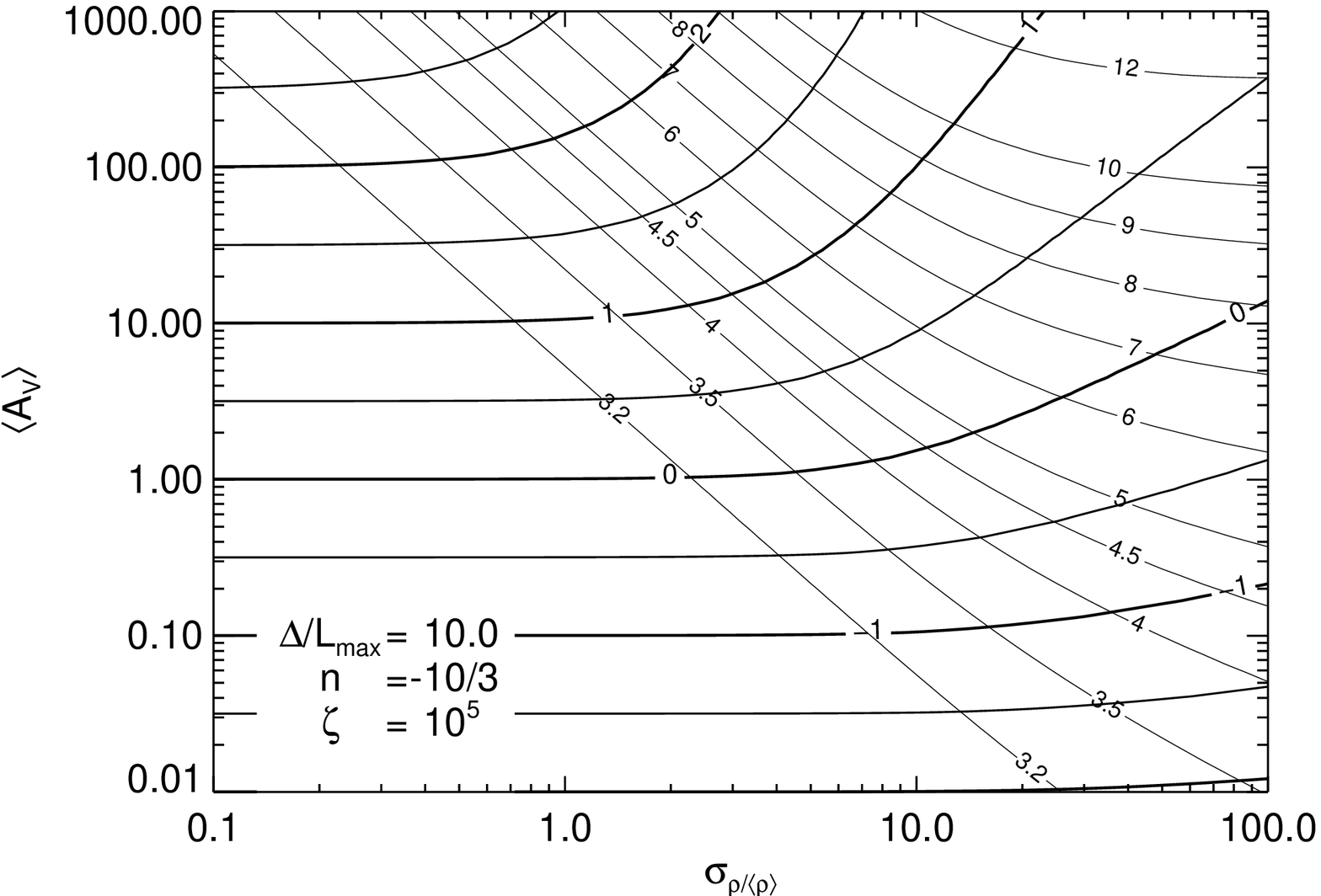}
  \vspace{0.1cm}
  \caption{\label{2dslice}
	Attenuation ${\rm A_V}$ in V as function of the standard deviation $\sigma_{\rho/\left<\rho\right>}$
	of the local density and the mean attenuation $\left<{\rm A_V}\right>$ for three slices of
	different thickness where
	$\Delta/L_{\rm max}$ is chosen to be 0.1, 1.0, and 10.
  }
\end{figure*}

\begin{figure}
  \includegraphics[width=0.90\hsize]{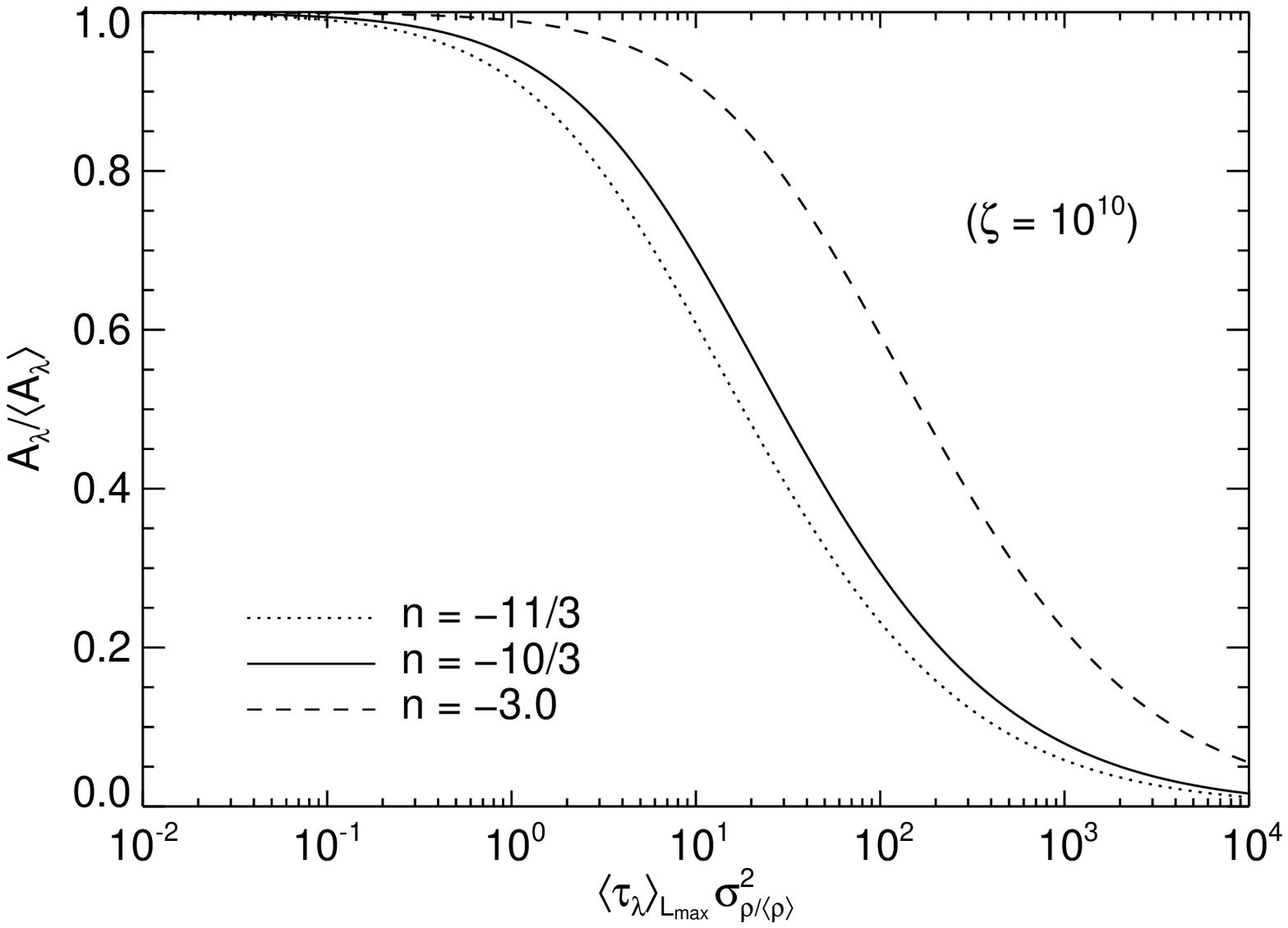}
  \vspace{0.1cm}
  
  \includegraphics[width=.90\hsize]{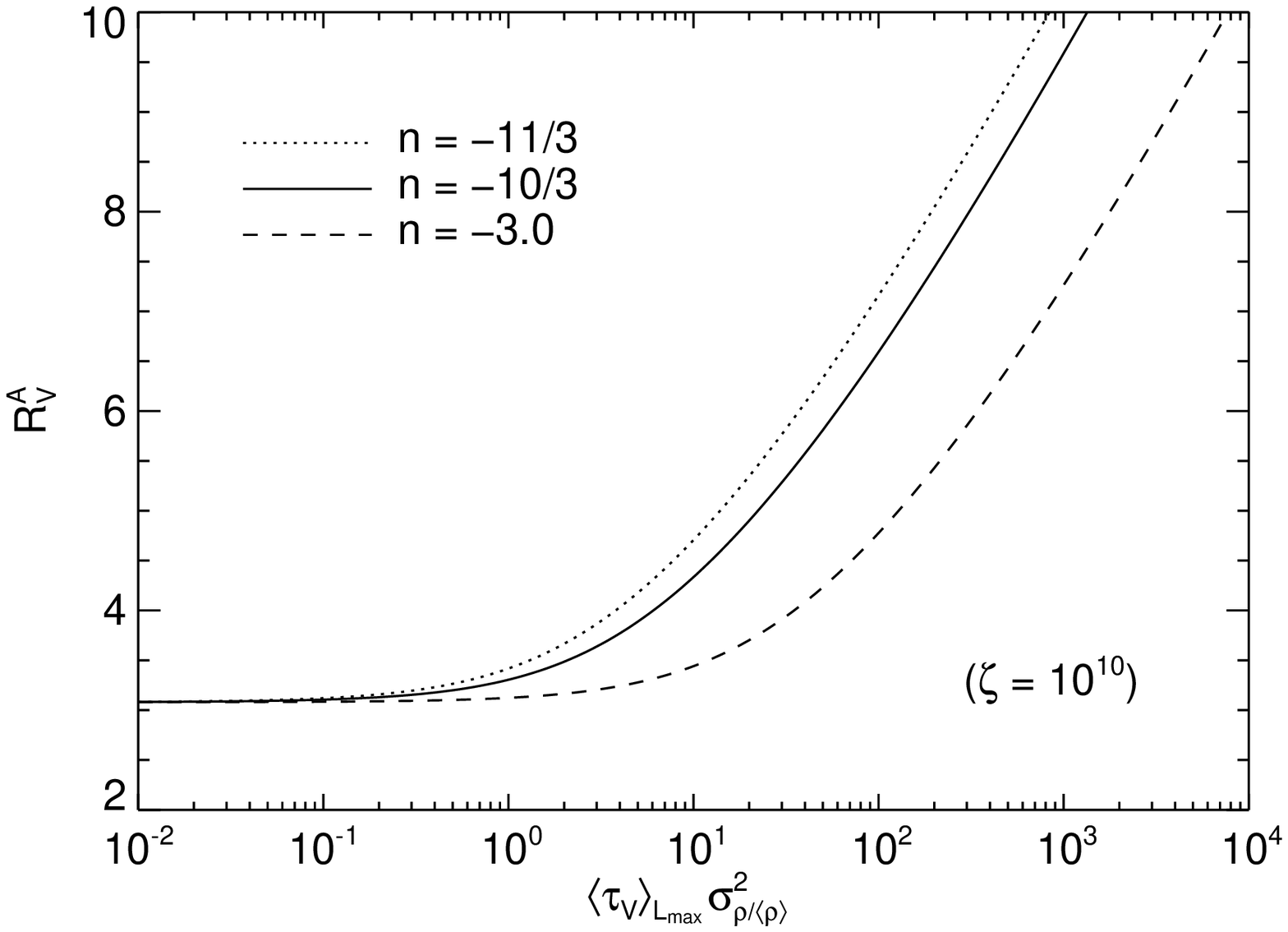}

  \caption{\label{av&rv_limit}
  Ratio $\rm A_{\lambda}/\left<A_{\lambda}\right>$  as a function of
  $\left<\tau_{\lambda}\right>_{L_{\rm max}}\sigma^2_{\rho/\left<\rho\right>}$ and ${\rm R^A_V}$
  as a function of
  $\left<\tau_{\rm V}\right>_{L_{\rm max}}\sigma^2_{\rho/\left<\rho\right>}$ in the limit of
  slices with $\sigma^2_{\tau/\left<\tau\right>}\ll 1$. The turbulence is assumed to extend over
  $\zeta=10^{10}$ scales.
  }
\end{figure}

\begin{figure*}
  \includegraphics[width=0.49\hsize]{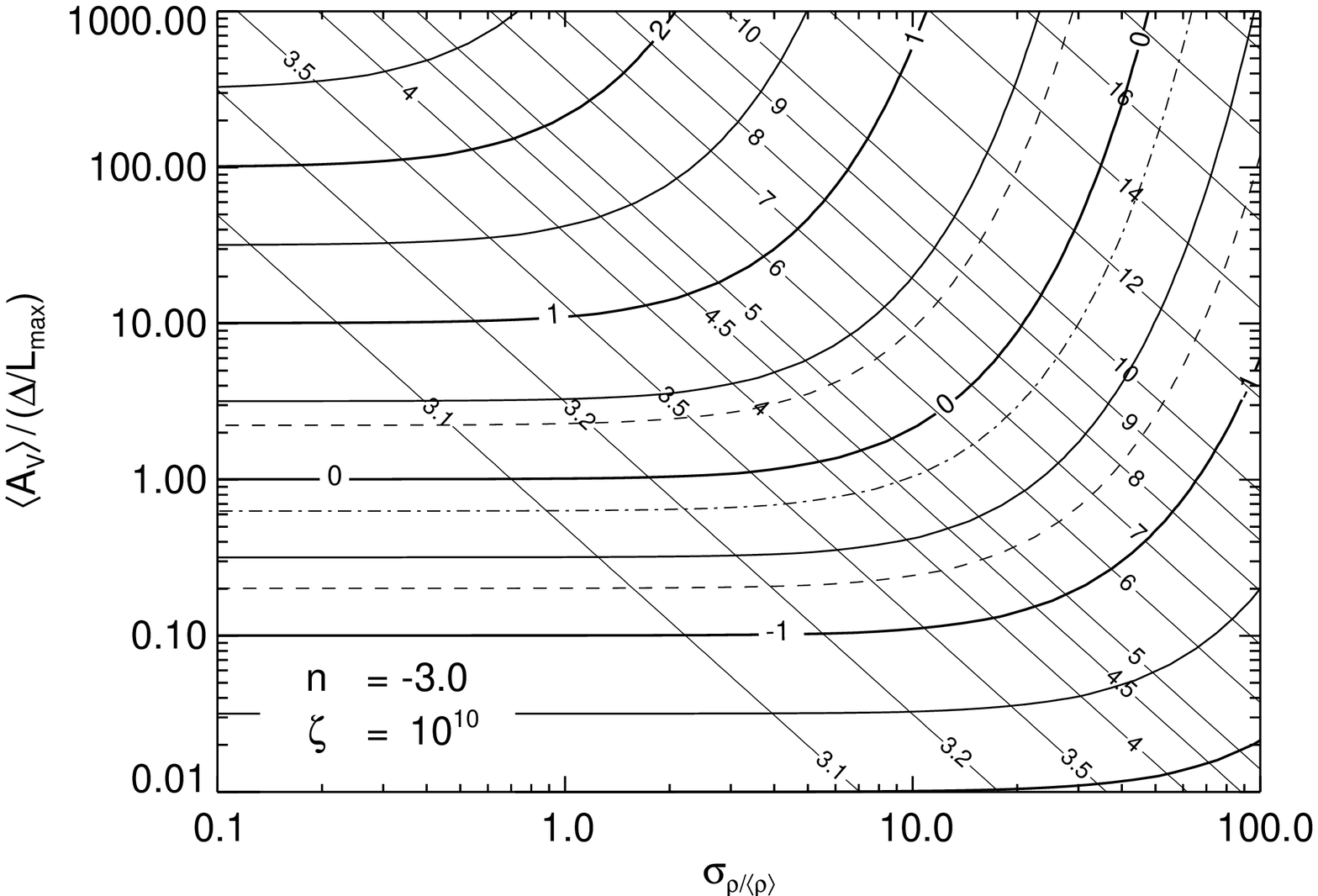}
  \hfill
  \includegraphics[width=0.49\hsize]{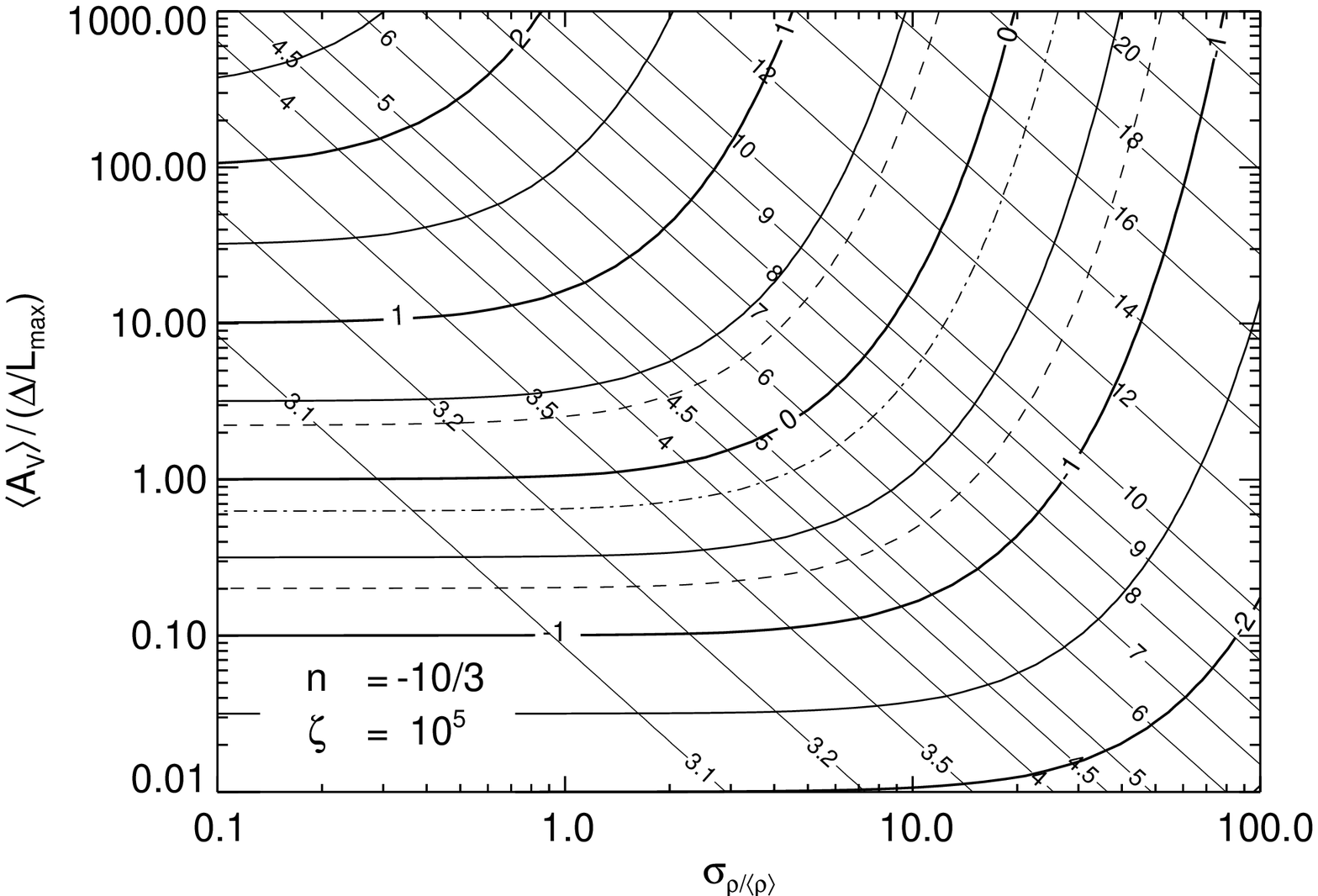}
  \caption{\label{rvavlimitcontour}
  Attenuation ${\rm A_{V}}/(\Delta/{L_{\rm max}})$ (thick lines) as function of
  $\sigma_{\rho/\left<\rho\right>}$ and $\left<{\rm A_V}\right>/(\Delta/{L_{\rm max}})$ in
  the limit of thick slices with $\Delta/{L_{\rm max}}\gg 1$. The absolute-to-relative attenuation
  ${\rm R_V^A}$ is shown as thin solid lines. Also shown are minimum, mean, and maximum
  values of the attenuation ${\rm A_V}$ derived for star-burst galaxies (dashed, dashed dotted,
  and dashed line) \citep{Calzetti2001}.
  } 
\end{figure*}

\begin{figure*}
  \includegraphics[width=0.49\hsize]{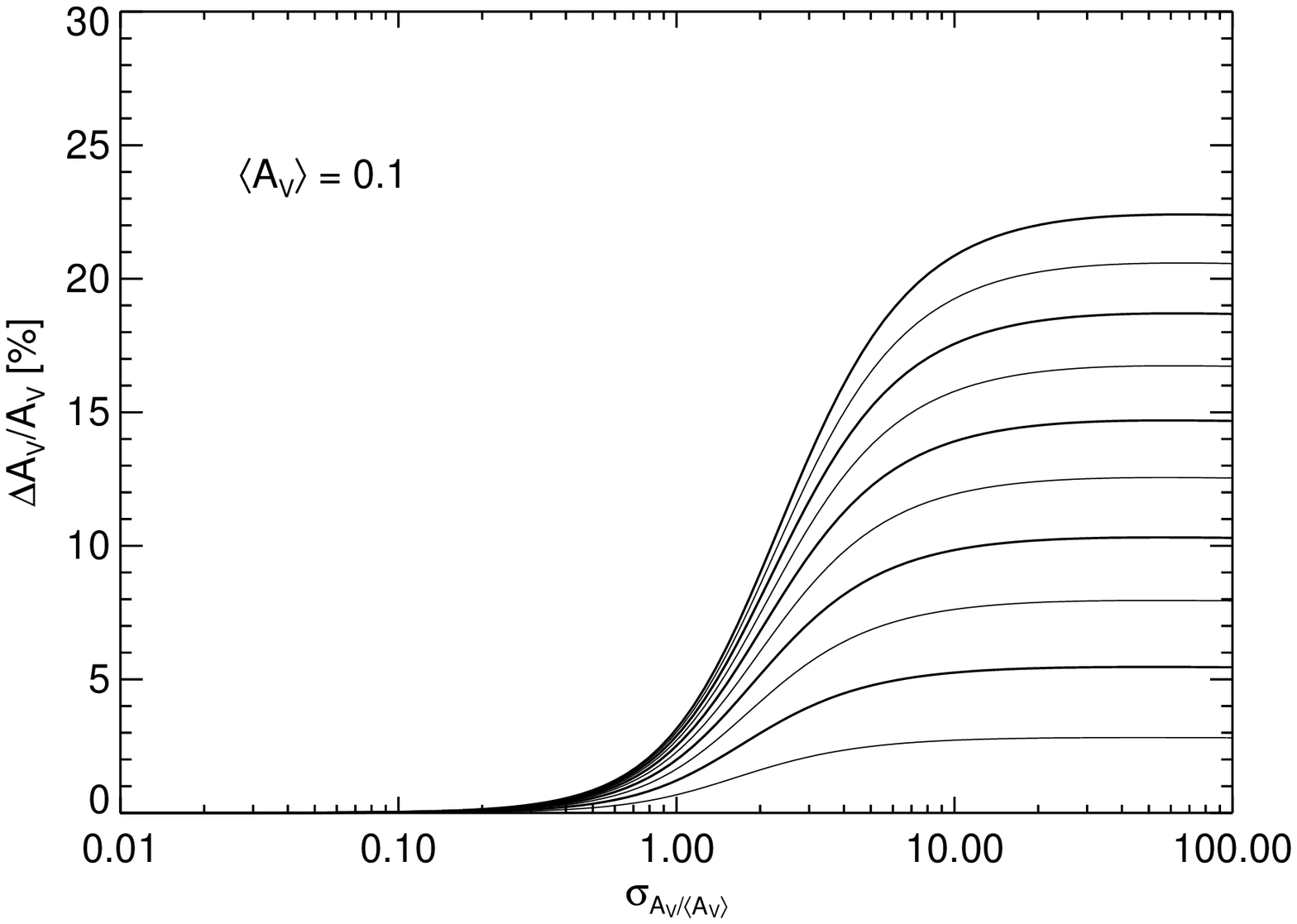}
  \hfill
  \includegraphics[width=0.49\hsize]{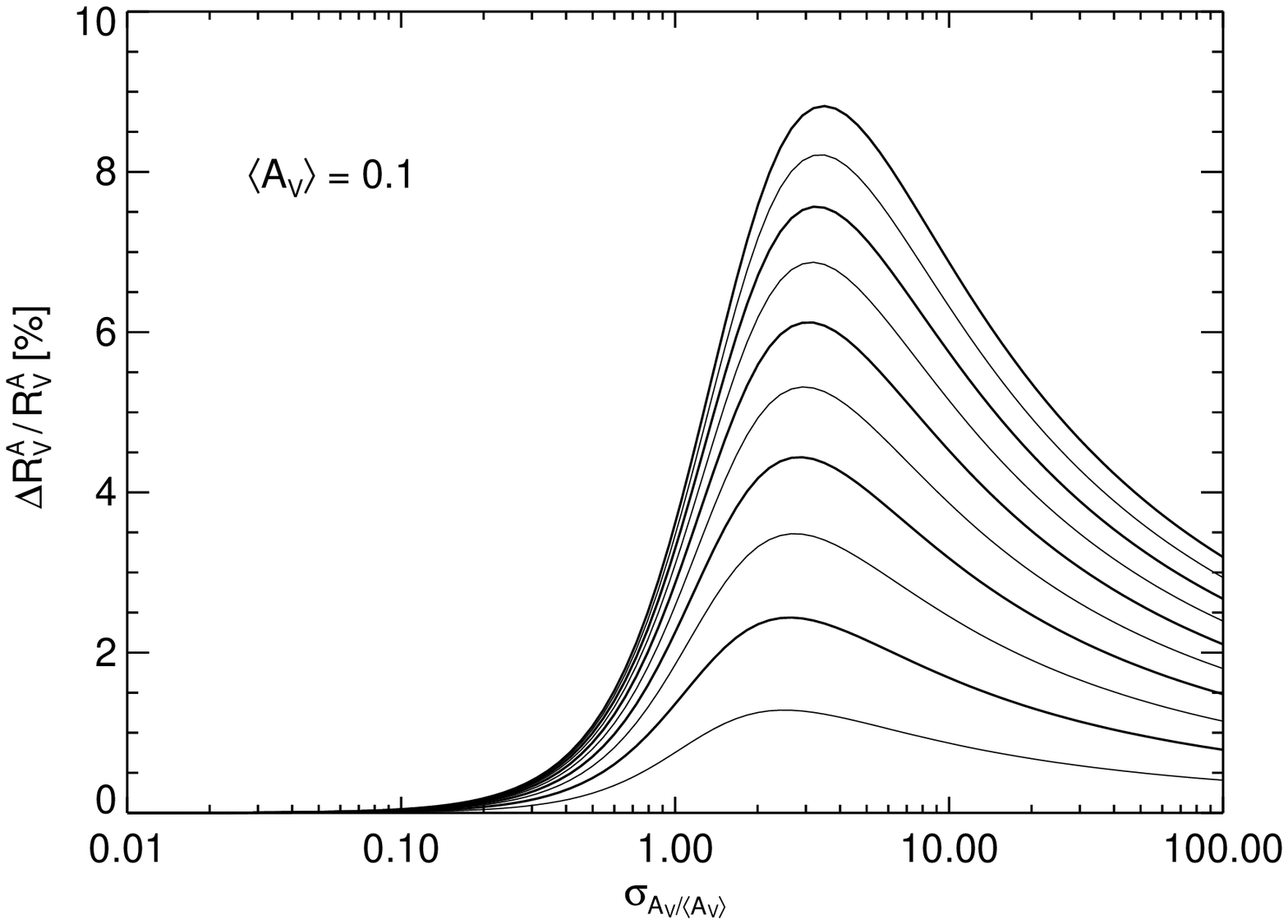}
  \vspace{0.1cm}

  \includegraphics[width=0.49\hsize]{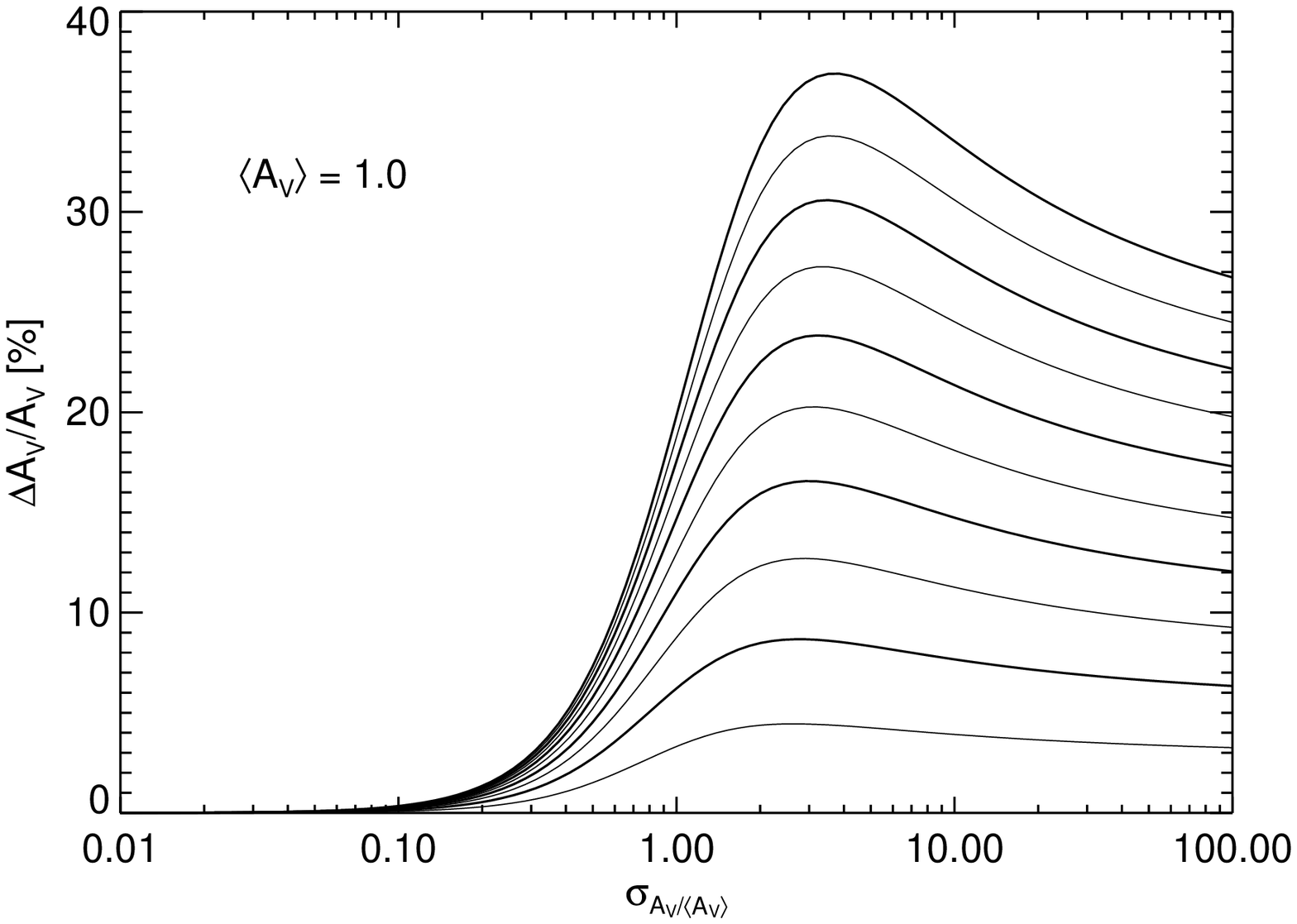}
  \hfill
  \includegraphics[width=0.49\hsize]{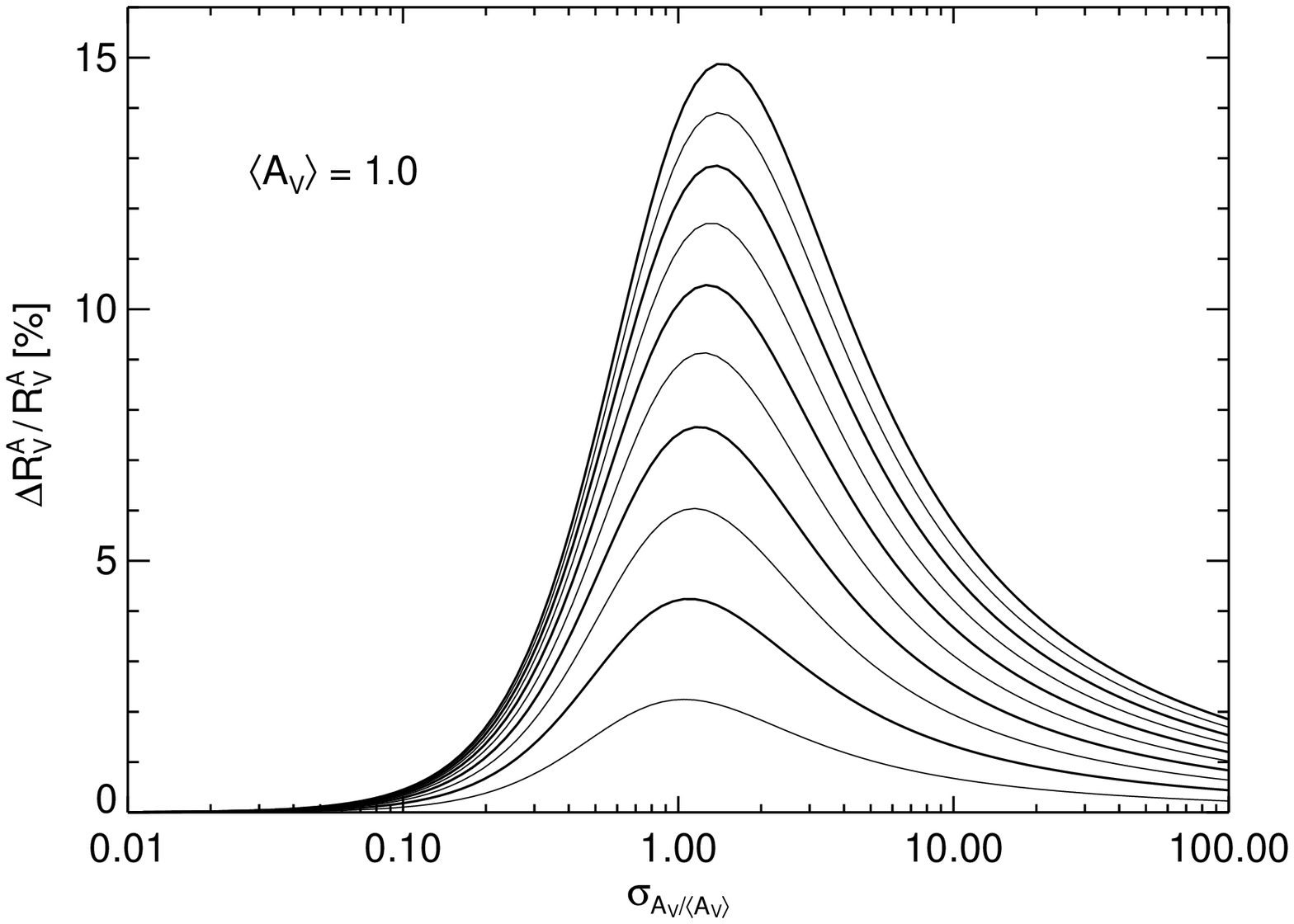}
  \vspace{0.1cm}

  \includegraphics[width=0.49\hsize]{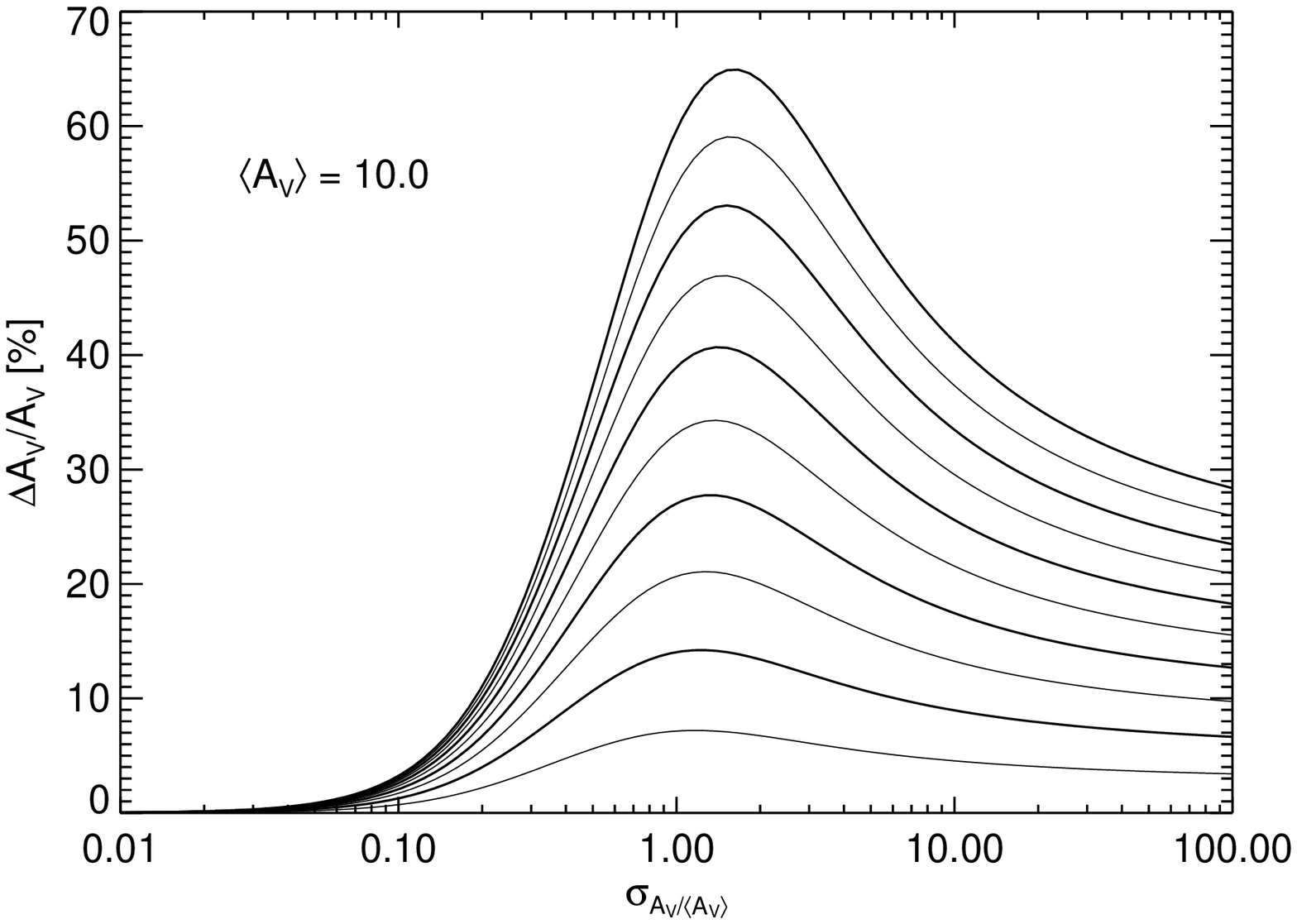}
  \hfill
  \includegraphics[width=0.49\hsize]{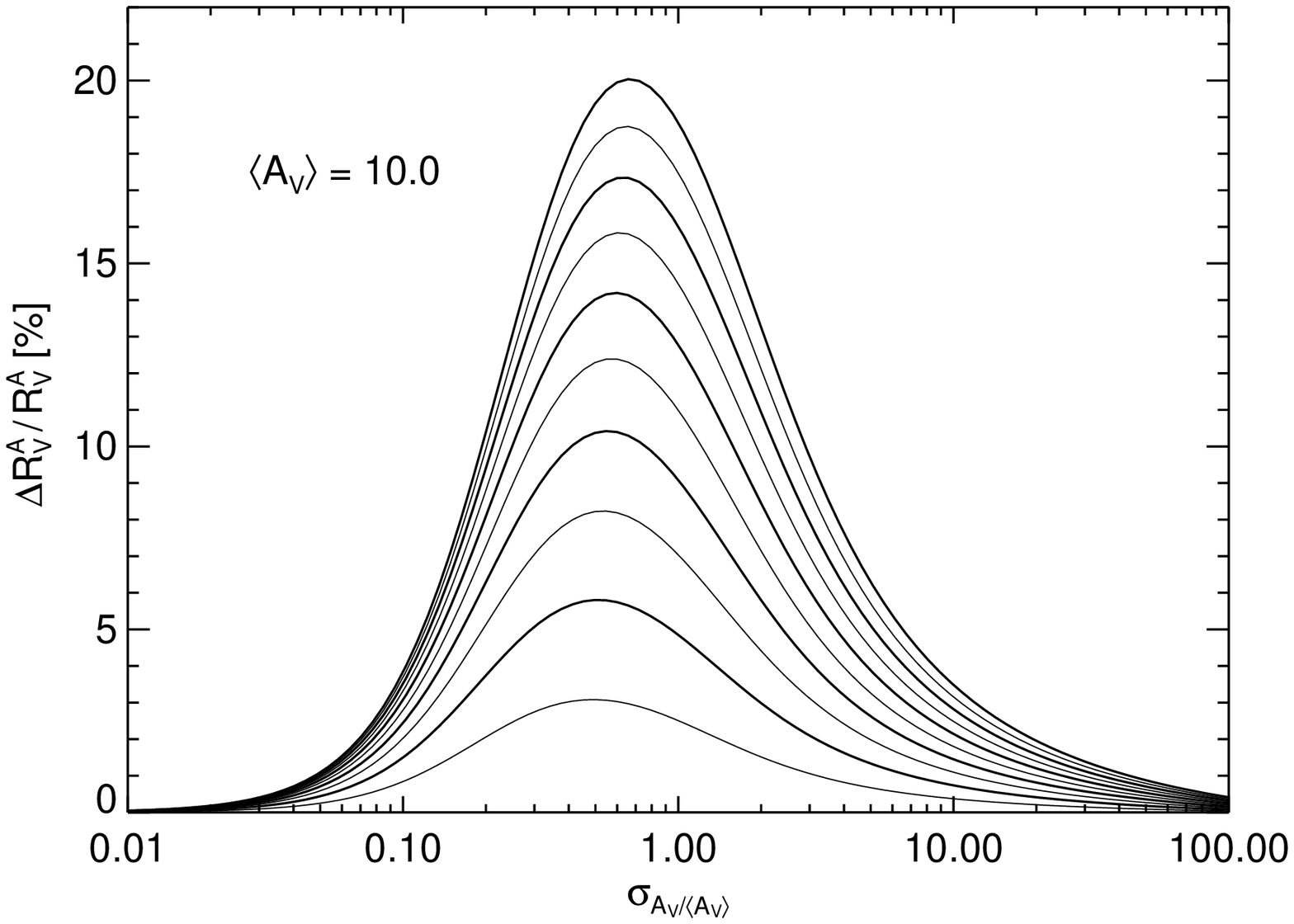}
  \vspace{0.1cm}
 
  \caption{\label{figaccuracy_b}
  Accuracy of the derived observable quantities ${\rm A_V}$ and ${\rm R_V^A}$
  for different assumptions of the uncertainty of the standard deviation $\sigma_{\rm A_V/\left<A_V\right>}$
  and the mean value $\left<{\rm A_V}\right>$.
  The standard deviation of the column density is assumed to be up to 100\% too large in comparison
  with the correct value. The difference in the assumed error for two neighboured curves is 10\%.
  The uncertainty of ${\rm A_V}$ and $\rm R_V^A$ increases with increasing error in
  the standard deviation $\sigma_{\rm A_V/\left<A_V\right>}$.
  }
\end{figure*}

\begin{figure*}
  \includegraphics[width=0.49\hsize]{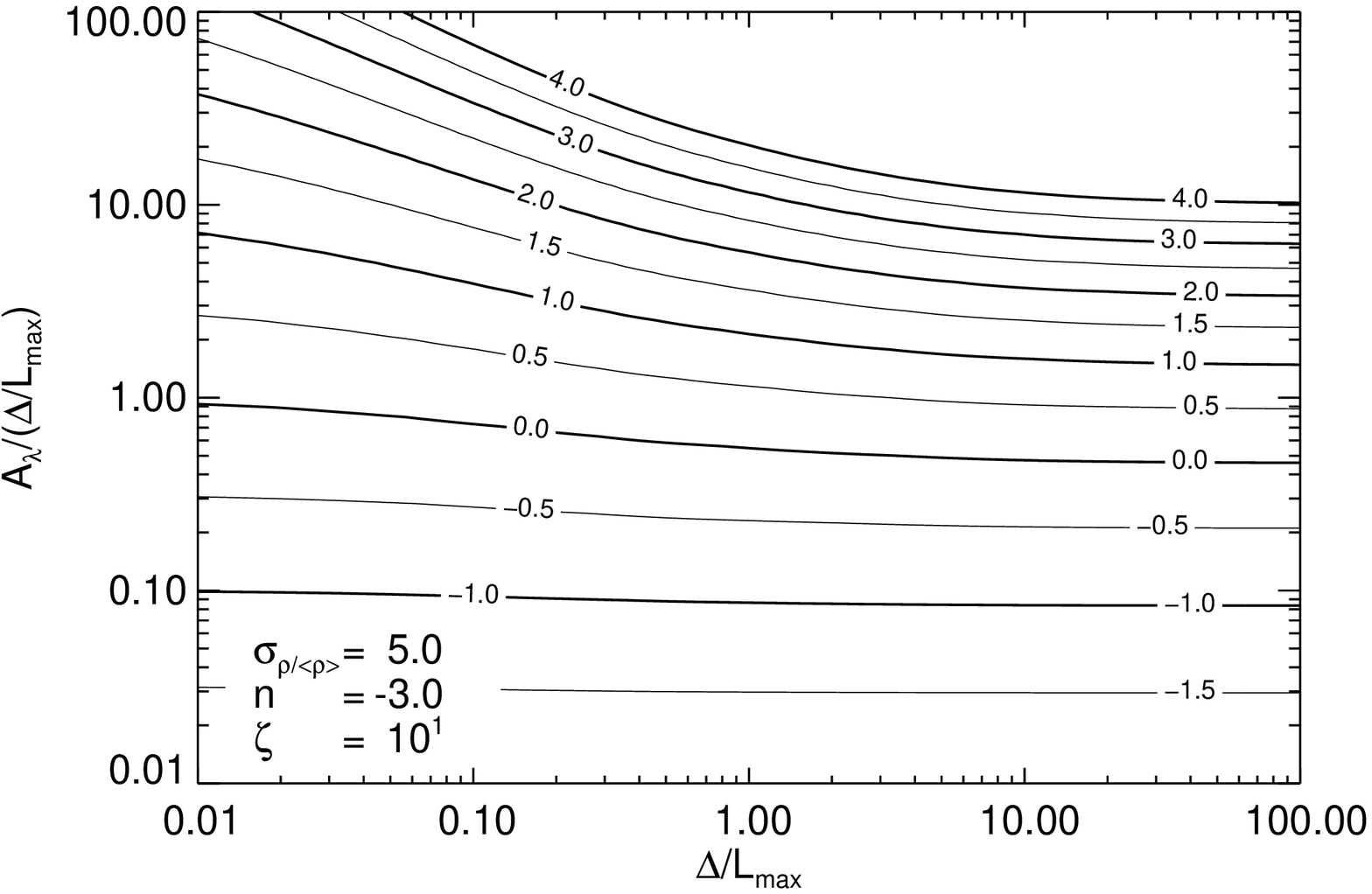}
  \hfill
  \includegraphics[width=0.49\hsize]{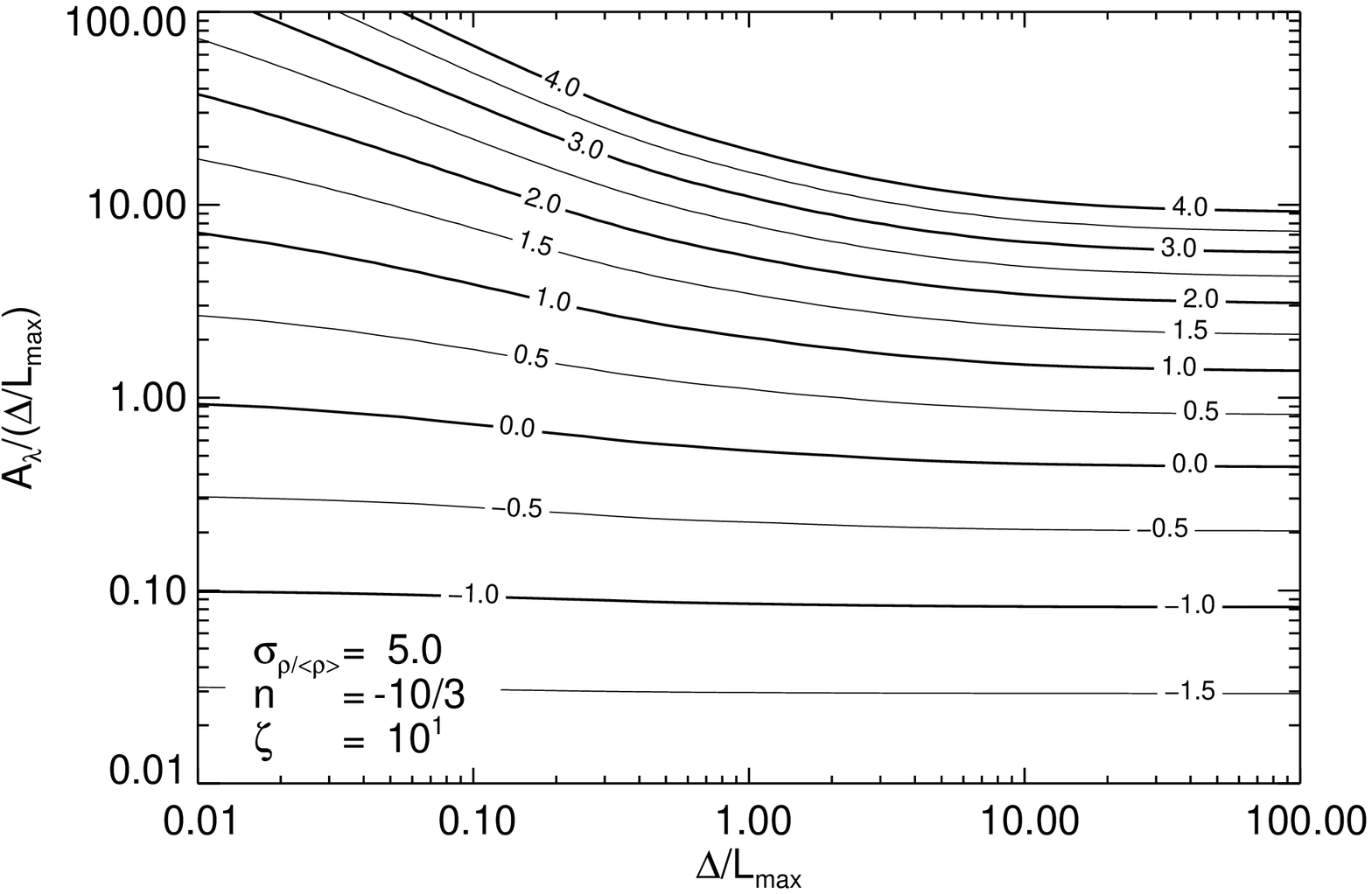}

  \caption{\label{figAlambda_2}
  Same as plot for $\sigma_{\rho/\left<\rho\right>}$ in Fig.~\ref{figAlambda} but
  assuming a turbulent medium with only $\zeta=10$.
  }
\end{figure*}

\begin{figure*}
   \includegraphics[width=0.49\hsize]{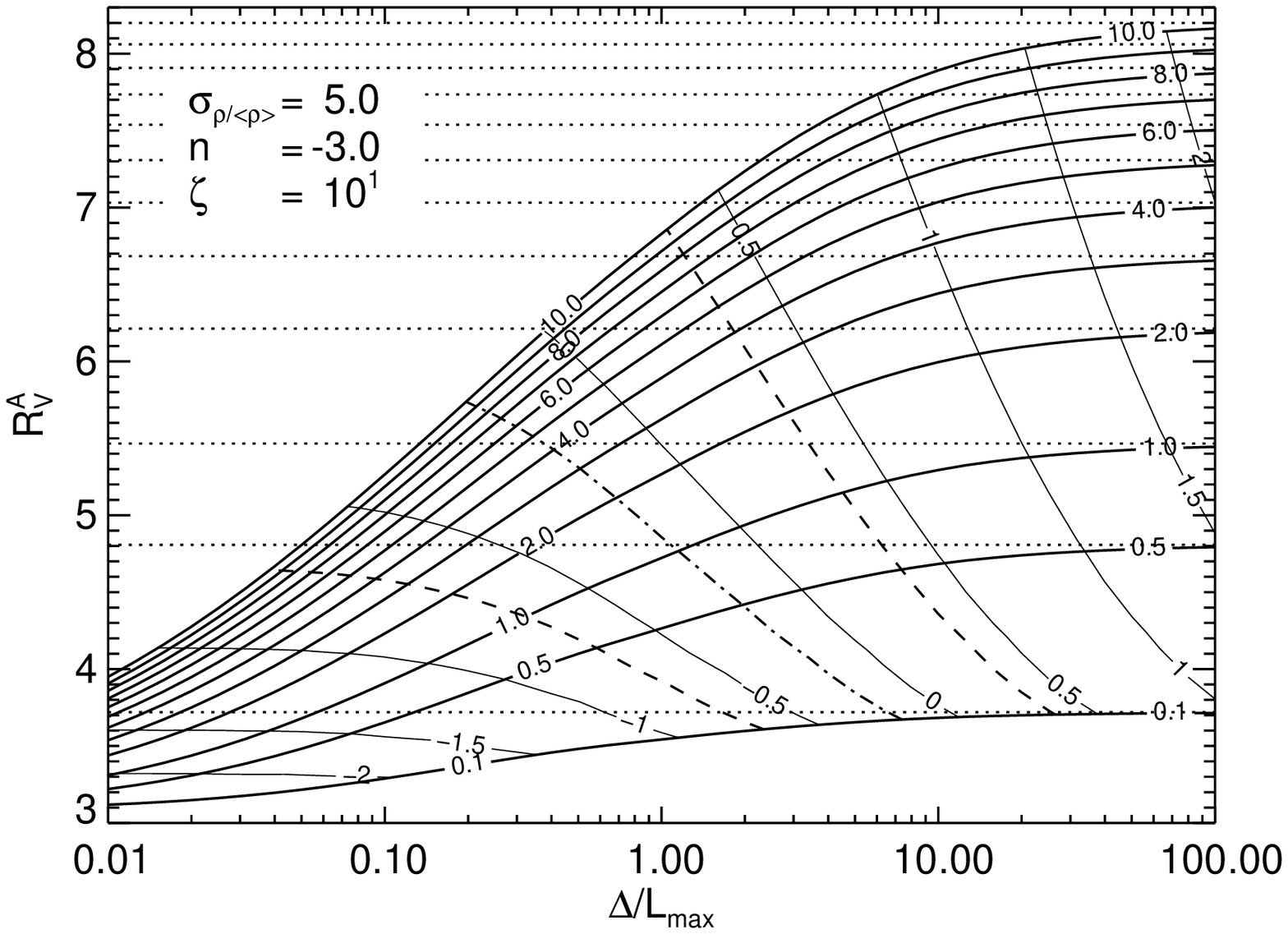}
  \hfill
  \includegraphics[width=0.49\hsize]{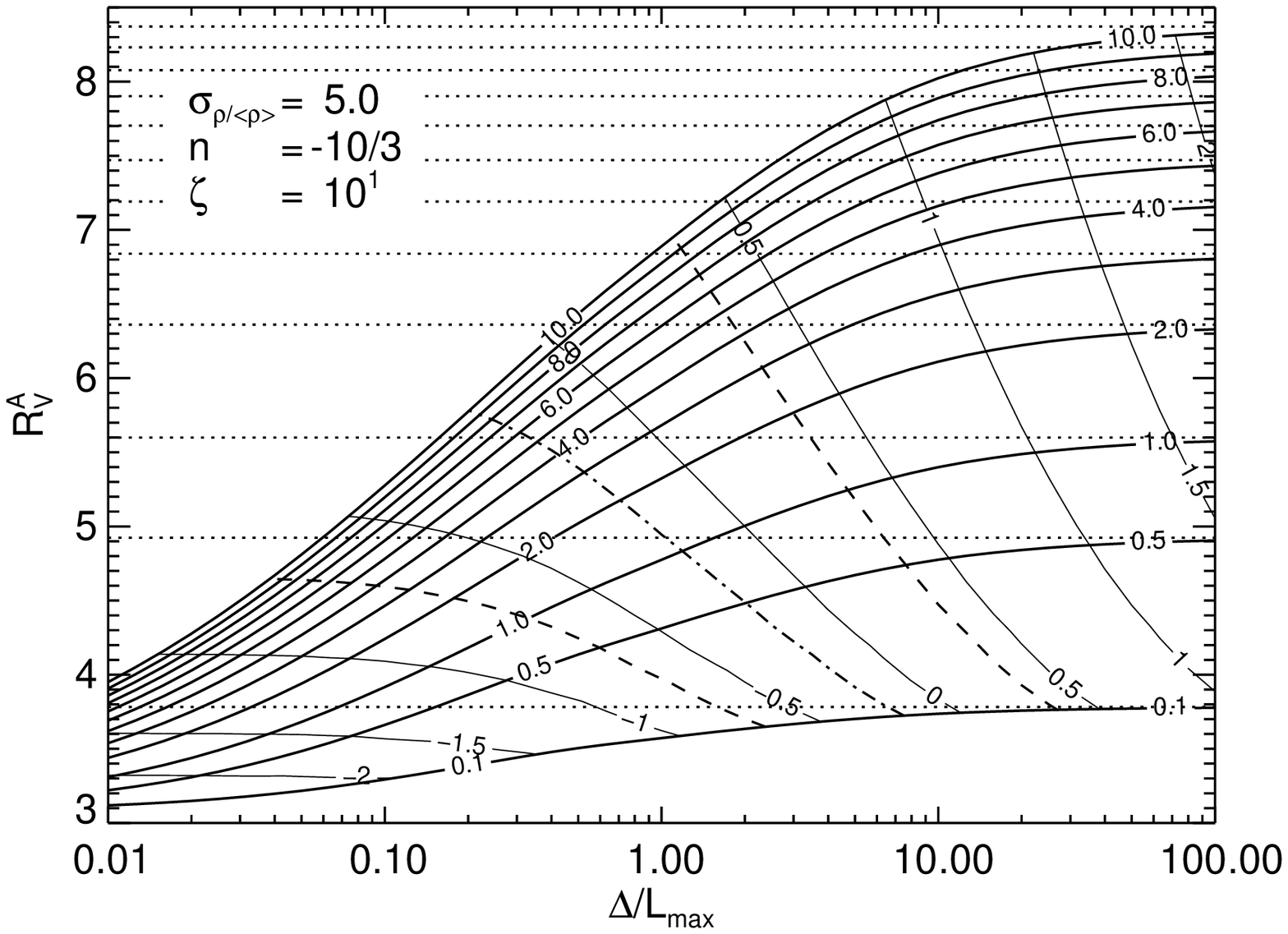}  
  \caption{\label{figrv3}
  Same as the plots for $\sigma_{\rho/\left<\rho\right>}=5$ in Fig.~\ref{figrv1} but
  assuming a power spectrum which extends only to $\zeta=10$.
  }
\end{figure*}

From the variation of $\rm R_V^A$ with slice thickness (Fig.~\ref{figrv1})
we must conclude that the turbulence must be characterised by a standard deviation $\sigma_{\rho/\left<\rho\right>} >1$ since none of the cases considered with $\sigma_{\rho/\left<\rho\right>}=1$ can reproduce the observational results of $\rm R_V^A=4.3$ and a mean attenuation of $\rm A_V=0.63$.

For $\sigma_{\rho/\left<\rho\right>}=2.5$ the slice thickness has to be very thin if the mean value of $\rm A_V=0.63$ is to be consistent with the observed value, $\rm R_V^A\sim 4.3$.
More likely we require a standard deviation of at least $\sigma_{\rho/\left<\rho\right>}\sim 5.0$ to explain the observed $\rm R_V^A$. With these parameters the thickness of the
dust screen is only slightly larger than one maximum scale and $\rm \left<A_V\right>_{L_{\rm max}}\sim 0.6$. The variation of the attenuation $\rm A_V$ can be explained by different values
of the slice thickness $\Delta/L_{\rm max}$ ranging from $\sim 0.4$ to $\sim 6$ and may
partly caused by the different inclination angles of the galaxies. 
As seen in Fig.~\ref{figrv1} the variation of the attenuation $\rm A_V$ should be connected with different attenuation curves with $\rm R_V^A$-values in the range from $\sim 4$ to $\sim 4.5$. 

Wider distributions of the local density require lower optical depths 
and thicker dust layers. In case of $\sigma_{\rho/\left<\rho\right>}=7.5$ the attenuation curve
of star-bursts can be reproduced if $\left<\rm A_V\right>_{L_{\rm max}}\approx 0.3$. The
slice thickness varies from $\sim 0.9$ to $\sim 11$ changing $\rm R_V^A$ from $\sim 4.1$ to $\sim 4.5$. The mean thickness is $\sim 3$. 

As can be seen in Fig.~\ref{figrv1} the `Calzetti-curve' can also be reproduced in
a turbulent medium with $\sigma_{\rho/\left<\rho\right>}=10$. In this case $\left<{\rm A_V}\right>_{L_{\rm max}}\approx 0.15$. The dust screen would have a thickness ranging from $\sim 2$ to $\sim 22$ with a mean of $\sim 6$ and $\rm R_V^A$ would vary from 4 to 4.4.
If we identify the lower limit of the attenuation with galaxies seen face on and require that its
dust screen is not larger than one maximum scale we can conclude that the standard deviation of the local density has to be smaller than $\sigma_{\rho/\left<\rho\right>}=10$. 

Based on this analysis the `Calzetti-curve' implies a standard deviation of the local density of order $\sim 7.5$, only slightly larger than 
the minimum value derived by using the thick slice approximation.
However, one has to bear in mind that due to the strong UV-radiation the dust properties in star-bursts may be different to the ones of the ISM of our galaxy. Nonetheless, the interpretation is probably still correct because the extinction curve up to $4400$~\AA{} should not be strongly effected by such variations, if mainly due to the presence or absence of PAH molecules.

A thin slice with a thickness much less than the largest scale would imply that the turbulent medium is mixed with the stars. In this case the model of a distant foreground screen is somewhat questionable and
another approach may be more appropriate. If the scattered light makes only a very small contribution it is still a reasonable assumption to consider the dust layers in front of each layer of stars as a distant foreground screen. The effective attenuation curve of such a medium is obtained by comparing the whole transmitted light with
the light emitted inside the screen.

By fitting the `Calzetti-extinction-curve' with the turbulent foreground screen model we found
that the minimum Mach number is in the range 1.3 to 22 if relation~\ref{densmachcorr} applies.
This is consistent with the Mach numbers found for the WNM and CNM in our own galaxy (see Sect.~\ref{model}).

The fact that the standard deviation $\sigma_{\rho/\left<\rho\right>}$ of the turbulent screen has to be significantly larger than unity to explain the `Calzetti-extinciton-law' would imply that the Mach number, assuming the relation~\ref{densmachcorr} applies, has to be larger than 2, i.e. larger than the Mach number of the WNM of our galaxy. On the other hand, the range for $\sigma_{\rho/\left<\rho\right>}$ which is consistent with the attenuation observed for star burst galaxies would imply Mach numbers in the range 10 to 15. This suggest that the attenuation of star burst galaxies is caused by the structure related to the turbulent CNM.


It is likely, that at the epoch of galaxy formation, galaxies were more violent systems characterised by a wider distribution of the local density (larger $\sigma_{\rho/\left<\rho\right>}$) and by a higher mean optical thickness $\left<\tau\right>_{L_{\rm max}}$. These systems will therefore probably have attenuation curves which are flatter than the curve derived for local star-bursts.

\subsection{Limitations}

An obvious limitation of these models is the fact that scattered light is not taken into account.
It may also be more realistic to consider other geometries where the stars are connected with the turbulent dust screen or mixed with the turbulent ISM. By including scattering we expect
that the attenuation curve will be slightly lower where the scattered light makes a large
contribution to the total extinction which is the case in the UV. But as we outlined in paper~I the light scattered out of the observed direction cannot be totally compensated by scattered light from other stars. To understand how far our simple model can be applied to explain the attenuation for different geometries it will still be important to study the effects of scattering quantitatively.

It may also be important to consider the variation of dust properties with local density, radiation field and stellar population and to understand how these variations affect the mean extinction curve. However, since these effects are not sufficiently understood it may still be better to assume a mean extinction curve of our own galaxy and to use the attenuation curves presented in this paper.

\section{Summary}
We have analysed in detail the attenuation curve caused by a distant isothermal turbulent 
dust screen. We used a simplified model of the turbulent medium based on its statistical properties where the PDF of the local density is log-normal and the power spectrum
a simple power law. The turbulence is assumed to extend from a maximum scale $L_{\rm max}$ down to a minimum scale $L_{\rm min}$. The optical depth is taken to be proportional to the column density and the PDF of the column density to be
log-normal. The attenuation curve is derived using the mean extinction curve of our galaxy as provided by \citet{Weingartner2001}.

The main results are:
\begin{enumerate}
  \item The attenuation curves in isothermal turbulent dusty media are 
  	determined by the absolute-to-relative attenuation ratio$\rm R_V^A$ and the absolute
	attenuation $\rm A_V$. The $\rm R_V^A$-value determines the curvature while
	$\rm A_V$ the scaling. A higher $\rm R_V^A$-value relates to a flatter attenuation curve.  
  \item The absolute-to-relative attenuation ratio$\rm R_V^A$ increases with slice thickness.
  \item For slices much thicker than the maximum scale $L_{\rm max}$ the 
  	$\rm R_V^A$-value converges towards a maximum value. The attenuation curve
	becomes independent on slice thickness.
\end{enumerate}
  
  The $\rm R_V^A$-value depends on the structure, the PDF of the local density and
  the mean value of the optical thickness of the dust screen. The dependence of
  single parameters characterising the turbulent slice is as follows:
  
\begin{enumerate}
   \item The $\rm R_V^A$-value increases with optical thickness.
   \item More turbulent media, characterised by a broader PDF or a higher standard deviation
   	$\sigma_{\rho/\left<\rho\right>}$ of the local density, are characterised by higher 
	$\rm R_V^A$-values
	(assuming that the structure is the same). The attenuation is smaller
	in such media.
   \item Turbulence characterised by a broader correlation function (steeper power spectrum)
      leads to a higher $\rm R_V^A$-value as a result of a broader PDF of the optical depths.
   \item The curvature of the attenuation curve depends on the ratio 
   	$\zeta=L_{\rm max}/L_{\rm min}$. For higher values of $\zeta$ the PDF of the column
	density becomes narrower and therefore the effect due to turbulence smaller: The 
	$\rm R_V^A$-value becomes smaller and the attenuation $\rm A_V$ larger. The effect is
	particularly important in turbulent media with $n\le-3$. 
	In the limit of $\log_{10}\zeta\gg 1$ the effect due to turbulence would disappear. For
	Kolmogorov turbulence (or general $n<-3$) the $\rm R_V^A$-value will reach a lower limit.
\end{enumerate}

\begin{acknowledgements}
M. Dopita acknowledges the support of the ARC and the Australian National University through his Australian Federation Fellowship. Both authors acknowledge financial support for this research through ARC Discovery project DP0208445.
\end{acknowledgements}

\clearpage
\clearpage

\clearpage
\clearpage

\clearpage

\clearpage
\clearpage

\clearpage
\begin{deluxetable}{l|c|ccccccccccc}

  \rotate
  \tabletypesize{\tiny}
  \tablewidth{.92\vsize}
  \tablecaption{
  Relative attenuation $\rm E(\lambda-V)/E(B-V)$ for various values of $\rm R_V^A$\label{tableattenuation}}
  \tablehead{\colhead{$\lambda[\mu{\rm m}]$}&\colhead{remark} & \colhead{${\rm R_V=3.08}$}& \colhead{${\rm R_V^A}=3.5$} & \colhead{${\rm R_V^A}=4.0$} & \colhead{${\rm R_V^A}=4.5$} & \colhead{${\rm R_V^A}=5.0$} & \colhead{${\rm R_V^A}=5.5$} & \colhead{${\rm R_V^A}=6.0$} & \colhead{${\rm R_V^A}=7.0$} &\colhead{${\rm R_V^A}=8.0$} & \colhead{${\rm R_V^A}=9.0$} & \colhead{${\rm R_V^A}=10$}} 
  \startdata
 	\multicolumn{13}{c}{${\rm A_V=1}$}  \\
   \hline
   0.0912 & ${\rm Ly}\infty$   & 12.066 & 9.716 & 8.675 & 8.113 & 7.750 & 7.490 & 7.294 & 7.014 &
   	6.822 & 6.680 & 6.570 \\
   0.1053 &                          & 9.694 & 8.105 & 7.349 & 6.932 & 6.659 & 6.463 & 6.314 & 6.100 &
   	5.953 & 5.844 & 5.759 \\
   0.1216 & ${\rm Ly}\alpha$  & 7.328 & 6.385 & 5.898 & 5.621 & 5.437 & 5.303 & 5.202 & 5.055 &
   	4.953 & 4.878 & 4.820 \\
   0.1550 &  C IV            & 5.018 & 4.573 & 4.3109& 4.169 & 4.068 & 3.994 & 3.937 &  3.854 &	
   	3.796 & 3.754 & 3.720 \\
   0.1906 &  C III            & 5.214 & 4.732 & 4.460 & 4.300 & 4.192 & 4.113 & 4.053 & 3.965 &
   	3.903 & 3.858 & 3.822 \\
   0.2175 & 2175                  & 6.937 & 6.088 & 5.643 & 5.388 & 5.219 & 5.096 & 5.002 & 4.866 &
   	4.772 & 4.702 & 4.648 \\
   0.2480 &                          & 4.573 & 4.206 & 3.992 & 3.865 & 3.779 & 3.715 & 3.667 & 3.596 &
   	3.546 & 3.509 & 3.480\\
   0.3000 &                          & 2.967 & 2.827 & 2.738 & 2.684 & 2.646 & 2.618 & 2.596 & 2.564 &
   	2.541 & 2.515 & 2.511 \\
   0.3650 & U                       & 1.924 & 1.880 & 1.849 & 1.830 & 1.817 & 1.807 & 1.799 & 1.787 &
   	1.779 & 1.773 & 1.768 \\
   0.440   & B 			    & 1.000 & 1.000 & 1.000 & 1.000 & 1.000 & 1.000 & 1.000 & 1.000 & 
   	1.000 & 1.000 & 1.000 \\
   0.548   & V 			    & 0.000 & 0.000 & 0.000 & 0.000 & 0.000 & 0.000 & 0.000 & 0.000 & 
   	0.000 & 0.000 & 0.000 \\
   0.720   & R 			    &-0.991 & -1.048 &-1.098 &-1.134 &-1.162 &-1.184 &-1.201 &-1.228 &
   	-1.248 & -1.263 & -1.274 \\
   1.030   & I 			    &-1.871 & -2.035 &-2.197 &-2.323 &-2.424 &-2.507 &-2.575 &-2.682 &
   	-2.761 & -2.823 & -2.872 \\
   1.239   & J 			    &-2.171 & -2.386 &-2.607 &-2.786 &-2.934 &-3.056 &-3.159 &-3.322 &
   	-3.445 & -3.542 & -3.619\\
   1.649   & H 			    &-2.498 & -2.777 &-3.078 &-3.336 &-3.557 &-3.746 &-3.909 &-4.173 &
   	-4.376 & -4.537 & -4.668 \\
   2.192   & K 			    &-2.716 & -3.044 &-3.413 &-3.741 &-4.034 &-4.293 &-4.521 &-4.900 &
   	-5.199 & -5.440 &-5.638 \\
   3.592   & L 			    &-2.932 & -3.312 &-3.759 &-4.179 &-4.575 &-4.944 &-5.283 &-5.877 &
   	-6.371 & -6.783 &-7.128\\
   4.777   & M 			    &-2.992 & -3.388 &-3.859 &-4.310 &-4.745 &-5.158 &-5.547 &-6.250 &
   	-6.852 & -7.364 &-7.800\\
   \hline
	\multicolumn{13}{c}{${\rm A_V=10}$} \\
   \hline
   	0.0912 	& Ly$\infty$	& 12.066 & 9.686 &	8.602 & 8.015 & 7.636 & 7.369 &  7.169 & 
   	6.888 & 6.698 & 6.561 & 6.456\\
	0.1053 	&			&  9.694  & 8.085 &	7.298 & 6.861 & 6.576 & 6.373 & 6.221 & 
   	6.005 & 5.859 & 5.753 & 5.673\\
	0.1216 	& Ly$\alpha$	&  7.328  & 6.374 &	5.866 & 5.575 & 5.383 & 5.244 &5.140 & 	
 	4.991 & 4.890 & 4.816 & 4.760\\
	0.1550 	& C IV		&  5.018 & 4.568 &	4.303 & 4.146 & 4.040 & 3.962 &3.903 & 
   	3.819 & 3.761 & 3.719 & 3.687\\
	0.1906 	& CIII		&  5.214 & 4.726 &	4.443 & 4.275 & 4.162 & 4.080 &4.017 & 
   	3.927 & 3.866 & 3.821 & 3.787\\
	0.2175 	& 2175		&  6.937 & 6.078 &	5.614 & 5.347 & 5.169 & 5.041 &4.945 & 
   	4.859 & 4.713 & 4.645 & 4.593\\
	0.2480 	&			&  4.573 & 4.202 &	3.979 & 3.846 & 3.755 & 3.689 &3.638 & 
   	3.566 & 3.516 & 3.480 & 3.452\\
	0.3000 	& 			&  2.967 & 2.826 &	2.733 & 2.676 & 2.636 & 2.606 &2.584 & 
   	2.551 & 2.528 & 2.511 & 2.499\\
	0.3650 	& U			&  1.924 & 1.879 &	1.848 & 1.827 & 1.813 & 1.803 &1.794 & 
   	1.782 & 1.774 & 1.768 & 1.763\\
	0.440	& B			&  1.000 & 1.000 &	1.000 & 1.000 & 1.000 & 1.000 &1.000 & 
   	1.000 & 1.000 & 1.000 & 1.000\\
	0.548	& V			&  0.000 & 0.000 &	0.000 & 0.000 & 0.000 & 0.000 &0.000 & 
   	0.000 & 0.000 & 0.000 & 0.000\\
	0.720	& R			&  -0.991 & -1.049 & -1.100 & -1.138 & -1.168 & -1.191 &-1.209 &
   	-1.238 &-1.258 &-1.273 &-1.285\\
	1.030	&	I		& -1.871  & -2.038 & -2.201 & -2.333 & -2.441 & -2.529 &-2.602 &
   	-2.716 &-2.799 &-2.863 &-2.913\\
	1.239	&	J		& -2.171 & -2.390 & -2.612 & -2.799 & -2.956 & -3.087 &-3.197 &
   	-3.372 &-3.502 &-3.602 &-3.681\\
	1.649	&	H		& -2.498 & -2.781 &	-3.084 & -3.350 & -3.583 & -3.786 &-3.961 &
   	-4.244 &-4.461 &-4.631 &-4.767\\
	2.192	&	K		& -2/716 & -3.049 &	-3.417 & -3.755 & -4.061 & -4.336 &-4.580 &
   	-4.990 &-5.312 &-5.569 &-5.777\\
	3.592	&	L		& -2.932 & -3.318 &	-3.761 & -4.188 & -4.595 & -4.979 &-5.339 &
   	-5.978 &-6.515 &-6.959 &-7.327\\
	4.777	& 	M		& -2.992 & -3.394 &	-3.860 & -4.316 & -4.758 & -5.185 &-5.593 &
   	-6.342 &-6.993 &-7.544 &-8.006\\
	\enddata
\end{deluxetable}
\clearpage

\clearpage

\clearpage

  \begin{deluxetable}{cccccc}
    \tablecaption{\label{tableattfitcoeff}Coefficients of the Polynomial Approximation of the Attenuation Curve}
   \tablehead{ & \colhead{$b_0$} & \colhead{$b_1$} & \colhead{$b_2$} & \colhead{$b_3$} &
   	\colhead{$b_4$} }
  \startdata
	& \multicolumn{5}{c}{$\rm A_V=1.0$}\\
  \hline
  $a_1$  & 0.00495676 & 3.42939 & -3.62163 & 1.64220 & --- \\
  $a_2$  & 0.0393932  & -0.708645 & -0.668874 & 5.11725 & -2.19900 \\
  $a_3$  & 0.0174931 & -0.484444 & 5.17489 & -19.2137 & 21.5465 \\
  \hline
	& \multicolumn{5}{c}{$\rm A_V=10.0$}\\
  \hline
  $a_1$  & 0.00117052 & 3.49788 & -3.87317 & 1.90329 & --- \\
  $a_2$  & 0.0717661  & -1.40474 & 2.76259 & -1.30223 & 1.96333 \\
  $a_3$  & -0.0163800 & 0.186388 & 1.30772 & -10.5378 & 14.8253   
  \enddata
  \end{deluxetable}

\clearpage
\clearpage

\clearpage
\clearpage

\clearpage
\clearpage

\clearpage
\clearpage

\clearpage
\clearpage

\clearpage
\clearpage

\clearpage
\clearpage

\clearpage
\clearpage

\clearpage

\clearpage

\end{document}